\documentclass[aps,prb,superscriptaddress,nofootinbib,notitlepage,10pt]{revtex4-2}

\usepackage{comment}
\usepackage{natbib}

\usepackage{amsmath,amssymb}
\usepackage{dsfont}

\usepackage{graphicx}
\usepackage{float}
\usepackage{epstopdf}

\usepackage[english]{babel}
\usepackage{setspace}
\usepackage{enumitem}
\usepackage{verbatim}

\usepackage{subfigure}
\usepackage{braket}

\usepackage{xcolor} 
\usepackage{soul}
\definecolor{myyellow}{RGB}{255,255,0} 
\sethlcolor{myyellow}
\usepackage[normalem]{ulem}

\usepackage[title]{appendix}

\usepackage{titlesec}
\titlespacing\subsection{0pt}{12pt plus 4pt minus 2pt}{12pt plus 2pt minus 2pt}

\usepackage[colorlinks=true,linkcolor=magenta,urlcolor=blue,citecolor=blue]{hyperref}

\definecolor{SMblue}{rgb}{0.5,0.8,1}

\definecolor{yellow}{rgb}{0.8,0.8,0.0}

\begin{document}

\title{Stealthy-Hyperuniform Wave Dynamics in Two-Dimensional Photonic Crystals}

\author{Maria Barsukova*}
\affiliation{Department of Physics, The Pennsylvania State University, University Park, PA, USA}
\thanks{These authors contributed equally}

\author{Zeyu Zhang*}
\affiliation{Department of Physics, The Pennsylvania State University, University Park, PA, USA}
\thanks{These authors contributed equally}

\author{Brian Gould*}
\affiliation{Department of Physics, The Pennsylvania State University, University Park, PA, USA}
\thanks{These authors contributed equally}

\author{Koorosh Sadri}
\affiliation{Department of Physics, The Pennsylvania State University, University Park, PA, USA}

\author{Christian Rosiek}
\affiliation{Department of Physics, The Pennsylvania State University, University Park, PA, USA}
\affiliation{Technical University of Denmark, Kongens Lyngby, Denmark}

\author{Søren Stobbe}
\affiliation{Technical University of Denmark, Kongens Lyngby, Denmark}

\author{Jonas Karcher}
\affiliation{Department of Physics, The Pennsylvania State University, University Park, PA, USA}
\affiliation{Princeton University, Department of Electrical and Computer Engineering, Princeton, NJ, USA}

\author{Mikael C. Rechtsman}
\affiliation{Department of Physics, The Pennsylvania State University, University Park, PA, USA}

\date{\today}

\begin{abstract}

Hyperuniform structures are spatial patterns whose fluctuations disappear on long length scales, making them effectively homogeneous when observed from afar.  Mathematically, this means that their spectral density, \( \tilde{\rho}({\bf k}) \), approaches zero for low wavenumber, $|\textbf{k}|$.  Crystalline lattices are hyperuniform, as are certain quasicrystals, maximally random jammed packing of spheres, and electrons in the fractional quantum Hall state.  Stealthy-hyperuniformity is an even stronger constraint on the spectral density: it requires that \( \tilde{\rho}({\bf k}) \) is strictly zero in a finite range of wavevectors around $\mathbf{k}=\mathbf{0}$, called the stealthy regime, or exclusion region.  Since the degree of scattering by disorder is, to leading order, proportional to \( \tilde{\rho}({\bf k}) \), waves propagating through such structures may do so without scattering for sufficiently long wavelengths and short distances.  
Here, we measure scattering by disorder in photonic crystal slabs with stealthy-hyperuniform disorder by measuring the linewidths of the photonic bands. We observe the transition between the stealthy and non-stealthy regimes, marked by a sharp increase in linewidth.  We also observe the effects of multiple scattering in the stealthy regime, which implies diminishing transparency.  Moreover, we show that residual single scattering in the stealthy regime arises from an intrinsically non-Hermitian effect: propagating light has a complex effective mass due to radiative loss out of the slab.

\end{abstract}

\maketitle

\twocolumngrid

In hyperuniform systems, the spectral density associated with the spatial structure of the material in question, \( \tilde{\rho}({\bf k}) \), approaches zero as the wavevector \( {\bf k} \) approaches zero \cite{torquato2003local, torquato2018hyperuniform}.  This contrasts with spatially uncorrelated disorder, for which the averaged spectral density is constant for all wavevectors. Because hyperuniform structures approach homogeneity at long length scales, scattering at leading order is much suppressed, leading to greater transparency at long wavelengths for system sizes on the scale of several scattering lengths or below \cite{leseur2016high, cheron2022wave, kim2023theoretical, kim2023effective}.  

Wave transport in hyperuniform systems is an active field of study \cite{florescu2009designer, man2013isotropic, muller2013silicon, froufe2016role, gkantzounis2017hyperuniform, aubry2020experimental, klatt2022wave, vynck2023light, siedentop2024stealthy} yet is challenging due to the intrinsically large system sizes needed (a result of the long wavelengths -- i.e., small wavevectors -- involved).  As an example, recent work \cite{klatt2022wave} used numerical simulations to probe the nature of a photonic band gap in a hyperuniformly disordered photonic medium, and in particular, the Lifshitz tails decaying into the band gap.  However, it is difficult to observe statistically rare Lifshitz tails in numerical simulations due to system size limitations, so an analytical approach was employed \cite{karcher2024effect}.  Experiments were also carried out in a related context, where the authors examined Anderson localization when the spectrum of the disorder was non-zero only for long wavelengths (the opposite to the stealthy case) \cite{dikopoltsev2022observation}.  In fact, a recent theoretical result posits that under certain conditions, stealthy-hyperuniform disorder may allow for Anderson localization in the vector wave context, whereas uncorrelated disorder may not \cite{monsarrat2021pseudo}.  On the other hand, it has been argued analytically and numerically that stealthy point patterns may exhibit greater transparency than expected from the single-scattering approximation \cite{leseur2016high}, as well as subdiffusive transport in three dimensions \cite{sgrignuoli2022subdiffusive}.  The common theme in much previous work is that a key limitation in understanding the hyperuniform wave dynamics is the ability to numerically simulate very large systems.  It is therefore clear that further progress must be made by the use of experiments, particularly where system sizes exceed those that can be calculated numerically.  Compared to other candidate platforms in solid-state physics and ultracold atoms, photonic crystals are ideal because they can be very large and highly structurally tailorable in fabrication.

Here, we directly probe the interplay between stealthy-hyperuniform disorder and photonic bands, observing the transition from the stealthy to the normal scattering regime.  We demonstrate the presence and measure the degree of multiple scattering in the stealthy regime through higher-order scaling as a function of disorder.  Furthermore, we show that intrinsic non-Hermiticity present in the photonic crystal slab diminishes the sharpness of the transition and allows for single-scattering events in the stealthy regime.  To do this, we experimentally probe large silicon photonic crystal slabs (with on the order of $10^6-10^7$ sites per sample) that have stealthy-hyperuniform disorder imposed as a perturbation upon a two-dimensional square lattice of holes. The structures used here are generated in a different way than typical stealthy-hyperuniform patterns \cite{vynck2023light,morse2023generating} in that they are perturbations to an underlying lattice, rather than being statistically isotropic.  However, they may still be considered stealthy because the spectral density in the stealthy region of the spectrum (also called the `exclusion region') is negligible in experiments, whereas the spectral density outside of it is significant and governs the observed wave dynamics.  Therefore, the geometry considered here represents a very different instantiation of stealthiness compared to the subwavelength point scatterers described in Ref. \cite{leseur2016high}, which found that transparency could be achieved even in the multiple-scattering regime for sufficiently small system size.       

We use an unconventional observable for experiments on disordered photonic systems -- linewidths associated with photonic crystal bands -- in order to quantify the degree of scattering as a function of wavevector and frequency.  This contrasts with the conventional approach, namely, to use transmission \cite{kim2023effective} through a medium to measure the degree of scattering.  We directly observe the transition between stealthy and non-stealthy behavior of the modes in terms of these linewidths.  Furthermore, we examine the dependence of scattering on disorder and find that higher-order scattering effects appear for increasing disorder.  The exact form of this dependence is not known and is very difficult to calculate numerically due to the large system sizes required.

We also observe a direct effect of the photonic non-Hermiticity (i.e., complex effective mass) intrinsic to the slab, namely single scattering within the stealthy regime, which scales with the square of disorder.  The extra scattering arises as a direct result of the fact that for the relevant band, the light propagating in the slab behaves differently than in free space: photonic modes radiate out of the slab with a rate that scales as the square of their momentum.  This is equivalent to a complex effective mass (in analogy with a massive quantum mechanical particle governed by the Schr\"odinger equation) and acts to broaden the linewidths of the photonic crystal modes, even without disorder.  As we demonstrate theoretically, this, in turn, gives rise to scattering within the stealthy regime that we measure directly.   



\begin{figure*}[ht]
  \begin{center}
    \includegraphics[width=18 cm]{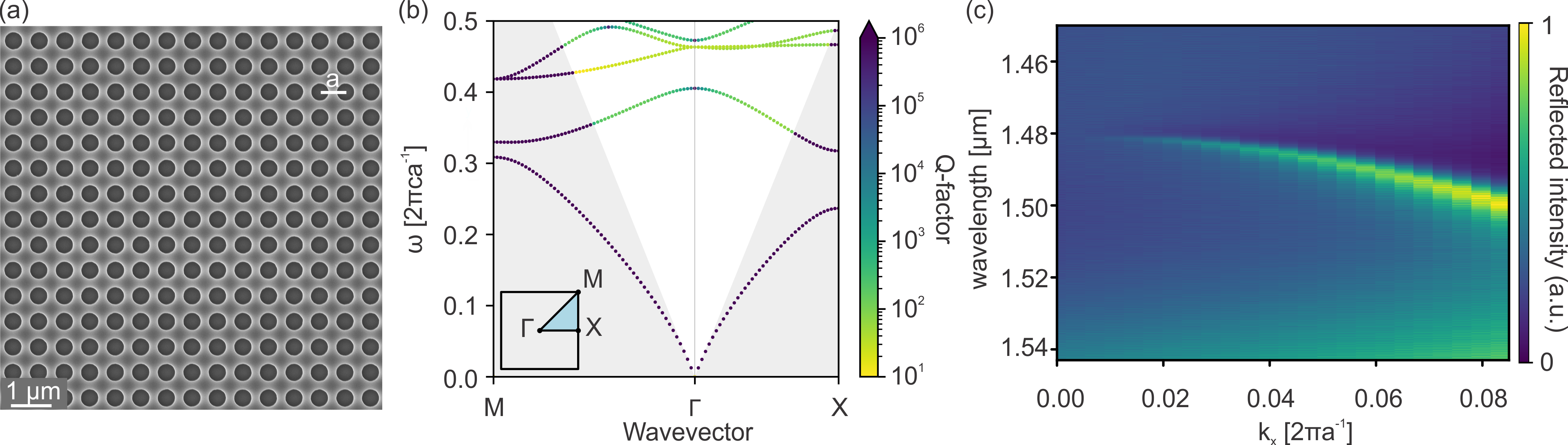}

    \caption{Properties of the clean (i.e., non-disordered) periodic photonic crystal slab. (a) SEM image of photonic crystal slab without disorder. The structure contains circular air holes in a square lattice. (b) Calculated band structure of the photonic crystal slab with circular holes in a square lattice, showing a quadratic TE-like band around $\Gamma$. The color represents the $Q$-factor of the mode. (c) Experimentally-observed band structure (along $k_y=0$) associated with the quadratic band in (b). Color represents the reflection intensity.} 
    \label{figure1}
  \end{center}
\end{figure*}

We start from a two-dimensional photonic crystal structure consisting of circular air holes in a square lattice configuration in a $h=220\ \rm{nm}$ thick silicon slab ($\varepsilon=12.11$) on top of a silica substrate ($\varepsilon=2.25$). The radii of the holes are $r_0=0.32a$ with lattice constant $a=629\ \rm{nm}$. The substrate breaks the up-down mirror symmetry of the structure, but this has a negligible effect on the observables in our experiment because the mode mostly resides in the silicon, and the dielectric constant of the substrate is low. We employed electron beam lithography and inductively coupled plasma reactive-ion etching to fabricate the patterns. Figure \ref{figure1}(a) shows a scanning electron microscope (SEM) image of a typical periodic sample used for the experiment. More details on fabrication methods can be found in Supplementary Information section 6 \cite{HU2025SI}. 

We numerically compute the photonic band structure in the transverse electric (TE)-like polarization using the guided mode expansion method, as implemented in the open-source software package \textsc{Legume} \cite{Legume}, which diagonalizes Maxwell's equations for slab geometry. There is an isolated quadratically dispersing band centered at $\Gamma$ ($\left|\mathbf{k}\right|=0$), as shown in Fig. \ref{figure1}(b).  As a matter of convention and mathematical convenience, below we will describe the dispersion of the band in terms of the eigenvalue of the Maxwell wave operator, namely $E=(\omega/c)^2$, rather than $\omega$ itself.  In analogy with the dispersion of a massive quantum mechanical particle, we refer to this quantity as the `energy' (although it has units $[a^{-2}]$).  The band in Fig. \ref{figure1}(b) is isotropic in the vicinity of $\Gamma$, such that:
\begin{equation}
E_{\mathbf{k}}=E_0-\frac{1}{2m}\left(k_x^2+k_y^2\right),
\end{equation}
where $E_\mathbf{k}$ is the energy of the band at wavevector $\mathbf{k}=\left(k_x,k_y\right)$ and $m$ is the (complex) effective mass. In simulations, the band tip energy is $E_0=6.48a^{-2}$, corresponding to frequency $\omega_0=0.405\ [2\pi c a^{-1}]$, and the effective mass is $\frac{1}{2m}=0.715+0.104i$. We emphasize that the effective mass is a complex number. This is because the band is not only quadratically dispersive in the real part of the energy (or frequency), but also has an $O(\left|\mathbf{k}\right|^2)$-dependent loss associated with the imaginary part (which is related to the out-of-plane radiative loss and proportional to linewidth or inverse quality factor).  This arises due to the symmetry-protected bound state in the continuum (BIC) at $\Gamma$-point \cite{hsu2016bound, koshelev2018asymmetric}, which pins the radiative $Q$-factor to zero at the $\Gamma$. The non-Hermiticity that arises from the imaginary part of the effective mass cannot be neglected since it dominates the linewidth broadening effect in the stealthy region, as we describe later.  


To experimentally characterize the photonic bands, we use angle- and frequency-resolved reflection measurements. The samples are illuminated by a collimated beam from a tunable continuous wave laser (Keysight 81606A) within the wavelength range of $\lambda=1.45-1.65\ \mathrm{\mu m}$. More details about the experimental setup can be found in Supplementary Information section 6 \cite{HU2025SI}. Figure \ref{figure1}(c) shows the measured band structure along the $\Gamma-X$ line in the Brillouin zone. The color represents the reflection intensity. We can see a clear quadratic band with tip energy $E_0=7.11a^{-2}$ (corresponding to $\omega_0=0.424\ [2\pi c a^{-1}]$) with effective mass defined by $\frac{1}{2m}=0.579+0.121i$. Due to the symmetry-protected BIC at $\Gamma$, the radiative linewidth near $\Gamma$ approaches zero, making it difficult to quantify the linewidth or even identify the band. 

\begin{figure*}[ht]
  \begin{center}
    \includegraphics[width=18 cm]{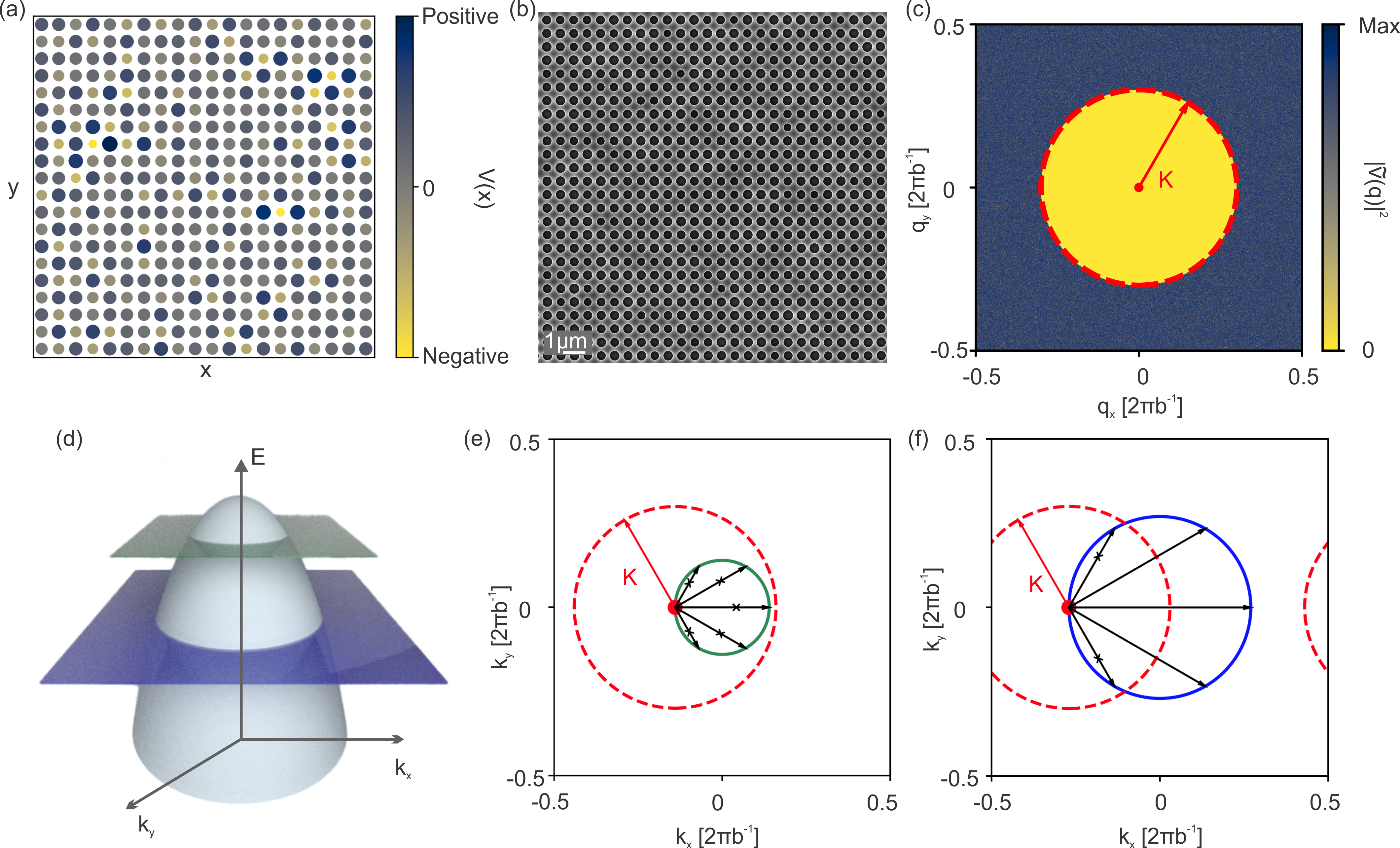}

    \caption{Description of the stealthy-hyperuniform photonic crystal. (a) Schematic diagram of disorder manifested by varying sizes of holes in the sample. The color and size of the holes represent the change of local potential. (b) Zoomed-out SEM image capturing the highly correlated nature of the stealthy-hyperuniform disorder at $K=0.5\left[2\pi b^{-1}\right]$. (c) Schematic diagram of the Fourier components of the local potential $|\tilde{V}(\mathbf{q})|^2$ in $\mathbf{q}$ space. (d) Iso-frequency contours of a quadratic band. (e) An illustration showing why, for small values $|\textbf{k}|$ ($|\textbf{k}| < \frac{K}{2}$), there is no scattering and therefore no linewidth broadening (for single-scattering). (f) An illustration showing that for $|\textbf{k}| > \frac{K}{2}$, scattering is possible to other points on the isofrequency contour, allowing for linewidth broadening due to scattering loss.  Arrows denote scattering events; those marked with an ``$\times$'' are forbidden due to stealthiness.}
    \label{figure2}
  \end{center}
\end{figure*}

Next, we introduce a stealthy-hyperuniform disorder pattern into the periodic photonic crystal slab. We add disorder to the system by randomly varying the size of the air holes.  Previous calculations of stealthy-hyperuniform configurations have not been generated as perturbations on a lattice \cite{morse2023generating}.  However, we see no fundamental distinction -- in terms of wave transport related to stealthiness -- between the perturbed periodic structures used here and others. Figure \ref{figure2}(a) shows an exaggerated version of how we add disorder by changing the radius of the hole at each site. The color and size of the holes in the figure represent the change in radius. We add disorder in such a way that within a $3\times3$ plaquette of holes, the radius change is uncorrelated, but it is correlated in a stealthy-hyperuniform way from plaquette to plaquette (further details in Supplementary Information section 5 \cite{HU2025SI}).  This allows us to carry out the experiment at smaller angles than otherwise; here we use $b=3a$ as the size of the plaquette.   

In analogy with a massive quantum mechanical particle, the disorder that we impose may be seen as a random potential, $V({\bf r}_\alpha)$ that varies in space (${\bf r}_\alpha$ denotes the position of the site index by $\alpha$).  We calibrate how a change in radius corresponds to this potential by calculating how the band edge energy $E_0$ changes with radius in a periodic system; this allows us to quantitatively assign a value to the local potential. To generate the stealthy-hyperuniform disorder, we start from a random and spatially uncorrelated disorder potential in real space.  We use a Fourier filtering method \cite{makse1996method} and set to zero the Fourier component of the local random potential $\tilde{V}(\mathbf{q})$ when $q<K$. This defines a circular exclusion region (yellow area) -- called the stealthy region -- quantified by the cutoff wavenumber $K$, as shown in Fig. \ref{figure2}(c). The spectral density $\tilde{\rho}({\bf q}) =\langle |\tilde{V}(\mathbf{q})|^2\rangle$, which is averaged over disorder configurations, is therefore zero within this region.
We constrain the variance of the disorder in real space (after Fourier filtering) to be $\mathrm{Var}\left(V\right)=\frac{1}{3}V_0^2w_0^2$, where $V_0=2.03a^{-2}$ is a constant describing the linear relation between the local potential and the radius of the corresponding hole.  The spectral density outside of the stealthy region is constant, $\tilde{\rho}({\bf q})=V_0^2 w_0^2/3\left(1-\pi\cdot\left(Kb/2\pi\right)^2 \right) $ for $q>K$. The variance $\mathrm{Var}(V)$ is independent of the cutoff wavenumber $K$, so we define $w_0$ as the degree of disorder, such that the variance of disorder does not change as a function of $K$ at fixed $w_0$. While the potential $V(\mathbf{r})$ is stealthy by construction, the photonic crystal structure, described by the dielectric profile $\varepsilon(\mathbf{r})$ is not.  However, we have confirmed that for the photonic crystal, the ratio of the spectral density in the stealthy to the non-stealthy region stays below a threshold of $10^{-2}$. More details about the Fourier filtering method can be found in Supplementary Information section 2 \cite{HU2025SI}. Figure \ref{figure2}(b) shows an SEM image of a sample with cutoff wavenumber $K=0.5\left[2\pi b^{-1}\right]$. 

While the disorder breaks the periodicity of the system and there is no longer a well-defined Bloch momentum ${\bf k}$, the spectral response function maintains the character of the band, but with additional linewidth broadening due to disorder-induced scattering loss; we use this below to experimentally probe the degree of scattering.  In other words, the bands can still be directly observed experimentally, but with an additional linewidth broadening given by the imaginary part of the self-energy:
\begin{equation}
\mathrm{Im}\left(\Sigma_\mathbf{k}\right)=\frac{1}{N_x N_y}\sum_\mathbf{q} \mathrm{Im}\left(\frac{\tilde{\rho}(\mathbf{q})}{\mathrm{Re}\left(E_{\mathbf{k}}\right)-E_{\mathbf{k+q}}+i0^{+}}\right)+O(\tilde{\rho}^2),
\label{a1}
\end{equation}
where $N_x$ and $N_y$ are the system size in $x$ and $y$ directions, 
and $E_{\mathbf{k}}=E_0-\frac{1}{2m} |{\bf k}|^2$ is the quadratic band.  Each individual term in the summation represents the contribution of a single-scattering event from $\mathbf{k}$ to $\mathbf{k+q}$. The imaginary part of the self-energy is related to the total scattering loss at $\mathbf{k}$ and is therefore proportional to the increase in linewidth for the mode at $\mathbf{k}$.  We use the excess linewidth, which is defined as the linewidth at a given $\mathbf{k}$ in a sample with disorder minus the intrinsic linewidth at $\mathbf{k}$ in a clean periodic sample, as our experimental observable to quantify the effect of disorder. For the sake of simplicity, we will first consider a quadratic Hermitian band $E_{\mathbf{k}}=E_0-\frac{1}{2m} \left|\mathbf{k}\right|^2$ where the effective mass $m$ is real, and leave the discussion about the non-Hermitian effect introduced by the imaginary part of the effective mass $\mathrm{Im}(m)$ for afterward.  

We can see from Eq. \eqref{a1} that the spectral density $\tilde{\rho}(\mathbf{q})$ dictates the scattering amplitude at leading order, which means all single-scattering events from $\mathbf{k}$ to $\mathbf{k+q}$ are forbidden in the stealthy region, where $\tilde{\rho}(\mathbf{q})=0$. As a direct consequence, we find that the excess linewidth has a transition at $\left|\mathbf{k}\right|=K/2$, such that the excess linewidth is zero when $\left|\mathbf{k}\right|<K/2$, but grows rapidly after $\left|\mathbf{k}\right|>K/2$.  This transition may be explained as follows. Here, only elastic scattering (by wavevector ${\bf q}$) from one point on an iso-frequency contour (IFC) to another is allowed.  Two iso-frequency contours (which are analogous to Fermi surfaces) are depicted in Fig. \ref{figure2}(d).  We first consider the green one, which is smaller, and depict this in Fig. \ref{figure2}(e) in momentum space.  If we consider the point marked by a red dot, at $(-k_x,0)$, the mode there may not scatter to any other point on the isofrequency contour (green circle) because it is within the stealthy region defined by $K$ (red-dashed circle).  As a result, the mode does not experience additional loss at leading order.  However, we see in Fig. \ref{figure2}(f) that for points on the larger blue isofrequency contour, elastic single-scattering to other points on the isofrequency contour is indeed possible, leading to additional scattering loss.  Therefore, the transition is predicted to occur when the diameter of the IFC is equal to the radius of the stealthy region, meaning at $\left|\mathbf{k}\right|=K/2$. 

\begin{figure*}[ht]
  \begin{center}
    \includegraphics[width=18 cm]{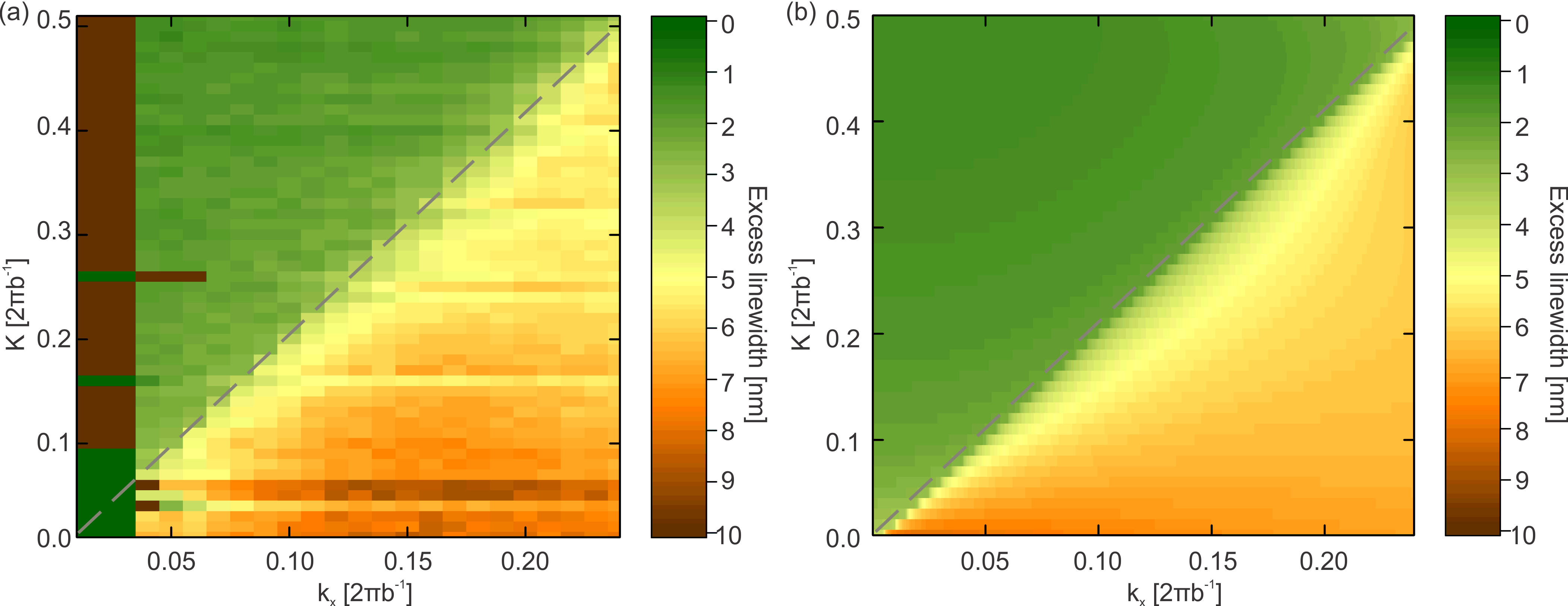}
    \caption{Dependence of excess linewidth on cutoff wavenumber $K$ and $k_x$. (a) Experimental and (b) theoretical phase diagrams showing the extracted excess linewidth (compared to the clean case) as a function of wavevector $k_x$ (along $k_y=0$) and cutoff wavenumber $K$ at the fixed degree of disorder $w_0=0.2$. The stealthiness transition is clearly observable (dashed line) and increases in wavevector linearly with $K$, as expected.}
    \label{figure3}
  \end{center}
\end{figure*}

As described above, we use the excess linewidth (compared to the clean case) as an experimental observable to measure the degree of scattering as a function of both wavevector ${\bf k}=(k_x,k_y)$ and cutoff wavenumber $K$, where we fabricate a new sample for every $K$ while keeping the degree of disorder $w_0=0.2$, so the variance of the disorder remains the same. Specifically, we experimentally extract the linewidth in the clean sample as a function of $k_x$ ($k_y$ is set to zero), do the same for the disordered samples, and then measure the difference in linewidth between them.  This quantity is plotted in Fig. \ref{figure3}(a) as a function of both $k_x$ and $K$.  There is clearly a sharp transition between the green and orange regions (low and high linewidth broadening, respectively) along the gray dashed line.  This is the stealthiness transition: as described above, we expect a sharp increase in scattering at precisely $k_x=K/2$, which is exactly the minimum wavenumber for which single-scattering event can occur as a result of the imposed disorder.  In other words, the stealthy-hyperuniform disorder does not support elastic scattering between modes where $|\mathbf{k}|<K/2$.  As a result, the linewidth does not broaden at the leading order in the scattering expansion (green region).  Past this critical value, scattering sets in, and we immediately observe strong linewidth broadening (orange region).  It is clear from the plot that there is a high degree of noise for low $k_x$; this is because the linewidth is very narrow or vanishing there as a result of the symmetry-protected BIC, as discussed above, making it difficult to extract the linewidth reliably.  The experimental results of Fig. \ref{figure3}(a) may be directly compared to our theoretical predictions based on the single-scattering theory described above, as shown in Fig. \ref{figure3}(b), and as discussed in Supplementary Information section 5 \cite{HU2025SI};  clear agreement is evident between theoretical predictions and experimental results.



\begin{figure*}[ht]
  \begin{center}
    \includegraphics[width=18 cm]{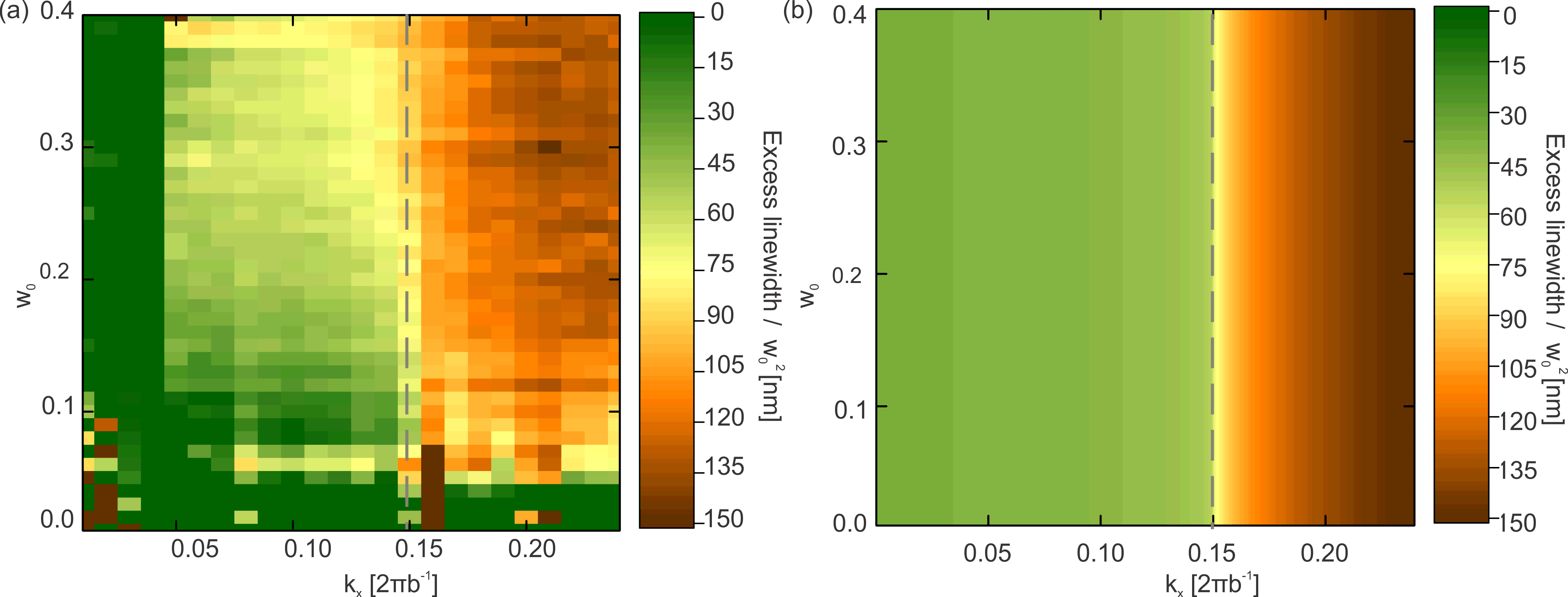}
    \caption{Dependence of excess linewidth on disorder $w_0$ and $k_x$.  (a) Experimental and (b) theoretical phase diagrams showing the extracted excess linewidth normalized by the square of the disorder $w_0^2$ as a function of wavevector $k_x$ (along $k_y=0$) and degree of disorder $w_0$ at fixed cutoff wavenumber $K=0.3\left[2\pi b^{-1}\right]$. The linewidth is proportional to $w_0^2$ to the leading order, and the stealthiness transition is clearly observable (dashed line).}
    \label{figure4}
  \end{center}
\end{figure*}

The spectral density dictates the degree of scattering to leading order in the disorder.  Fig. \ref{figure3} clearly shows that the stealthy-hyperuniformity of the disorder largely governs its scattering behavior as a function of wavevector ${\bf k}$ and cutoff wavenumber $K$ for fixed disorder strength $w_0$. Examining the behavior of the excess linewidth as a function of degree of disorder gives insight into both the effect of stealthiness as well as the extent to which multiple scattering enters (i.e., beyond the leading-order scattering term in Eq. \eqref{a1}).  In Fig. \ref{figure4}(a), we plot excess linewidth as a function of disorder $w_0$ and the component of the wavevector in the $x$ direction, $k_x$.  The linewidth is normalized by the square of the disorder, $w^2_0$ because we expect a quadratic dependence on the disorder for leading-order scattering.  For increasing $k_x$, we clearly observe the phase boundary (dashed vertical line), crossing over from weak scattering (green region) to stronger scattering, associated with larger linewidth (orange).  As in Fig. \ref{figure3}, in the regime of small $k_x$, it is difficult to extract the linewidth due to the BIC, causing the data to be noisy and unreliable there.  Furthermore, for weak disorder ($w_0<0.05$) the imposed disorder is on the scale of the intrinsic fabrication disorder, leading to increased relative error; this is enhanced in Fig. \ref{figure4} because we normalize the enhanced linewidth by $w^2_0$.  In the stealthy regime at left, there is a clear increase of normalized scattering with increasing $w_0$ (becoming more yellow), indicating the effects of multiple scattering.  This increase implies multiple scattering because it corresponds to scaling with disorder at orders higher than $O(w^2_0)$.  This is to be directly compared to Fig. \ref{figure4}(b), the theoretical prediction using single-scattering theory (Supplementary Information section 5 \cite{HU2025SI}), in which there is no such increase of scattering.  The theoretical calculation of the dependence of linewidth on disorder based on multiple scattering is beyond the scope of this paper and represents an open question to theorists of hyperuniform disordered systems.

\begin{figure*}[ht]
  \begin{center}
    \includegraphics[width=18 cm]{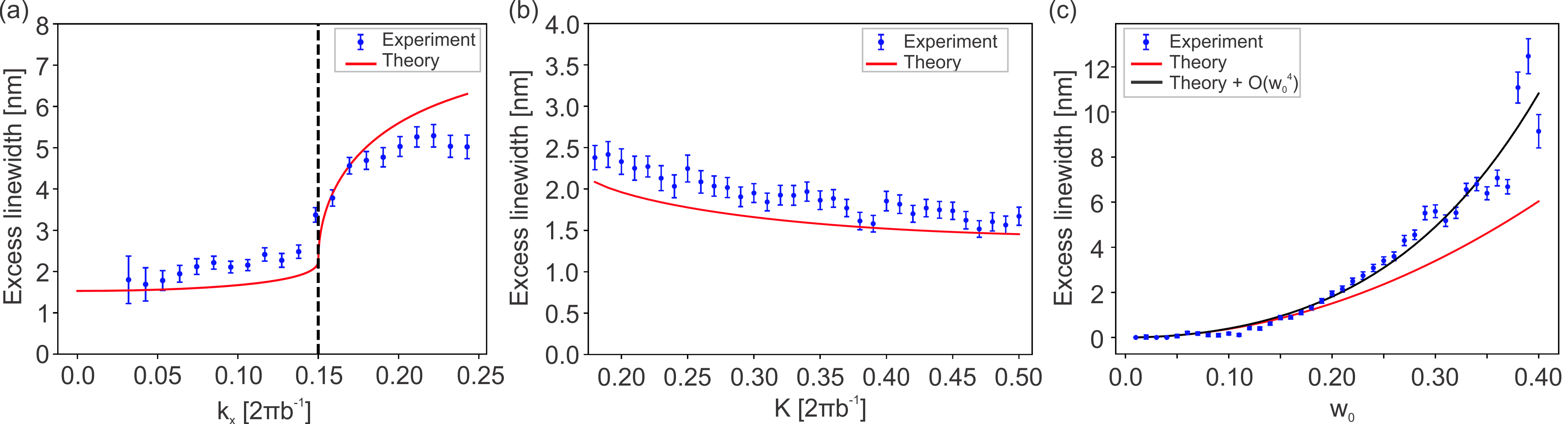}

    \caption{Non-Hermitian effects on excess linewidth due to scattering by the disorder.  Experimental (blue dots)  and theoretical (red line) excess linewidth as a function of (a) wavevector $k_x$ (along $k_y=0$), (b) cutoff wavenumber $K$ and (c) degree of disorder $w_0$. We fix $w_0=0.2$ and $K=0.3\left[2\pi b^{-1}\right]$ in (a), $w_0=0.2$ in (b) and $K=0.3\left[2\pi b^{-1}\right]$ in (c). The excess linewidth is averaged over a finite range $k_x\in \left[0.045,0.09\right]\left[2\pi b^{-1}\right]$ in the stealthy region in (b) and (c). The higher excess linewidth in the experimental results is attributable to higher-order scattering.}
    \label{figure5}
  \end{center}
\end{figure*}

We now turn our attention to the effect of non-Hermiticity, specifically the fact that the effective mass is not purely real. Figure \ref{figure5}(a) shows the experimental data and theoretical results of the excess linewidth at $K=0.3 \left[2\pi b^{-1}\right]$ for $w_0=0.2$. As predicted by the leading-order theory in Eq. \eqref{a1}, the excess linewidth before the transition $k_x<K/2$ is strictly zero when the effective mass is real and grows sharply after $k_x>K/2$. However, we observe in both experiment (blue dots) and theory (red line) that the excess linewidth is not zero but finite in the stealthy regime. Moreover, the transition in the experiment is not as sharp as in theoretical calculations. These are caused by non-Hermicity arising from the imaginary part of the effective mass, as we explain below. 

Recall that in the Hermitian case, allowed single-scattering events are from one point on the IFC to another with the same energy.  In the non-Hermitian case, however, the isofrequency contour is effectively broadened because energies will, in general, be complex and, therefore, have some finite linewidth.  Therefore, there may be scattering between states that have different real parts of their energy.  An alternative way to understand the origin of the non-Hermitian effect is directly from Eq. \eqref{a1}: 
when the effective mass is real, both $E_\mathbf{k}$ and $E_\mathbf{k+q}$ in Eq. \eqref{a1} are real, so only if $E_\mathbf{k}=E_\mathbf{k+q}$ can we obtain a non-zero imaginary part of the self-energy. However, when the effective mass is complex, we get a finite contribution to the imaginary part of the self-energy even when $\mathrm{Re}\left(E_\mathbf{k}\right) \neq \mathrm{Re}\left(E_\mathbf{k+q}\right)$. We show more detailed calculations in the Supplementary Information section 4 \cite{HU2025SI} that the imaginary part of the self-energy in the stealthy regime is proportional to the imaginary part of the effective mass $\mathrm{Im}\left(m\right)$, clearly showing the effect of the non-Hermiticity. 

To quantify the effect of non-Hermiticity in the stealthy regime, we extract the excess linewidth and compare the experimental data to the theoretical predictions, which are discussed in Supplementary Information section 5 \cite{HU2025SI}. We average the excess linewidth over a small interval near $\Gamma$: $k_x \in \left[0.045, 0.09\right]\left[2\pi b^{-1}\right]$ in order to eliminate the influence of the BIC and reduce fluctuations in the experimental data. Figure \ref{figure5}(b) shows a comparison between the experimental data and the leading-order theoretical prediction (calculated using Eq. \eqref{a1} with the complex effective mass extracted from the clean lattice) when we sweep the cutoff wavenumber $K$ with fixed degree of disorder $w_0=0.2$. Here, only the results for $K>0.18\left[2\pi b^{-1}\right]$ are shown because we require the averaging interval to be completely within the stealthy region with $k_x<K/2$. We observe strong quantitative agreement between the experimental data (blue dots) and the theoretical prediction (red line). The experimental excess linewidth is generally slightly higher than the theoretical prediction, which we attribute to the effects of higher-order scattering. Figure \ref{figure5}(c) shows the results of sweeping the degree of disorder $w_0$ while keeping a constant value for $K=0.3\left[2\pi b^{-1}\right]$. The experimental data are consistent with leading-order theory for small disorder ($w_0<0.2$). For large disorder, the experimental excess linewidth grows faster than $O(w_0^2)$ due to higher-order scattering. Upon incorporating a fitted $O(w_0^4)$ term (next order in the scattering expansion, scaling as $O(\tilde{\rho}^2)$) within the leading-order theory, we again observe very good quantitative agreement between experimental data (blue dots) and theory (black line). 



In conclusion, we used large silicon photonic crystal slabs exhibiting stealthy-hyperuniform disorder to experimentally probe the transition to stealthiness and the effect of non-Hermiticity in the form of complex effective mass.  By using linewidth as an experimental probe of the self-energy associated with scattering, we could directly measure the strength of scattering as a function of momentum and energy.  We observed scattering in the stealthy regime that was directly proportional to the imaginary part of the effective mass as well as to the square of the disorder, indicating that this was indeed a non-Hermitian leading order effect.  Furthermore, we found that as we increased disorder, we observed increased scattering in the stealthy regime over and above the non-Hermitian contribution, which we attribute to multiple scattering within the slab.  Due to their large system sizes, photonic crystals can be used to probe further effects of hyperuniformity on wave transport, including in the non-stealthy case when scattering goes smoothly to zero at long wavelength.  They may also be useful in observing Lifshitz tails for measuring rare-region statistics.  The exploration of strongly spatially correlated disorder can potentially lead to an understanding of new phenomena in the transport and localization properties of waves across a wide range of settings, including electronic, photonic, acoustic, and hybrid wave systems, such as polaritons of different types. Because of the combination of large system sizes and customizability, photonic crystals in two and three dimensions provide an ideal testing ground for new physics of this kind that lies beyond the capabilities of numerical simulation.


\section*{Acknowledgments}
We gratefully acknowledge the Nanofabrication Laboratory at the Penn State Materials Research Institute for assistance with sample fabrication (lithography, etching, and deposition) and characterization (field emission scanning electron microscopy, atomic force microscopy, and focused ion beam), particularly Michael Labella, Bangzhi Liu, Guy Lavallee, Shane Miller, Bill Mahoney, and Chad Eichfeld, as well as Kathleen Gehoski and Andrew Fitzgerald, Tim Tighe and Jordan Meyet.  We thank Salvatore Torquato, Paul Steinhardt, and Sachin Vaidya for useful discussions.  We acknowledge the Army Research Office under the MURI program, grant number W911NF-22-2-0103.  We also acknowledge the ONR and AFOSR under grant numbers N00014-20-1-2325 and FA9550-22-1-0339, respectively.  

\onecolumngrid
\clearpage

\clearpage

\section*{Supplemental Material}



\label{supplementary}

\tableofcontents

\renewcommand{\theequation}{S\arabic{equation}}
\renewcommand{\thefigure}{S\arabic{figure}}
\setcounter{equation}{0}
\setcounter{figure}{0}

\clearpage
\section*{Overview} 
\addcontentsline{toc}{section}{Overview}

This Supplementary Information section acts as a companion to the main text, where we provide additional details on theoretical background and experimental methods. \\

In Section 1, we give a brief introduction to the general formalism for disorder. We present the definition of the spectral function $A_\mathbf{k}(E)$ and explain how it is broadened by the imaginary part of self-energy in a disordered system. Next, we show the relation between the spectral function and the linewidth of a mode in a photonic crystal system.\\

In Section 2, we provide more details on how we use the Fourier filtering method to generate a stealthy-hyperuniform disorder configuration in the context of photonic crystals. The local potential distribution forms a stealthy-hyperuniform configuration. The stealthy-hyperuniform pattern is described by a cutoff wavenumber $K$, such that the spectral density $\tilde{\rho}(\mathbf{q})=0$ when $k<K$ ($k=\left|\mathbf{k}\right|$ represents the the magnitude of the wavevector). The stealthiness parameter of the pattern, $\chi$, is then defined as the ratio of constrained degrees of freedom and the total degrees of freedom $\chi=\pi \left(K\frac{a}{2\pi}\right)^2$, where $a$ is the lattice constant. Due to the fact that the local potential and the radius of the hole in our photonic crystal slab are not perfectly linearly related, the spectral density $\tilde{\rho}(\mathbf{q})$ in the stealthy region is only zero at leading order in the degree of disorder. We then quantitatively estimate this effect and prove in Section 3 that it is negligible under experimental conditions.\\

In Section 3, we apply the leading-order scattering theory to a Hermitian quadratic band in a lattice system. We analytically derive the linewidth broadening effect at different wavevectors $\mathbf{k}$. We find that a transition occurs at $k=\frac{K}{2}$, where $K$ is the cutoff wavenumber of the stealthy-hyperuniform pattern. When $k<\frac{K}{2}$, the excess linewidth is strictly zero and then grows rapidly after $k>\frac{K}{2}$. We also estimate the effect of non-zero $\tilde{\rho}(\mathbf{q})$ in the stealthy region and show that it is negligible.\\

In Section 4, we further extend our leading-order theory to a non-Hermitian quadratic band with complex effective mass and complex tip energy. The linewidth transition still occurs at $k=\frac{K}{2}$, but the excess linewidth is no longer zero before the transition $k<\frac{K}{2}$. We explain the origin of this non-Hermitian effect, and show that the transition becomes smoother than in the Hermitian case.\\

In Section 5, we introduce a technique called period-tripling that we used when generating the disorder configuration. The motivation of period-tripling is to observe the transition for large stealthiness $\chi$, while still at small wavevector $k$ and cutoff wavenumber $K$. We add disorder in such a way that within a $3\times3$ plaquette of holes, the hole radii are uncorrelated, but they are correlated in a stealthy-hyperuniform way from plaquette to plaquette. We find that the transition still occurs at $k=\frac{K}{2}$, but the stealthiness parameter increases to $\chi=\pi \left(K\frac{b}{2\pi}\right)^2$ with $b=3a$ defined as the lattice constant of the plaquette. Moreover, a second transition emerges at $k_x=\frac{\pi}{3a}-\frac{K}{2}$ along the $k_y=0$ line in the Brillouin zone.\\

In Section 6, we introduce the fabrication method for the photonic crystal slabs. Then, we describe the experimental setup that characterizes the photonic modes of the sample by measuring its reflection intensity. Finally, we present the method for extracting linewidth from the reflection intensity spectrum obtained in the experiment.

\clearpage

\section*{Section 1: Formalism and notation: from disorder to photonic crystal} 
\addcontentsline{toc}{section}{Section 1: Formalism and notation: from disorder to photonic crystal}
In this section, we will introduce the notation of disorder and the basic formalism for adding disorder in the context of photonic crystals.

In a 2D photonic crystal slab, the eigenmodes are described by Maxwell's equations:
\begin{equation}
\mathcal{L} \mathbf{H_k}=E_\mathbf{k} \mathbf{H_k},
\label{a01}
\end{equation}
where $\mathcal{L}=\mathbf{\nabla}\times \left[\varepsilon(\mathbf{x})^{-1}\mathbf{\nabla}\times\cdot\right]$ is the Maxwell operator with $\varepsilon(\mathbf{x})$ the dielectric function, $\mathbf{k}=\left(k_x, k_y\right)$ is the 2D lattice quasi-momentum, $\mathbf{H_k}$ is the magnetic field profile of the eigenmode, and $E_\mathbf{k}=\left(\frac{\omega_\mathbf{k}}{c}\right)^2$ is the energy of the eigenmode, where $\omega_\mathbf{k}$ is the eigenfrequency of the mode and $c$ is the speed of light.  Two-dimensional photonic crystal slabs have both guided (i.e., bound) and radiative (i.e., resonant) modes that are below and above the light line respectively \cite{photonicbook}. 
 In our work, we focus on radiative modes that are above the light line, centered on $\textbf{k}=\Gamma=(0,0)$.  The radiative loss associated with these modes (as well as absorption loss, which is significantly smaller in our system) may be treated as an imaginary part of the energy $E_\mathbf{k}$ in the 2D description.  From this point onward, we refer this imaginary part of $E_\mathbf{k}$ in the clean (periodic without disorder) case as ``intrinsic loss" to distinguish it from the extra scattering loss caused by the disorder.            


The Green's function in the clean case can be defined as:
\begin{equation}
G^0_{\mathbf{k},\mathbf{k'}}(E)=\left<\mathbf{H_k}\left|\frac{1}{E-\mathcal{L}+i0^{+}}\right|\mathbf{H_{k'}}\right>,
\label{a2}
\end{equation}
and the spectral function is:
\begin{equation}
A_{\mathbf{k}}^{0}(E)=\mathrm{Im}\left(G^0_{\mathbf{k},\mathbf{k}}(E)\right).
\label{a3}
\end{equation}

For simplicity, the $i0^{+}$ term is omitted from this point onward. From Eq. \eqref{a01}, we can see that the spectral function can be written in terms of the real and imaginary parts of the energy $E_\mathbf{k}$:
\begin{equation}
A_{\mathbf{k}}^{0}(E)=\mathrm{Im}\left(\frac{1}{E-E_\mathbf{k}}\right)=\frac{\mathrm{Im}\left(E_\mathbf{k}\right)}{\left(E-\mathrm{Re}\left(E_\mathbf{k}\right)\right)^2+\left(\mathrm{Im}\left(E_\mathbf{k}\right)\right)^2},
\label{a4}
\end{equation}
which means $A^{0}_{\mathbf{k}}(E)$ forms a Lorentzian lineshape where the peak is centered at $E=\mathrm{Re}\left(E_\mathbf{k}\right)$, with linewidth $-2\mathrm{Im}\left(E_\mathbf{k}\right)$.

Then, a weak disorder $V(\mathbf{x})$ that breaks the translational symmetry is added to the system; in other words, $\mathbf{k}$ is no longer a good quantum number. The spectral density of the system can be defined as:
\begin{equation}
\tilde{\rho}(\mathbf{q})=\left<\tilde{V}(\mathbf{q})\tilde{V}(\mathbf{-q})\right>_{\mathrm{dis}},
\label{a5}
\end{equation}
where $\tilde{V}(\mathbf{q})$ is the Fourier component of $V(\mathbf{x})$ corresponding to momentum $\mathbf{q}$, and $\left<\cdot\right>_{\mathrm{dis}}$ represents the average over disorder configurations.

According to disorder theory \cite{lee1985disordered}, the averaged Green’s function can always be written in terms of the self-energy:
\begin{equation}
\left<G_{\mathbf{k},\mathbf{k}}\left(E\right)\right>_{\mathrm{dis}}=\frac{1}{\left(E-E_\mathbf{k}\right)-\Sigma_\mathbf{k}(E)},
\label{a6}
\end{equation}
where $\Sigma_\mathbf{k}(E)$ is the self-energy:
\begin{equation}
\Sigma_\mathbf{k}(E)=\frac{1}{N_x N_y}\sum_\mathbf{q} G^0_{\mathbf{k+q},\mathbf{k+q}}(E) \tilde{\rho}(\mathbf{q})+O(\tilde{\rho}^2),
\label{a7}
\end{equation}
where $N_x$ and $N_y$ are the numbers of unit cells in $x$ and $y$ directions. The first term is the leading-order contribution proportional to the spectral density $\tilde{\rho}(\mathbf{q})$, which represents a single-scattering event from $\mathbf{k}$ to $\mathbf{k+q}$. The next term is $O(\tilde{\rho}^2)$, which represents double-scattering events, and so on.

From Eq. \eqref{a6}, the spectral function in the disorder system becomes:
\begin{equation}
A_{\mathbf{k}}(E)
=\mathrm{Im}\left<G_{\mathbf{k},\mathbf{k}}\left(E\right)\right>_{\mathrm{dis}}
=\frac{\mathrm{Im}\left(E_\mathbf{k}\right)+\mathrm{Im}\left(\Sigma_\mathbf{k}(E)\right)}
{\left(E-\mathrm{Re}\left(E_\mathbf{k}\right)-\mathrm{Re}\left(\Sigma_\mathbf{k}(E)\right)\right)^2+\left(\mathrm{Im}\left(E_\mathbf{k}\right)+\mathrm{Im}\left(\Sigma_\mathbf{k}(E)\right)\right)^2},
\label{a8}
\end{equation}
After adding disorder, the spectral function still has a Lorentzian lineshape, but the center energy is shifted to $E=\mathrm{Re}\left(E_\mathbf{k}\right)+\mathrm{Re}\left(\Sigma_\mathbf{k}(E)\right)$, with linewidth $-2\mathrm{Im}\left(E_\mathbf{k}\right)-2\mathrm{Im}\left(\Sigma_\mathbf{k}(E)\right)$. Compared with Eq. \eqref{a4}, we can see that the width of the peak is broadened from $-2\mathrm{Im}\left(E_\mathbf{k}\right)$ in the non-disordered case (which we call the intrinsic linewidth) to $-2\mathrm{Im}\left(E_\mathbf{k}\right)-2\mathrm{Im}\left(\Sigma_\mathbf{k}(E)\right)$ after adding disorder. We refer the extra linewidth, $-2\mathrm{Im}\left(\Sigma_\mathbf{k}(E)\right)$, as the linewidth broadening or excess linewidth, which is directly caused by the scattering loss from disorder.


The energy $E_\mathbf{k}$ and the frequency $\omega_\mathbf{k}$ can be written in their real part and imaginary part:
$E_\mathbf{k}=E_r+iE_i$ and $\omega_\mathbf{k}=\omega_r+i\omega_i$. Given that $E_\mathbf{k}=\left(\frac{\omega_\mathbf{k}}{c}\right)^2$ and assuming that $\omega_r\gg \omega_i$, we get:
\begin{equation}
E_r=\frac{\omega_r^2}{c^2},
\label{a9}
\end{equation}
\begin{equation}
E_i=\frac{2\omega_r\omega_i}{c^2}.
\label{a10}
\end{equation}
In order to find the center wavelength ($\lambda_0$) and linewidth ($\Delta \lambda$) of the mode, we notice that the ratio of the linewidth to the resonant value of the wavelength and frequency are the same:
\begin{equation}
\frac{\Delta \lambda}{\lambda_0}=\left|\frac{\Delta \omega}{\omega_0}\right|=2\left|\frac{\omega_i}{\omega_r}\right|,
\label{a11}
\end{equation}
\begin{equation}
\lambda_0=\frac{2\pi c}{\omega_0}.
\label{a12}
\end{equation}
Writing the linewidth of wavelength $\Delta \lambda$ in terms of the energy, we find: 
\begin{equation}
\Delta \lambda=-\frac{2\pi E_i}{E_r^{\frac{3}{2}}}.
\label{a13}
\end{equation}
In the limit of weak disorder, the real part of the intrinsic energy is much larger than the real part of the self-energy; that is, $\mathrm{Re}\left(E_\mathbf{k}\right)\gg \mathrm{Re}\left(\Sigma_\mathbf{k}(E)\right)\propto \left\|V\right\|^2$ where $\left\|V\right\|$ represents the degree of disorder. In this limit, the excess linewidth in wavelength can be calculated from the self-energy and the intrinsic energy as: 
\begin{equation}
\Delta \lambda_\mathbf{k}=-\left.\frac{2\pi\mathrm{Im} \left(\Sigma_\mathbf{k}\left(E\right)\right)}{E^{\frac{3}{2}}}\right|_{E=\mathrm{Re}\left(E_\mathbf{k}\right)}.
\label{a14}
\end{equation}

Next, we will use Eq. \eqref{a7} and Eq. \eqref{a14} to calculate the excess linewidth, for different band dispersions, spectral densities, and approximations.

\clearpage

\section*{Section 2: Fourier filtering and spectral density} 
\addcontentsline{toc}{section}{Section 2: Fourier filtering and spectral density}

In this section, we will introduce how the stealthy-hyperuniform disorder configuration is generated by the Fourier filtering method and calculate the spectral density $\tilde{\rho}(\mathbf{q})$ as a function of disorder.

In the experiment, our goal is to generate a stealthy-hyperuniform disorder pattern such that:
\begin{equation}
\tilde{\rho}(\mathbf{q})
\left\{
\begin{aligned}
=0 & , & q<K\\
\neq 0 & , & q>K
\end{aligned}
\right.
\label{e1}
\end{equation}
where $q=\left|\mathbf{q}\right|$ is the magnitude of $\mathbf{q}$. In other words, we want to create a circular scattering-free region in the Brillouin zone described by all $\mathbf{q}$ satisfying:
\begin{equation}
0<q<K.
\label{e2}
\end{equation}

Here, $K$ describes the size of the stealthy region in the Brillouin zone; we therefore call it the cutoff wavenumber. For this section, only the case $K\leq \frac{1}{2}\left[\frac{2\pi}{a}\right]$ is considered, so the stealthy region is always a complete circle in the first Brillouin zone.
\begin{figure}[H]
    \centering
    \subfigure[]{\includegraphics[width=0.23\textwidth]{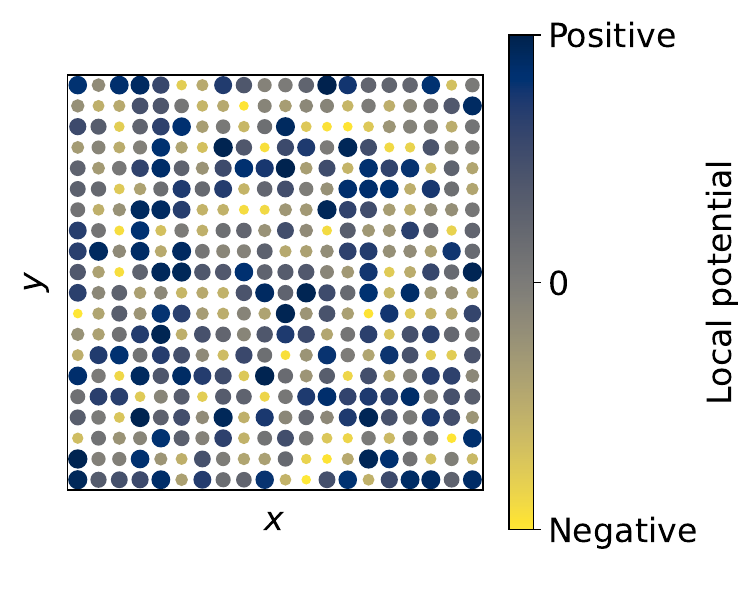}}
    \subfigure[]{\includegraphics[width=0.23\textwidth]{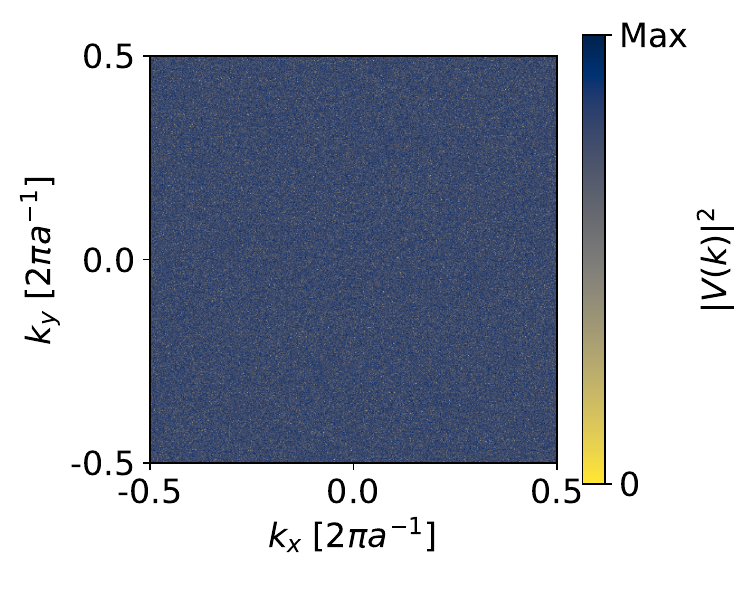}}
    \subfigure[]{\includegraphics[width=0.23\textwidth]{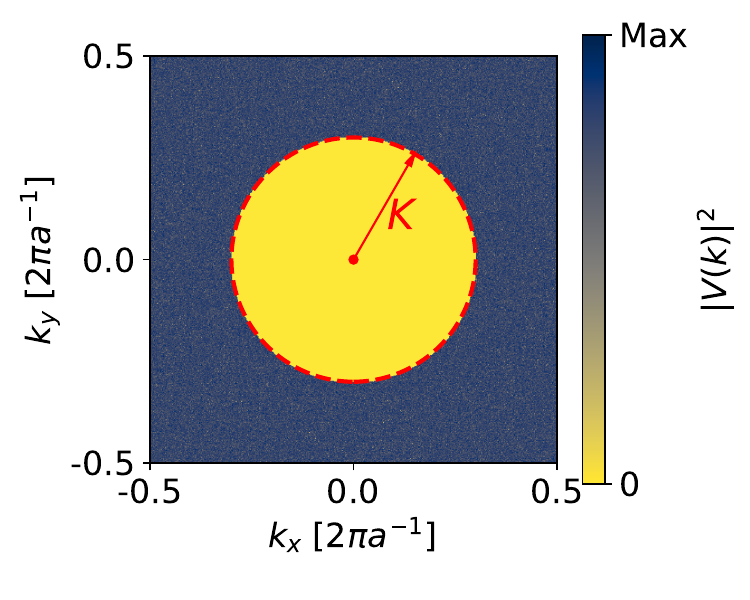}}
    \subfigure[]{\includegraphics[width=0.23\textwidth]{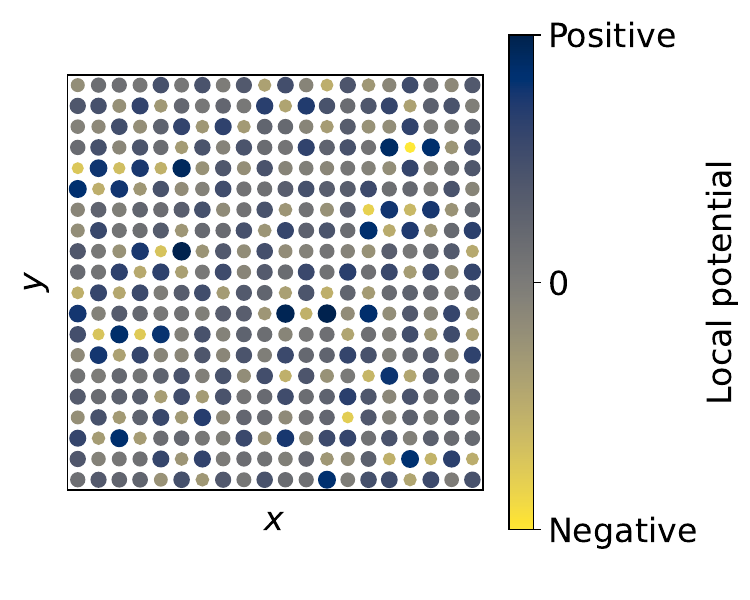}}

\caption{The Fourier filtering method. (a) Local potential configuration with uncorrelated disorder. The color and the size of the holes represent the local potential change at each site with respect to the periodic (without disorder) case. Only an area of $20\times 20$ unit cells is shown. (b) The discrete Fourier transform of the disorder pattern in (a). (c) The Fourier filtering method is used to manually filter out the Fourier components in the stealthy region and create a circular scattering-free region (yellow area). (d) Inverse Fourier transformation is used to convert (c) into a real space disorder configuration.}
\label{figs5}
\end{figure}

Figure \ref{figs5} shows the Fourier filtering method \cite{makse1996method} used in the experimental samples to generate such stealthy-hyperuniform disorder patterns in Eq. \eqref{e1}. We start from a random discrete potential disorder configuration in a square lattice such that the potential difference $V_{l,j}$ at each site $\left(l,j\right)$ is randomly and uniformly selected from the range $\left[-V,V\right]$, where $V$ represents the degree of disorder:
\begin{equation}
V_{l,j} \sim U\left(-V, V\right),
\label{e3}
\end{equation}
where $U$ denotes a uniform distribution. Then, the Fourier component of the local potential $V_{l,j}$ is calculated by discrete Fourier transform as shown in Fig. \ref{figs5}(b):
\begin{equation}
\tilde{V}(\mathbf{k})=\frac{1}{\sqrt{N_x N_y}}\sum_{l,j}e^{-i \left(l k_x a+j k_y a\right)} V_{l,j}.
\label{e4}
\end{equation}

Note that $V_{l,j}\in\mathbb{R}$, which means $\tilde{V}(-\mathbf{k})=\left(\tilde{V}(\mathbf{k})\right)^{*}$, or equivalently, $\tilde{V}(-\mathbf{k})\tilde{V}(\mathbf{k})=\left|\tilde{V}(\mathbf{k})\right|^2$. 

Since the local potential $V_{l,j}$ at different sites $(l,j)$ are independent and drawn from a uniform distribution, $V_{l,j}\sim U\left(-V, V\right)$, it is straightforward to prove that both the real part and the imaginary part of $\tilde{V}\left(\mathbf{k}\right)$ satisfy a normal distribution for large system size $N_x\times N_y$:
\begin{equation}
\mathrm{Re} \left(\tilde{V}(\mathbf{k})\right)\sim N\left(0,\frac{1}{6}V^2\right),
\label{e5}
\end{equation}
\begin{equation}
\mathrm{Im} \left(\tilde{V}(\mathbf{k})\right)\sim N\left(0,\frac{1}{6}V^2\right),
\label{e6}
\end{equation}
where $N$ denotes a normal distribution. Therefore,
\begin{equation}
\tilde{\rho}(\mathbf{q})
=\left<\tilde{V}(-\mathbf{q}) \tilde{V}(\mathbf{q)}\right>=
\left<\left|\tilde{V}(\mathbf{q})\right|^2\right>=\frac{1}{3}V^2.
\label{e7}
\end{equation}

Next, all the Fourier components $\tilde{V}(\mathbf{k})$ in the stealthy region $k<K$ are manually set to be zero, so a circular scatter-free region is created as shown in Fig. \ref{figs5}(c):
\begin{equation}
\tilde{V}^{\prime}(\mathbf{k})=
\left\{
\begin{aligned}
0 & , & k<K\\
\tilde{V}(\mathbf{k}) & , & k>K
\end{aligned}
\right.
\label{e8}
\end{equation}

Here,  $\tilde{V}^{\prime}(\mathbf{k})$ denotes for the Fourier component after the Fourier filtering. After the filtering, $\tilde{\rho}(\mathbf{q})$ remains the same outside of the stealthy region, but becomes zero when $q<K$:
\begin{equation}
\tilde{\rho}(\mathbf{q})=
\left\{
\begin{aligned}
0 & , & q<K\\
\frac{1}{3}V^2 & , & q>K
\end{aligned}
\right.
\label{e9}
\end{equation}

Finally, the inverse discrete Fourier transformation is performed to transform $\tilde{V}^{\prime}(\mathbf{k})$ back into real space as shown in Fig. \ref{figs5}(d):
\begin{equation}
V^{\prime}_{l,j}=\frac{1}{\sqrt{N_x N_y}}\sum_{\mathbf{k}}e^{i \left(l k_x a + j k_y a\right)} \tilde{V}^{\prime}(\mathbf{k}).
\label{e10}
\end{equation}

Since the filtering in Eq. \eqref{e8} does not change the fact that $\tilde{V}^{\prime}(-\mathbf{k})=\left(\tilde{V}^{\prime}(\mathbf{k})\right)^{*}$, the local potential after the filtering is still a real number $V^{\prime}_{l,j}\in\mathbb{R}$.

For a stealthy-hyperuniform pattern, the stealthiness parameter $\chi$ is defined as the ratio between the constrained degrees of freedom and the total degrees of freedom. For the stealthy-hyperuniform pattern we generated by the Fourier filtering method described in Eqs. \eqref{e9} and \eqref{e10}, it's straightforward to prove that:
\begin{equation}
\chi=\pi\left(K\frac{a}{2\pi}\right)^2 .
\label{e10-2}
\end{equation}

Since we require $K<\frac{1}{2}\left[\frac{2\pi}{a}\right]$, only $\chi<\frac{\pi}{4}$ is considered in this paper.

\begin{figure}[H]
    \centering
    \subfigure[]{\includegraphics[width=0.23\textwidth]{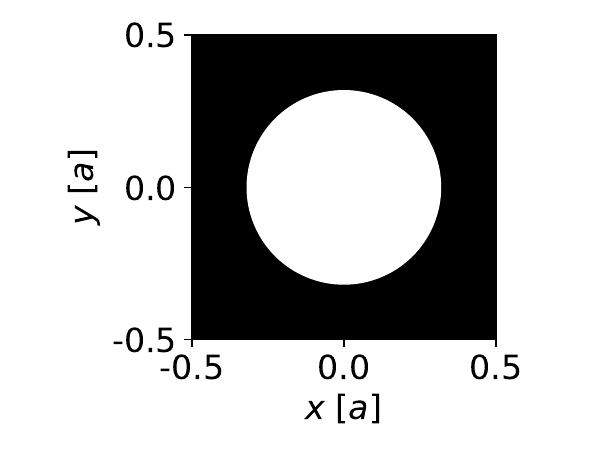}}
    \subfigure[]{\includegraphics[width=0.23\textwidth]{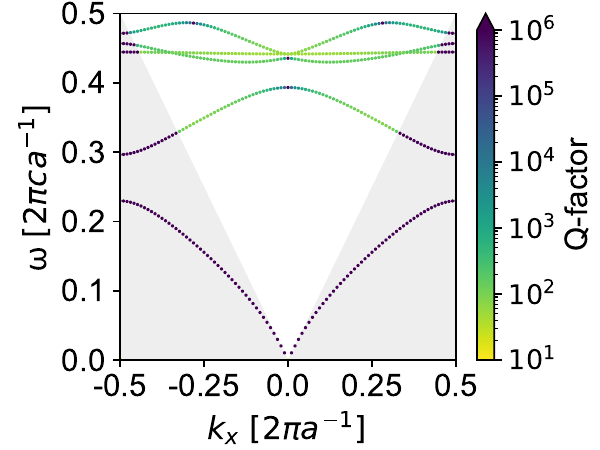}}
    \subfigure[]{\includegraphics[width=0.23\textwidth]{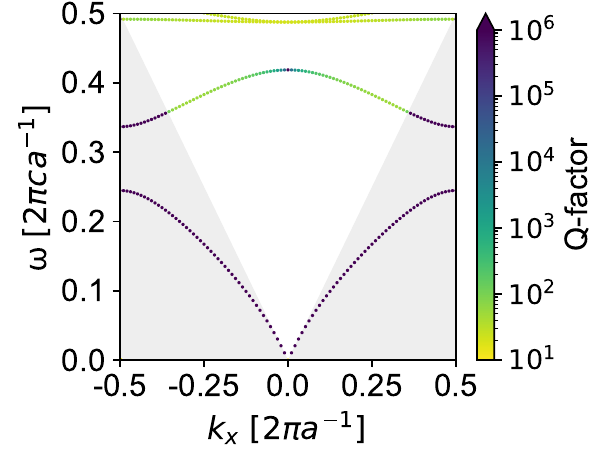}}
    \subfigure[]{\includegraphics[width=0.23\textwidth]{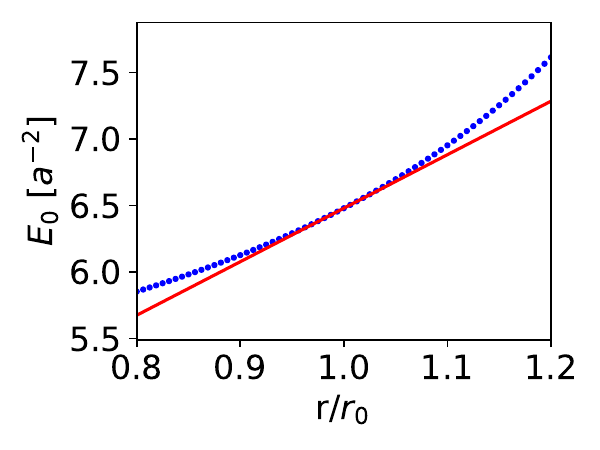}}

\caption{The method to add disorder into a photonic crystal slab. (a) The unit cell (in the $x-y$ plane, $z$ is the out-of-plane direction) of our photonic crystal slab. The structure contains a circular air hole ($\varepsilon=1.0$) with $r_0=0.32a$ in a square lattice, where $a$ is the lattice constant. The slab is made of silicon ($\varepsilon=12.11$) with thickness $h=0.35a$ and sits on top of a silica substrate ($\varepsilon=2.25$). (b) The simulated band structure of (a) along the $k_y=0$ line, but with $r=0.9r_0=0.288a$. Only the transverse electric (TE)-like modes are shown. (c) The simulated band structure of (a), but with $r=1.1r_0=0.352a$. (d) The tip energy ($E=\left(\frac{\omega}{c}\right)^2$) of the quadratic band as a function of air hole radius (blue dots). The red line shows the linear approximation near $r=r_0$.}
\label{figs6}
\end{figure}

Next, we translate the idea of a``local potential" outlined in the first part of Section 2 to the context of the photonic crystal with dielectric function $\varepsilon(\textbf{x})$. Figure \ref{figs6}(a) shows the photonic crystal slab structure used in the experiment. The structure contains circular air holes arranged to form a square lattice in a silicon slab. The structure hosts an isolated quadratic band in the vicinity of $\textbf{k}=\Gamma$. We find that the band tip energy changes as we vary the radius of the air hole. In Fig. \ref{figs6}(b), the radius of the hole is reduced to $r=0.9r_0$. The band tip energy decreases to $E_0=6.13a^{-2}$ (compared to $E_0=6.48a^{-2}$ at $r=r_0$). While in Fig. \ref{figs6}(c), as the radius expands to $r=1.1r_0$, the band tip energy increases to $E_0=6.95a^{-2}$. More generally, we show in Fig. \ref{figs6}(d) that the tip energy increases as the radius of the hole increases. In Fig. \ref{figs6}(d), we can see that for weak disorder (i.e., $\frac{r_{l,j}-r_0}{r_0}\ll1$), the tip energy approximately increases as a linear function of $r$ (red line). The simulation results are obtained by using the guided mode expansion method as implemented in the open-source software package \textsc{Legume} \cite{Legume}. 

The above result indicates that a potential disorder can be added to a photonic crystal by varying the radii of the holes $r_{l,j}$ at each site $(l,j)$, such that the local potential (which is defined as the tip energy corresponding to $r_{l,j}$ in the periodic structure) $V_{l,j}$ forms a pattern that is generated by Fourier filtering in Eq. \eqref{e10}. In the experiment, the degree of disorder $w$ is defined as the variation range of the local potential before the Fourier filtering:
\begin{equation}
V_{l,j}\in\left[E_0-wV_0,E_0+wV_0\right],
\label{e13}
\end{equation}
where $V_0$ describes the linear relation between the local potential and the radius of the hole in the photonic crystal slab: 
\begin{equation}
V_0=\lim_{\Delta r \to 0}\frac{\Delta V}{2\Delta r/r_0} =2.03a^{-2},
\label{e14}
\end{equation}
so the radii are in the range of $r_{l,j}\in\left[r_0\left(1-\frac{w}{2}\right),r_0\left(1+\frac{w}{2}\right)\right]$. 

Comparing Eq. \eqref{e13} to Eq. \eqref{e3}, we directly obtain the spectral density under disorder $w$ by replacing the $V$ in Eq. \eqref{e9} to $wV_0$:
\begin{equation}
\tilde{\rho}(\mathbf{q})=
\left\{
\begin{aligned}
0 & , & q<K\\
\frac{1}{3}w^2V_0^2 & , & q>K
\end{aligned}
\right.
\label{e15}
\end{equation}

Due to the properties of the discrete Fourier transform, $\tilde{\rho}(\mathbf{q})$ should be periodic in both the $q_x$ and $q_y$ directions: $\tilde{\rho}\left(q_x+\frac{2\pi}{a},q_y\right)=\tilde{\rho}\left(q_x,q_y+\frac{2\pi}{a}\right)=\tilde{\rho}\left(q_x,q_y\right)$. The variance of the potential after the filtering is:
\begin{equation}
\mathrm{Var}\left(V_{l,j}\right)=\frac{1}{3}\left( 1-\pi\left(K\frac{a}{2\pi}\right)^2\right) w^2 V_0^2.
\label{e11}
\end{equation} 

The variance depends on the cutoff wavenumber $K$. Therefore, in order to keep the variance the same for different values of $K$, the disorder $w$ at each $K$ should be different, such that:
\begin{equation}
w(K)=\frac{1}{\sqrt{1-\pi\left(K\frac{a}{2\pi}\right)^2}}w_0,
\label{e12}
\end{equation}
where $w_0$ is the degree of disorder normalized at $K=0$ to have the same variance. In the experiment, when $K$ is swept, the variance of the potential in real space remains the same (that is, keep $w_0$ the same) for each
individual $K$ sample, such that $w(K)=\frac{1}{\sqrt{1-\pi\left(K\frac{a}{2\pi}\right)^2}}w_0$ is larger for larger $K$. We explain the reason in further detail in Section 5. From this point onward, the normalized value $w_0$ is used to quantify the degree of disorder unless otherwise specified. The spectral density in Eq. \eqref{e15} becomes:
\begin{equation}
\tilde{\rho}(\mathbf{q})=
\left\{
\begin{aligned}
0 & , & q<K\\
\frac{w_0^2V_0^2}{3\left(1-\pi\left(K\frac{a}{2\pi}\right)^2\right)} & , & q>K
\end{aligned}
\right.
\label{e15-2}
\end{equation}

In the above derivation, a discrete local potential configuration that is strictly stealthy-hyperuniform is generated using a discrete Fourier transformation. However, the photonic crystal is a continuous medium that has the dielectric function $\varepsilon(\mathbf{x})$ where $\mathbf{x}$ can take continuous values. This means that additional approximations have been made when transferring the radius of the hole $r_{l,j}$ to the local potential $V_{l,j}$. For example, not only the tip energy is shifted, but also the effective mass (both the real part and imaginary part) is inevitably changed when the radius is varied. Next, this effect can be estimated by directly calculating the Fourier transform of $\varepsilon(\mathbf{x})$ in the stealthy region, instead of the discrete Fourier transform of local potential $V_{l,j}$. We define:
\begin{equation}
\tilde{\varepsilon}(\mathbf{k})
=\iint d\mathbf{x}\ e^{-i \mathbf{k}\cdot\mathbf{x}} \varepsilon(\mathbf{x}),
\label{e16}
\end{equation}
and
\begin{equation}
\varepsilon(\mathbf{x})=
\left\{
\begin{aligned}
\varepsilon_1 & , & \mathop{\min}_{l, j}\left|\mathbf{x}-\mathbf{x}_{l,j}\right|<r_{l,j}\\
\varepsilon_2 & , & \mathrm{else}
\end{aligned}
\right.,
\label{e17}
\end{equation}
where $\varepsilon_1=1.0$ is the dielectric constant of air, $\varepsilon_2=12.11$ is the dielectric constant of silicon, and $\mathbf{x}_{l,j}=(l\cdot a,j\cdot a)$ is the center position of the air hole at site $(l,j)$

Hence, we get:
\begin{equation}
\tilde{\varepsilon}(\mathbf{k})
=\frac{2\pi\left(\varepsilon_1-\varepsilon_2\right)}{k}\sum_{l,j} r_{l,j}\cdot J_{1}\left(r_{l,j}k\right)\cdot e^{-i \mathbf{k}\cdot\mathbf{x}_{l,j}},
\label{e18}
\end{equation}
where $J_n(x)$ is the Bessel function of the first kind. We know from Fig. \ref{figs6}(d) that the local potential and the radius of the hole are linearly related for small disorder:
\begin{equation}
r_{l,j}-r_0=\frac{1}{2V_0}\left(V_{l,j}-V\left(r_0\right)\right)+O\left(w_0^2\right).
\label{e19}
\end{equation}

Moreover, according to the Taylor expansion:
\begin{equation}
r_{l,j}\cdot J_{1}\left(r_{l,j}k\right)
=r_0J_1\left(r_0k\right)
+\left(2J_1\left(r_0k\right)-r_0kJ_2\left(r_0k\right)\right)\left(r_{l,j}-r_0\right)
+O\left(r_{l,j}-r_0\right)^2.
\label{e20}
\end{equation}

Therefore,
\begin{equation}
\tilde{\varepsilon}(\mathbf{k})
=\frac{\pi\left(\varepsilon_1-\varepsilon_2\right)\left(2J_1\left(r_0k\right)-r_0kJ_2\left(r_0k\right)\right)}{V_0k}\sum_{l,j} V_{l,j}\cdot e^{-i \mathbf{k}\cdot\mathbf{x}_{l,j}}+O\left(w_0^2\right)
=O\left(w_0^2\right).
\label{e21}
\end{equation}

We can see that, if the discrete Fourier transform of the local potential is zero (that is, $\sum_{l,j} V_{l,j}\cdot e^{-i \mathbf{k}\cdot\mathbf{x}_{l,j}}=0$), the Fourier transform of the dielectric function, $\tilde{\varepsilon}(\mathbf{k})$, is also zero in the leading order of disorder. However, it is no longer strictly zero if higher-order terms $O\left(w_0^2\right)$ are considered. In other words, $\left|\tilde{\varepsilon}(\mathbf{k})\right|^2$ in the stealthy region is on the order of $O\left(w_0^4\right)$. The higher-order contribution mainly comes from the nonlinear relation between the local potential and the radius of the hole in Eq. \eqref{e19}, and between the Bessel function and the linear function in Eq. \eqref{e20}. 

Despite the fact that $\tilde{\varepsilon}(\mathbf{k})$ is not strictly zero in the stealthy case due to the higher-order effects, the claim can still be made that the photonic crystal structure is effectively stealthy because the whole theory starting from Eq. \eqref{a7} only considers the leading order.  Furthermore, we have confirmed that the maximum of the spectral density in the stealthy region is less than $10^{-2}$ of the spectral density in the non-stealthy region among all the samples in the experiment as mentioned below, making the transition clearly observable and the non-stealthy fluctuations negligible.  

\begin{figure}[H]
    \centering
    \subfigure[]{\includegraphics[width=0.3\textwidth]{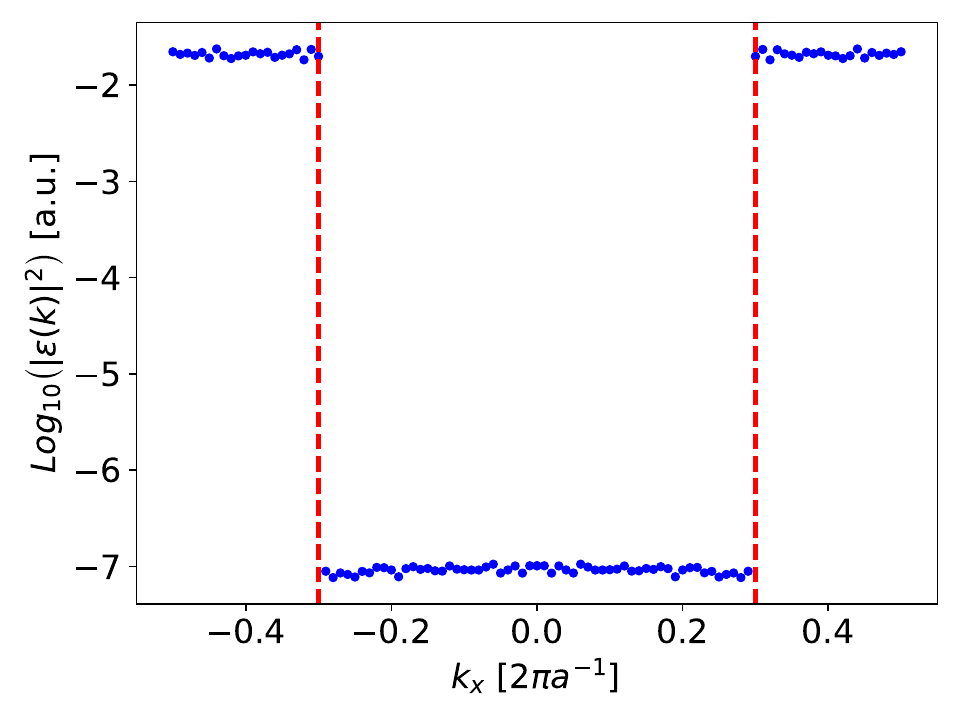}}
    \subfigure[]{\includegraphics[width=0.3\textwidth]{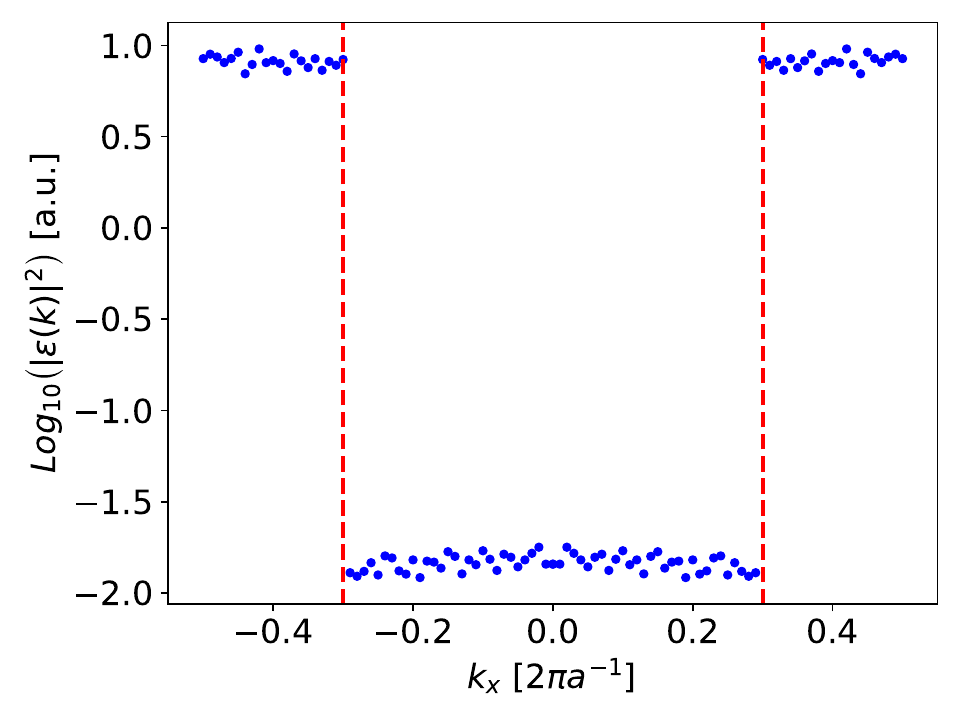}}
    \subfigure[]{\includegraphics[width=0.3\textwidth]{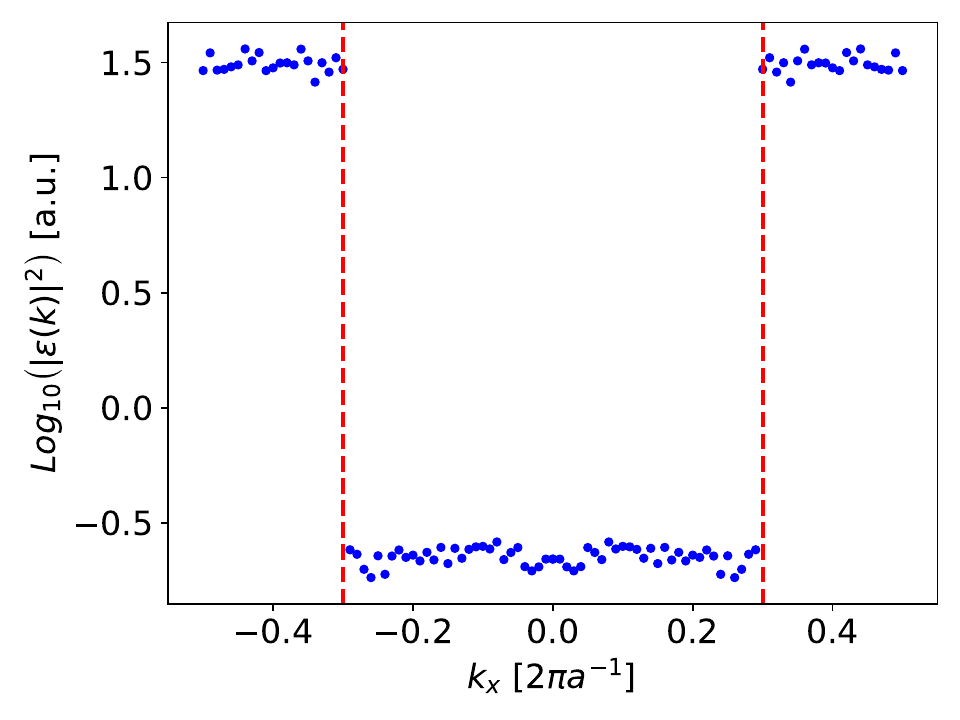}}

\caption{The quantitative estimates of $\tilde{\varepsilon}(\mathbf{k})$. A stealthy-hyperuniform potential disorder from Eq. \eqref{e10} is generated. Then instead of the discrete Fourier transform, $\tilde{\varepsilon}(\mathbf{k})$ is directly calculated from Eq. \eqref{e18}. The constant pre-factor in Eq. \eqref{e18} is ignored since we only focus on the relative scale of the $\tilde{\varepsilon}(\mathbf{k})$. The cutoff wavenumber is fixed at $K=0.3 \left[\frac{2\pi}{a}\right]$ (corresponding to stealthiness $\chi=0.283$) and the wavevector in $y$ direction is $k_y=0$. The values of disorder are: (a) $w_0=0.01$, (b) $w_0=0.2$, (c) $w_0=0.4$. The red dashed line shows the boundary of the stealthy region at $k_x=\pm\frac{K}{2}$. The system size is $N_x\times N_y=400\times400$. The figure shows the averaged results over $100$ different disorder configurations.}
\label{figs7}
\end{figure}

The $\tilde{\varepsilon}(\mathbf{k})$ is quantitatively estimated in Fig. \ref{figs7}. As predicted in Eq. \eqref{e15-2}, the Fourier component $\left|\tilde{\varepsilon}(\mathbf{k})\right|^2$ outside of the stealthy region is a constant. The fluctuation in the figures only arises due to the finite number of configurations taken in the average. In Fig. \ref{figs7}(a), the disorder $w_0=0.01$ is very small. The $\left|\tilde{\varepsilon}(\mathbf{k})\right|^2$ in the stealthy region is around $10^{-9}$ of the Fourier component outside the stealthy region. As the disorder $w_0$ increases, $\left|\tilde{\varepsilon}(\mathbf{k})\right|^2$ outside of the stealthy region grows as $O\left(w_0^2\right)$, but the $\left|\tilde{\varepsilon}(\mathbf{k})\right|^2$ inside the stealthy region grows faster, as $O\left(w_0^4\right)$. As a result, the ratio between the stealthy region and the scattering region increases. In Fig. \ref{figs7}(b), the disorder is $w_0=0.2$, which is the disorder of some of the samples used in the experiment. The ratio is around $10^{-3}$. The disorder further increases to $w_0=0.4$ in Fig. \ref{figs7}(c), which is the largest disorder among all samples fabricated in the experiment. The ratio is less than $10^{-2}$ in this case. 

From Fig. \ref{figs7}, Eq. \eqref{e15-2} is no longer strictly valid if higher-order effects are considered. It needs to be modified as:
\begin{equation}
\tilde{\rho}(\mathbf{q})=
\left\{
\begin{aligned}
F\cdot\frac{w_0^2V_0^2}{3\left(1-\pi\left(K\frac{a}{2\pi}\right)^2\right)} & , & q<K\\
\frac{w_0^2V_0^2}{3\left(1-\pi\left(K\frac{a}{2\pi}\right)^2\right)} & , & q>K
\end{aligned}
\right.
\label{e22}
\end{equation}
where $F$ is a factor that quantifies the quality of the stealthiness. $F=1$ represents a completely uncorrelated system while $F=0$ means perfectly stealthy. We refer $F$ as the ``stealthiness spectral density threshold" in the later text. In general, $F$ is a function of the degree of disorder $w_0$ (more specifically, $F\propto w_0^2$). $F\approx10^{-3}$ for most of the samples in the experiment where $w_0=0.2$. $F\approx10^{-2}$ for the most extreme case in our experiment where $w_0=0.4$. We will show later in Section 3 that even for the most extreme case where $F\approx10^{-2}$, the effect of non-zero $F$ is still negligible in the experiment.

In this section, we have explained in detail how the Fourier filtering method, applied as a perturbation to a periodic photonic crystal, can yield structures that behave as stealthy-hyperuniform systems.  As we mention in the main text, previous examples of stealthy two-phase structures have not been constructed as perturbations on periodic systems, and therefore may be unfamiliar to the reader as candidates for stealthy-hyperuniformity.  However, we have shown above that by using the technique outlined in this work, the photonic structures employed in the experiment can indeed be classified as stealthy.  This is because the excess spectral density in the stealthy region of momentum is completely negligible in experiments compared to that in the non-stealthy region.  Therefore, this allows us to experimentally probe the wave dynamics of stealthy systems.  

\clearpage

\section*{Section 3: Effects of disorder for a Hermitian quadratic band}
\addcontentsline{toc}{section}{Section 3: Effects of disorder for a Hermitian quadratic band}
In this section, we analytically calculate the excess linewidth due to scattering loss by disorder when we have a Hermitian quadratic band in a lattice system.

For a Hermitian quadratic band in a lattice, the energy dispersion is:
\begin{equation}
E_\mathbf{k}=E_0-\frac{1}{2m}\left(k_x^2+k_y^2\right)\quad 
\left(-\frac{\pi}{a}<k_x, k_y<\frac{\pi}{a}\right),
\label{b1}
\end{equation}
where $a$ is the lattice constant, $E_0\in \mathbb{R}$ is the energy of the tip of the quadratic band, and $m\in\mathbb{R}$ is the effective mass. In the experiment, $\mathrm{Re}\left(m\right)>0$, so we also require $m>0$ in the following derivation. Since we are in a lattice system, the band should be periodic under $k_x\mapsto k_x+\frac{2\pi}{a}$ and $k_y\mapsto k_y+\frac{2\pi}{a}$.

We can see from Fig. \ref{figs6}(b) that the photonic band in our experimental structure is an isolated and isotropic quadratic band in the vicinity of $\mathbf{k}=\Gamma$, so from this point onward, we will make the approximation to treat our photonic band in the experiment as a quadratic band described in Eq. \eqref{b1} in the whole first Brillouin zone. However, our photonic band is not a perfect quadratic band, especially near the edge of the first Brillouin zone. In other words, the quadratic approximation is only valid in the vicinity of $\mathbf{k}=\Gamma$, but breaks down when $\mathbf{k}$ is away from $\Gamma$. In the experiment we only focus on the excess linewidth near $\mathbf{k}=\Gamma$, so the above quadratic approximation is always valid.

The Fourier filtering method outlined in Section 2 is used to generate the stealthy-hyperuniform potential disorder configurations. Although the most precise way to describe the spectral density in this case is Eq. \eqref{e22}, we will start from the simpler case in Eq. \eqref{e15-2}, and examine the influence of non-zero stealthiness spectral density threshold $F$ in Eq. \eqref{e22} at the end of this section. The spectral density is: 
\begin{equation}
\tilde{\rho}(\mathbf{q})=
\left\{
\begin{aligned}
0 & , & q<K\\
\frac{w_0^2 V_0^2}{3\left(1-\pi\left(K\frac{a}{2\pi}\right)^2\right)} & , & q>K
\end{aligned}
\right.
\label{b8}
\end{equation}
where $w_0$ represents the degree of disorder, $K$ is the cutoff wavenumber, and $V_0$ describes the linear relation between the local potential and the radius of the hole in the photonic crystal slab.

Note that for large system size:
\begin{equation}
\frac{1}{N_x N_y}\sum_\mathbf{q}=\left(\frac{a}{2\pi}\right)^2\iint d\mathbf{q}.
\label{b11}
\end{equation}

From Eqs. \eqref{a7}, \eqref{b8}, and \eqref{b11}, we get:
\begin{equation}
\left.\Sigma_\mathbf{k}(E)\right|_{E=\mathrm{Re}\left(E_\mathbf{k}\right)}
=\frac{w_0^2 V_0^2}{3\left(1-\pi \left(K\frac{a}{2\pi}\right)^2\right)} \left(\frac{a}{2\pi}\right)^2\left(\Sigma_0-\Sigma_{00}\right),
\label{b12}
\end{equation}
where:
\begin{equation}
\Sigma_0=\iint_{-\frac{\pi}{a}<q_x,q_y<\frac{\pi}{a}}\frac{1}{\mathrm{Re}\left(E_\mathbf{k}\right)-E_{\mathbf{k+q}}} d q_x d q_y,
\label{b13}
\end{equation}

\begin{equation}
\Sigma_{00}=\iint_{0<q<K}\frac{1}{\mathrm{Re}\left(E_\mathbf{k}\right)-E_{\mathbf{k+q}}} d q_x d q_y.
\label{b14}
\end{equation}

The $\Sigma_0$ in Eq. \eqref{b13} has a square integration region. In order to get a simple analytical result, we make the approximation of changing the square region into a circular region:
\begin{equation}
-\frac{\pi}{a}<q_x,q_y<\frac{\pi}{a}\mapsto 
0<q<q_{\mathrm{max}},
\label{b15}
\end{equation}
where $q_{\mathrm{max}}=\frac{1}{\sqrt{\pi}}\left[\frac{2\pi}{a}\right]$ so the area of the region remains the same. This approximation is valid because we only focus on small $k$ in the experiment, where the imaginary part and most of the real part in the integral in Eq. \eqref{b13} arise from the points of $\mathbf{q}$ satisfying $\left|\mathbf{k+q}\right|=k$, which are far away from the boundary of either the circular region or the square region.

Hence, Eq. \eqref{b13} becomes:
\begin{equation}
\Sigma_0=\iint_{0<q<q_{\mathrm{max}}}\frac{1}{\mathrm{Re}\left(E_\mathbf{k}\right)-E_{\mathbf{k+q}}} d q_x d q_y.
\label{b16}
\end{equation}

We then analytically calculate the integrals in Eqs. \eqref{b14} and \eqref{b16}. The final result is:
\begin{equation}
\begin{aligned}
\mathrm{Im}\left(\left.\Sigma_\mathbf{k}(E)\right|_{E=\mathrm{Re}\left(E_\mathbf{k}\right)}\right)
=\left\{
\begin{aligned}
0 & , & k<\frac{K}{2}\\
-\frac{w_0^2 V_0^2}{3\left(1-\pi \left(K\frac{a}{2\pi}\right)^2\right)} \left(\frac{a}{2\pi}\right)^2 4\pi m\cdot \mathrm{arccos}\left(\frac{K}{2k}\right) & , & k>\frac{K}{2}
\end{aligned}
\right.
\label{b26}
\end{aligned}
\end{equation}

Then, according to Eq. \eqref{a14}, we finally get:
\begin{equation}
\Delta \lambda_\mathbf{k}=
\left\{
\begin{aligned}
0 & , & k<\frac{K}{2}\\
\frac{2m\cdot w_0^2 V_0^2 a^2}{3\left(1-\pi \left(K\frac{a}{2\pi}\right)^2\right)E_0^{\frac{3}{2}}} \mathrm{arccos}\left(\frac{K}{2k}\right) & , & k>\frac{K}{2}
\end{aligned}
\right.
\label{b27}
\end{equation} 

Here, since in our experiment, $E_0\gg \frac{k^2}{2m}$, we replace $\rm{Re}\left(E_\mathbf{k}\right)$ in Eq. \eqref{b27} with $E_0$. Next, our analytical result is verified by comparing Eq. \eqref{b27} with the numerical summation result calculated from Eq. \eqref{a7}. The parameters in the numerical summation are set as realistic parameters in the experiment: lattice constant $a=629nm$; band tip energy $E_0=7.11a^{-2}$; potential coefficient $V_0=2.03a^{-2};$ and effective mass $\frac{1}{2m}=0.579$.

\begin{figure}[H]
    \centering
    \subfigure[]{\includegraphics[width=0.3\textwidth]{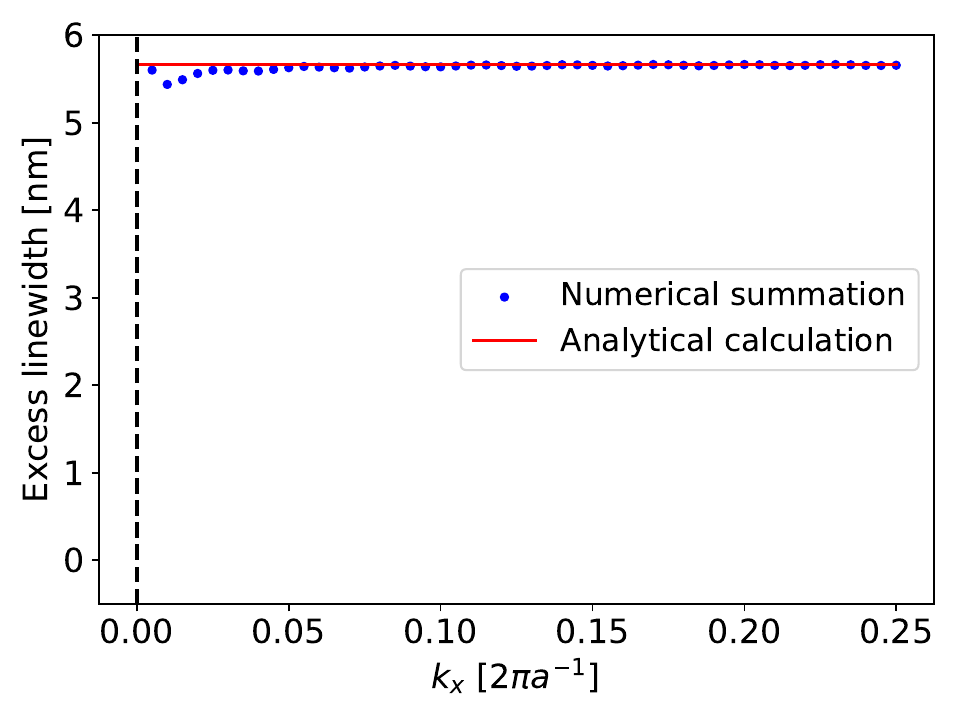}}
    \subfigure[]{\includegraphics[width=0.3\textwidth]{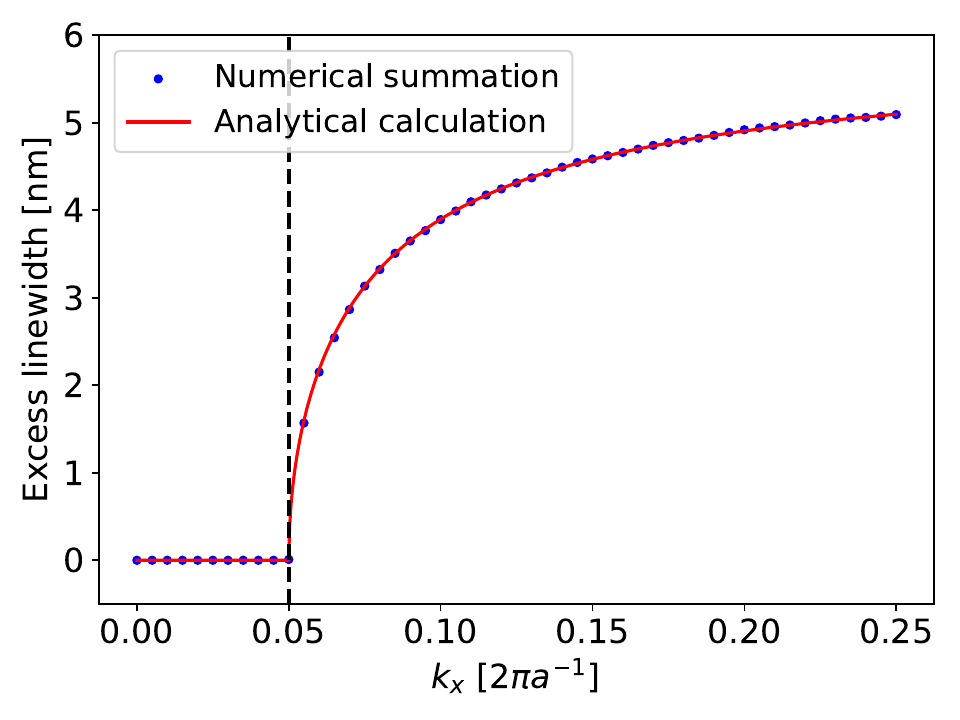}}
    \subfigure[]{\includegraphics[width=0.3\textwidth]{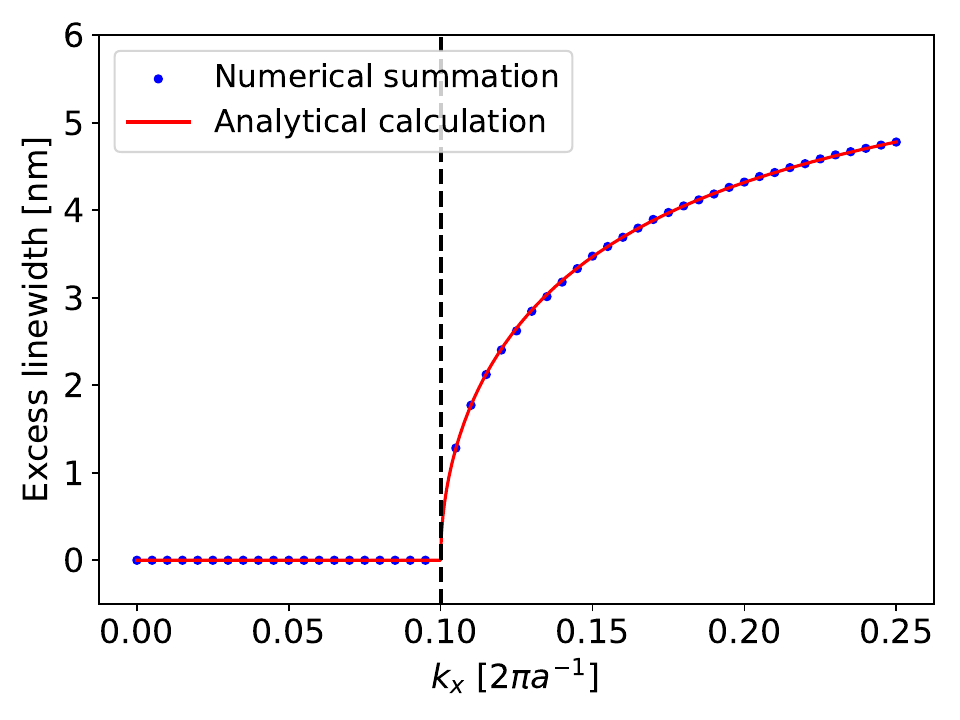}}\\
    \subfigure[]{\includegraphics[width=0.3\textwidth]{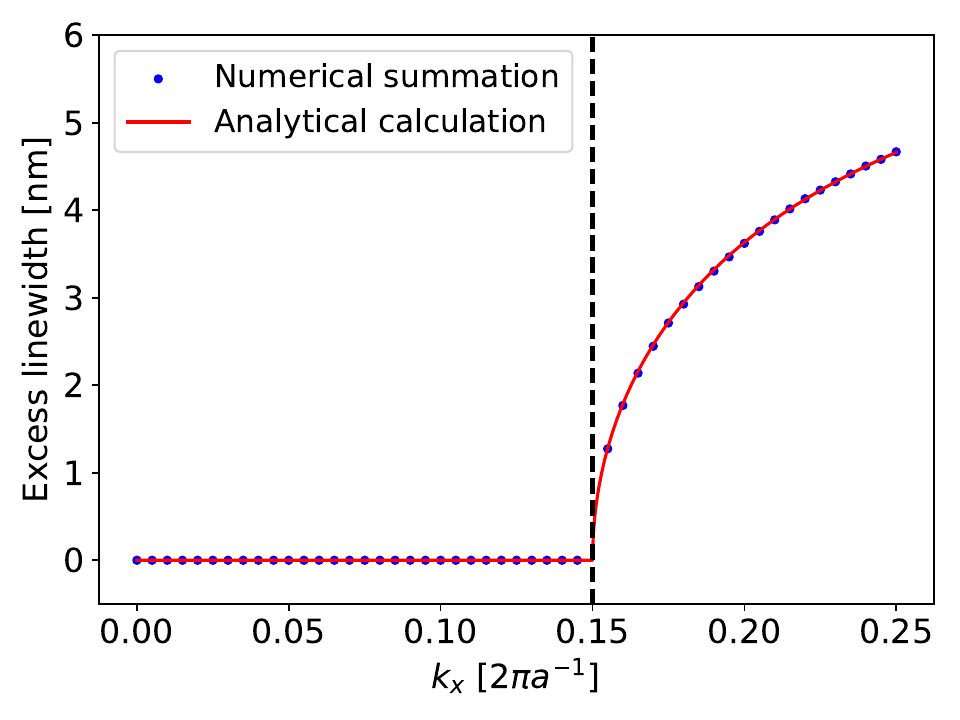}}
    \subfigure[]{\includegraphics[width=0.3\textwidth]{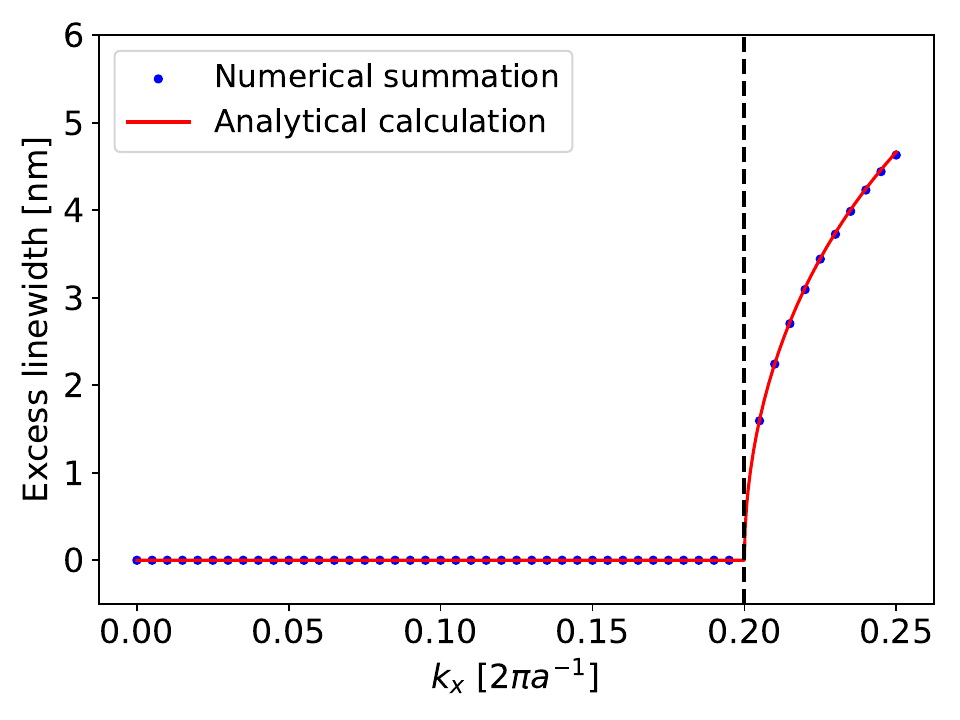}}
    \subfigure[]{\includegraphics[width=0.3\textwidth]{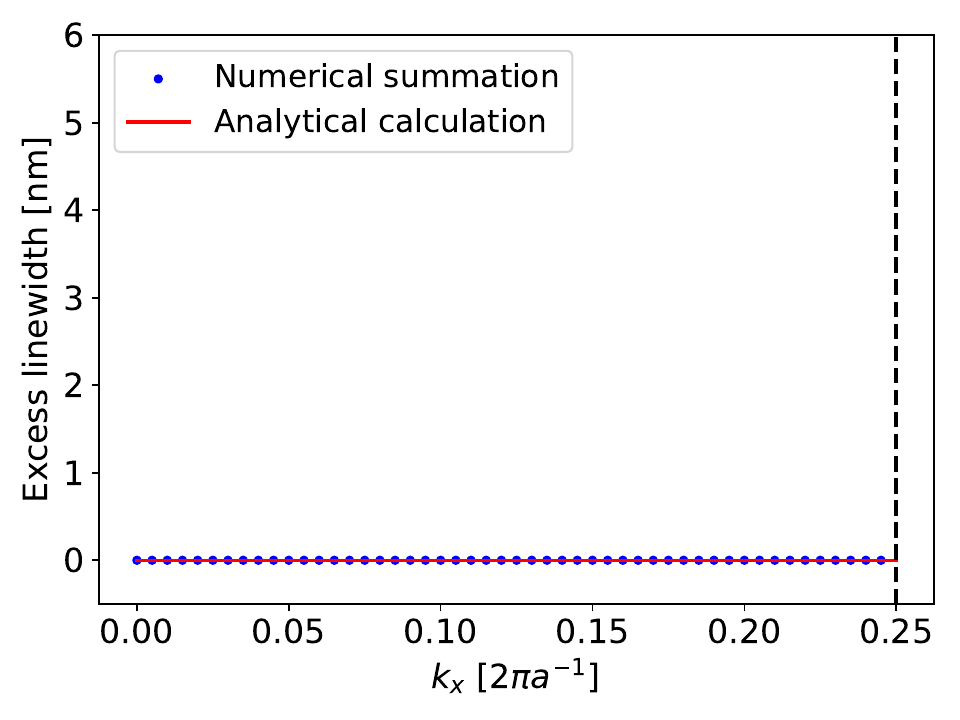}}

\caption{The comparison between analytical calculation (obtained from Eq. \eqref{b27}) and numerical summation (calculated from Eq. \eqref{a7}, with system size $N_x=N_y=400$) along the $k_y=0$ line in the Brillouin zone. The degree of disorder is fixed at $w_0=0.2$. The values of cutoff wavenumber $K$ are: (a) $K=0$, (b) $K=0.1\left[\frac{2\pi}{a}\right]$, (c) $K=0.2\left[\frac{2\pi}{a}\right]$, (d) $K=0.3\left[\frac{2\pi}{a}\right]$, (e) $K=0.4\left[\frac{2\pi}{a}\right]$, (f) $K=0.5\left[\frac{2\pi}{a}\right]$. The black dashed line signifies the excess linewidth transition at $k=\frac{K}{2}$.}
\label{figs1}
\end{figure}

Figure \ref{figs1} shows a comparison across different $K$ at a fixed degree of disorder, $w_0$. We find good quantitative agreement between the analytical and numerical summation results. A transition in the excess linewidth at $k=\frac{K}{2}$ has been observed: The excess linewidth is strictly zero when $k<\frac{K}{2}$ and then increases sharply after $k>\frac{K}{2}$ as an $\arccos$ function. It is also worth mentioning that when $K=0$, the disorder is uncorrelated, so it reduces to the random disorder case. 

Finally, the effect of stealthiness spectral density threshold $F$ in Eq. \eqref{e22} is examined. The spectral density in Eq. \eqref{b8} must be modified as:
\begin{equation}
\tilde{\rho}(\mathbf{q})=
\left\{
\begin{aligned}
F\cdot\frac{w_0^2 V_0^2}{3\left(1-\pi\left(K\frac{a}{2\pi}\right)^2\right)} & , & q<K\\
\frac{w_0^2 V_0^2}{3\left(1-\pi\left(K\frac{a}{2\pi}\right)^2\right)} & , & q>K
\end{aligned}
\right.
\label{b28}
\end{equation}

As a result, Eq. \eqref{b12} must be modified as:
\begin{equation}
\left.\Sigma_\mathbf{k}(E)\right|_{E=\mathrm{Re}\left(E_\mathbf{k}\right)}
=\frac{w_0^2 V_0^2}{3\left(1-\pi \left(K\frac{a}{2\pi}\right)^2\right)} \left(\frac{a}{2\pi}\right)^2\left(\Sigma_0-\left(1-F\right)\Sigma_{00}\right),
\label{b29}
\end{equation}
with $\Sigma_0$ and $\Sigma_{00}$ defined in Eqs. \eqref{b16} and \eqref{b14}. The final result in Eq. \eqref{b27} changes to:
\begin{equation}
\Delta \lambda_\mathbf{k}=
\left\{
\begin{aligned}
F\cdot \frac{\pi m\cdot w_0^2 V_0^2 a^2}{3\left(1-\pi \left(K\frac{a}{2\pi}\right)^2\right)E_0^{\frac{3}{2}}} & , & k<\frac{K}{2}\\
\frac{2m\cdot w_0^2 V_0^2 a^2}{3\left(1-\pi \left(K\frac{a}{2\pi}\right)^2\right)E_0^{\frac{3}{2}}} \left(\mathrm{arccos}\left(\frac{K}{2k}\right)+F\cdot \mathrm{arcsin}\left(\frac{K}{2k}\right)\right) & , & k>\frac{K}{2}
\end{aligned}
\right.
\label{b30}
\end{equation}

\begin{figure}[H]
    \centering
    \subfigure[]{\includegraphics[width=0.23\textwidth]{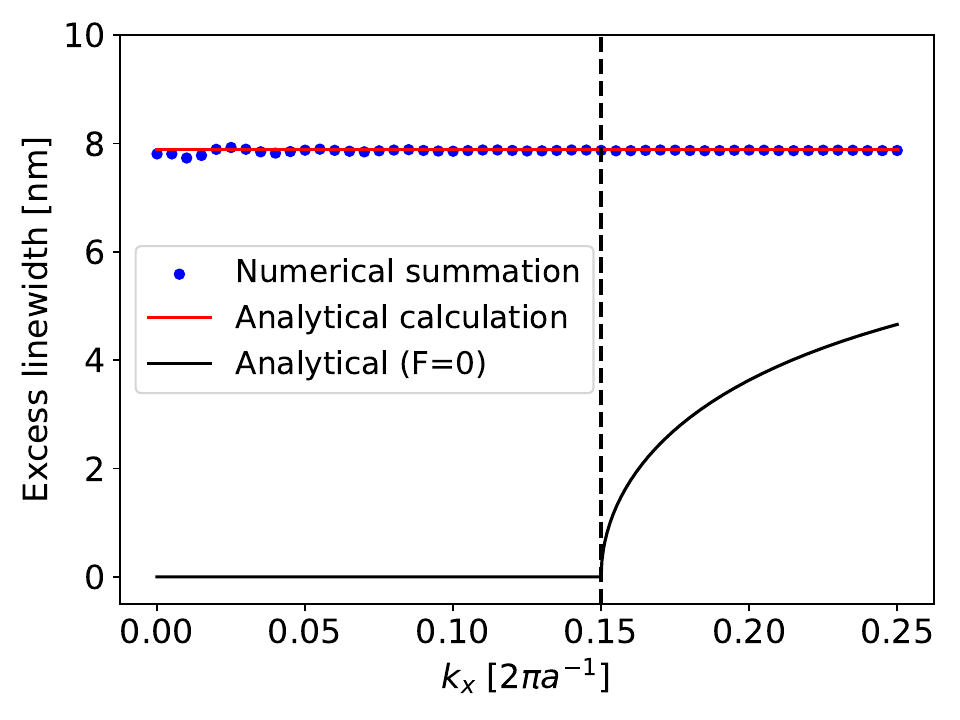}}
    \subfigure[]{\includegraphics[width=0.23\textwidth]{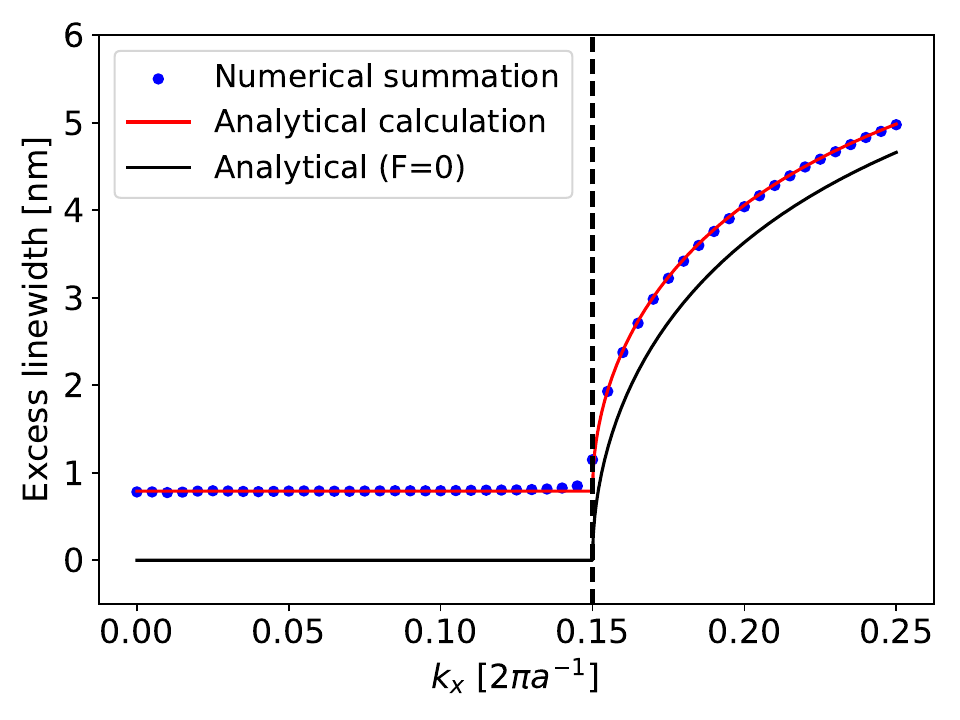}}
    \subfigure[]{\includegraphics[width=0.23\textwidth]{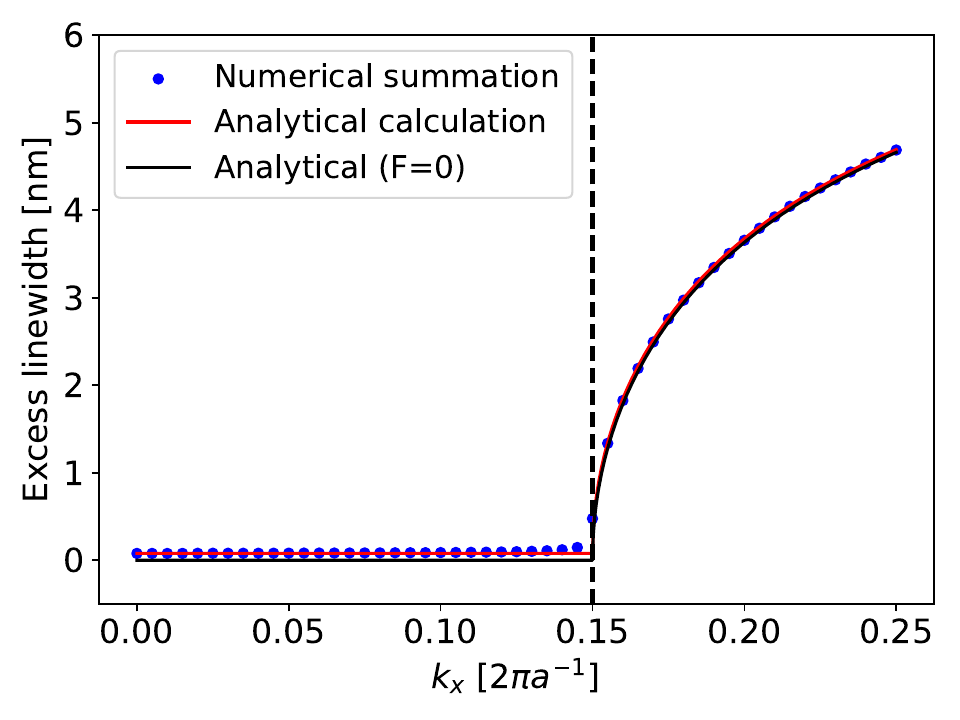}}
    \subfigure[]{\includegraphics[width=0.23\textwidth]{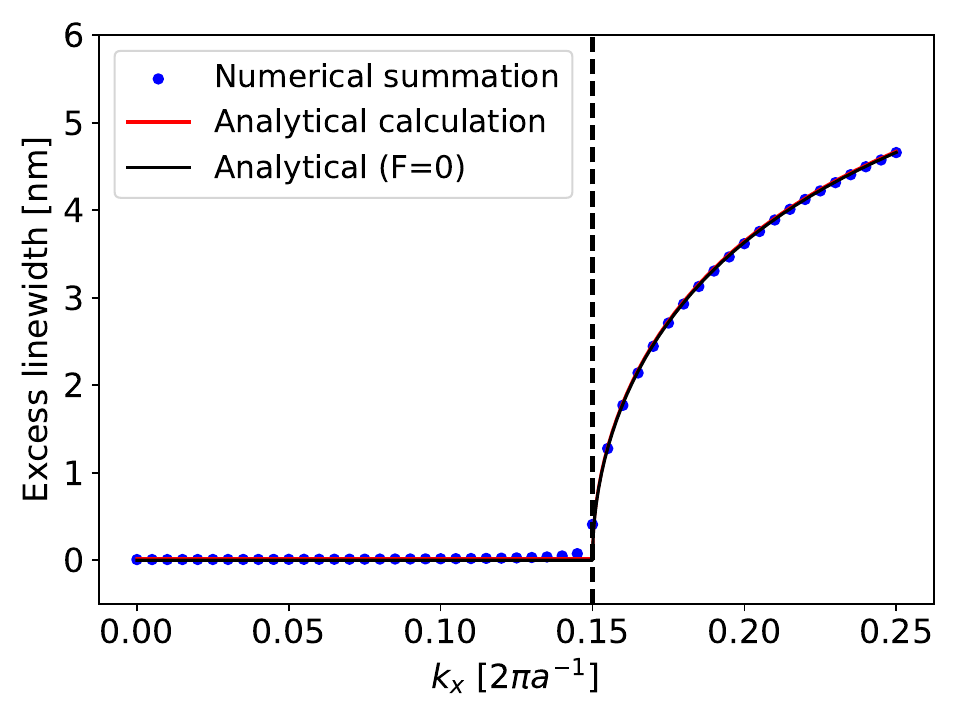}}

\caption{The comparison between analytical calculation (red line, obtained from Eq. \eqref{b30}) and numerical summation (blue dots, calculated from Eq. \eqref{a7}, with system size $N_x=N_y=400$) along the $k_y=0$ line in the Brillouin zone. The analytical result of $F=0$ (black line, obtained from Eq. \eqref{b27}) is also listed. The degree of disorder and the cutoff wavenumber are fixed at $w_0=0.2$ and $K=0.3\left[\frac{2\pi}{a}\right]$. The values of the quality threshold $F$ are: (a) $F=1$, (b) $F=10^{-1}$, (c) $F=10^{-2}$, (d) $F=10^{-3}$. The black dashed line marks the excess linewidth transition at $k=\frac{K}{2}$.}
\label{figs8}
\end{figure}

The calculation in Eq. \eqref{b30} is verified by comparing it with the numerical summation results in Fig. \ref{figs8}. We observe good quantitative agreement between them. Due to the fact that $F\neq 0$, the excess linewidth in the stealthy region is no longer strictly zero, but a finite value proportional to $F$. When $F=1$, the disorder is uncorrelated so the excess linewidth remains the same for all $\mathbf{k}$, similar to the $K=0$ result in Fig. \ref{figs1}(a). The absolute value of excess linewidth is different between Fig. \ref{figs8}(a) and Fig. \ref{figs1}(a) because the normalized degree of disorder is fixed at $w_0=0.2$, but the values of $K$ are different in the two cases. As the stealthiness spectral density threshold $F$ decreases, the analytical line approaches the result of $F=0$, which is the perfect stealthy case. For most of the samples in our experiment, $F\approx 10^{-3}$, the difference between Eqs. \eqref{b30} and \eqref{b27} are negligible, as shown in Fig. \ref{figs8}(d). In fact, the excess linewidth difference is less than $0.01nm$. The largest value of $F$ that we have among all our experimental samples is approximately $F\approx 10^{-2}$, which is shown in Fig. \ref{figs8}(c). The difference between Eqs. \eqref{b30} and \eqref{b27} is less than $0.08nm$, which is smaller than nearly all of the linewidth uncertainties measured in the experiment. 

Based on the results in Fig. \ref{figs8}, we can safely state that the effect arising from non-zero stealthiness spectral density threshold $F$ is insignificant and has a negligible influence on our experimental observables. The photonic crystal pattern we generate by the Fourier filtering method is therefore effectively stealthy.

\clearpage

\section*{Section 4: Effects of disorder for a non-Hermitian quadratic band with complex effective mass and energy}
\addcontentsline{toc}{section}{Section 4: Effects of disorder for a non-Hermitian quadratic band with complex effective mass and energy}

In this section, we will analytically calculate the effects of disorder under a non-Hermitian quadratic band with a complex effective mass.

In the previous section, we consider a Hermitian quadratic band in a lattice system:
\begin{equation}
E_\mathbf{k}=E_0-\frac{1}{2m}\left(k_x^2+k_y^2\right)\quad 
\left(-\frac{\pi}{a}<k_x, k_y<\frac{\pi}{a}\right),
\label{c1}
\end{equation}
where the tip energy $E_0\in \mathbb{R}$ and effective mass $m\in \mathbb{R}$. However, this is not the case for our photonic band. Due to the symmetry-protected bound state in the continuum (BIC) at $\Gamma$ in our photonic crystal structure, the photonic modes in the vicinity of $\Gamma$ experience an out-of-plane radiative loss on the order $O\left(k^2\right)$ \cite{koshelev2018asymmetric}. This means that the eigenenergy of the photonic mode is no longer a purely real number, but has a finite imaginary part proportional to $k^2$, which is equivalent to treating the effective mass $m$ as a complex number with a non-zero imaginary part. Moreover, we also find an absorption loss that is uniform for all $\mathbf{k}$ in the experiment, indicating that the tip energy $E_0$ is also no longer a real number but a complex number. In this section, we will consider this scenario, i.e.,
\begin{equation}
E_0=E_r+iE_i,
\label{c2}
\end{equation}
\begin{equation}
m=m_r+im_i.
\label{c3}
\end{equation}

We assume $E_i\ll E_r$ and $m_i\ll m_r$ so only the leading-order effect of the imaginary part is considered. In our experiment, $E_r>0$, $E_i<0$, $\mathrm{Re}\left(\frac{1}{2m}\right)>0$ and $\mathrm{Im}\left(\frac{1}{2m}\right)>0$, so $m_r>0$ and $m_i<0$.

We follow the same derivation as from Eq. \eqref{b11} to Eq. \eqref{b16}, but we make an additional approximation in Eq. \eqref{b14} and \eqref{b16}:
\begin{equation}
\mathrm{Re}\left(E_\mathbf{k}\right)
=E_\mathbf{k}-i\left(E_i-\mathrm{Im}\left(\frac{k^2}{2m}\right)\right)
\approx E_\mathbf{k}.
\label{c4}
\end{equation}

The approximation in Eq. \eqref{c4} has been made mainly in order to obtain a tractable and interpretable analytical result. The approximation is valid because $E_i=-0.0075a^{-2}$ is very small in our experiment (corresponding to $\left|\frac{E_i}{E_r}\right|=1.05\times 10^{-3}\ll1$), and we only focus on the excess linewidth near $\Gamma$ (which means small $k$) so the term $\mathrm{Im}\left(\frac{k^2}{2m}\right)$ is also small.

Then we apply the approximation in Eq. \eqref{c4} to Eqs. \eqref{b14} and \eqref{b16}, and analytically calculate the integrals with the band parameters in Eqs. \eqref{c1}, \eqref{c2} and \eqref{c3}. The final result is:
\begin{equation}
\mathrm{Im}\left(\Sigma_\mathbf{k}\right)
=\left\{
\begin{aligned}
-\frac{w_0^2 V_0^2 a^2}{3\pi\left(1-\pi \left(K\frac{a}{2\pi}\right)^2\right)} m_i\cdot \ln\left[\frac{\frac{K}{2k}+\sqrt{\left(\frac{K}{2k}\right)^2-1}}{\frac{q_{\mathrm{max}}}{2k}+\sqrt{\left(\frac{q_{\mathrm{max}}}{2k}\right)^2-1}}\right] & , & k<\frac{K}{2}\\
-\frac{w_0^2 V_0^2 a^2}{3\pi\left(1-\pi \left(K\frac{a}{2\pi}\right)^2\right)}  \left(m_r \cdot \arccos\left(\frac{K}{2k}\right)- m_i \cdot \ln\left[\frac{q_{\mathrm{max}}}{2k}+\sqrt{\left(\frac{q_{\mathrm{max}}}{2k}\right)^2-1}\right]\right) & , & k>\frac{K}{2}
\end{aligned}
\right.
\label{c5}
\end{equation}

According to Eq. \eqref{a14}, we find:
\begin{equation}
\Delta \lambda_\mathbf{k}=
\left\{
\begin{aligned}
-\frac{2 w_0^2 V_0^2 a^2}{3\left(1-\pi \left(K\frac{a}{2\pi}\right)^2\right)E_r^{\frac{3}{2}}}m_i\cdot \ln\left[\frac{\frac{q_{\mathrm{max}}}{2k}+\sqrt{\left(\frac{q_{\mathrm{max}}}{2k}\right)^2-1}}{\frac{K}{2k}+\sqrt{\left(\frac{K}{2k}\right)^2-1}}\right] & , & k<\frac{K}{2}\\
\frac{2 w_0^2 V_0^2 a^2}{3\left(1-\pi \left(K\frac{a}{2\pi}\right)^2\right)E_r^{\frac{3}{2}}} \left(m_r\cdot \arccos\left(\frac{K}{2k}\right)-m_i\cdot \ln\left[\frac{q_{\mathrm{max}}}{2k}+\sqrt{\left(\frac{q_{\mathrm{max}}}{2k}\right)^2-1}\right]\right) & , & k>\frac{K}{2}
\end{aligned}
\right..
\label{c6}
\end{equation}

Next, our analytical result is verified by comparing Eq. \eqref{c6} with the numerical summation result calculated from Eq. \eqref{a7}. All the parameters are taken from the experiment: lattice constant $a=629nm$, band tip energy $E_0=\left(7.11-0.0075i\right)a^{-2}$, potential coefficient $V_0=2.03a^{-2}$, and effective mass $\frac{1}{2m}=0.579+0.121i$.

\begin{figure}[H]
    \centering
    \subfigure[]{\includegraphics[width=0.3\textwidth]{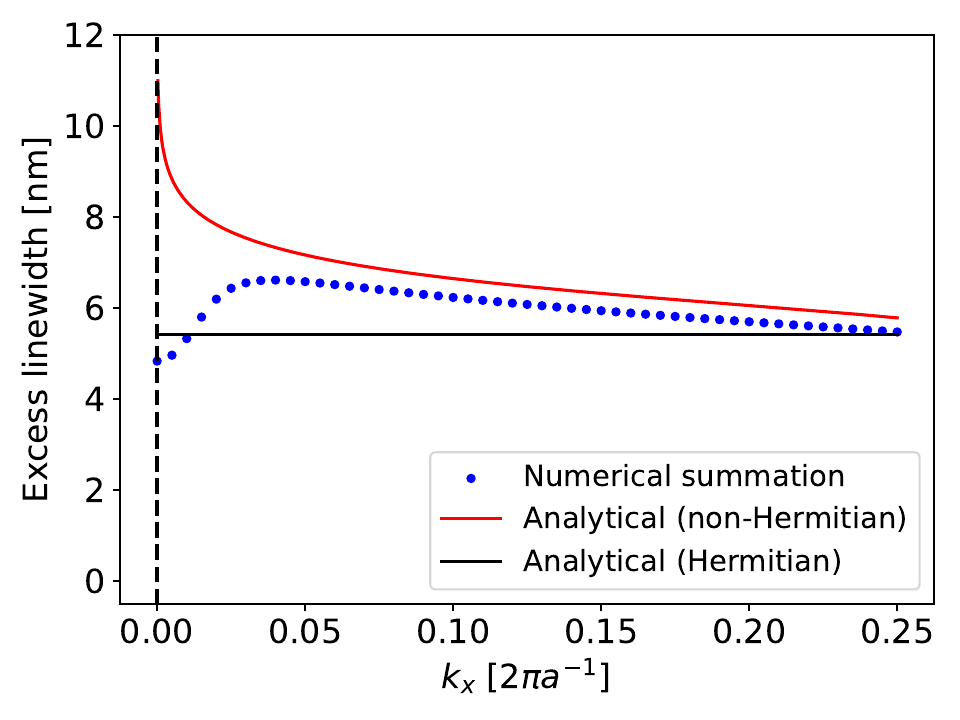}}
    \subfigure[]{\includegraphics[width=0.3\textwidth]{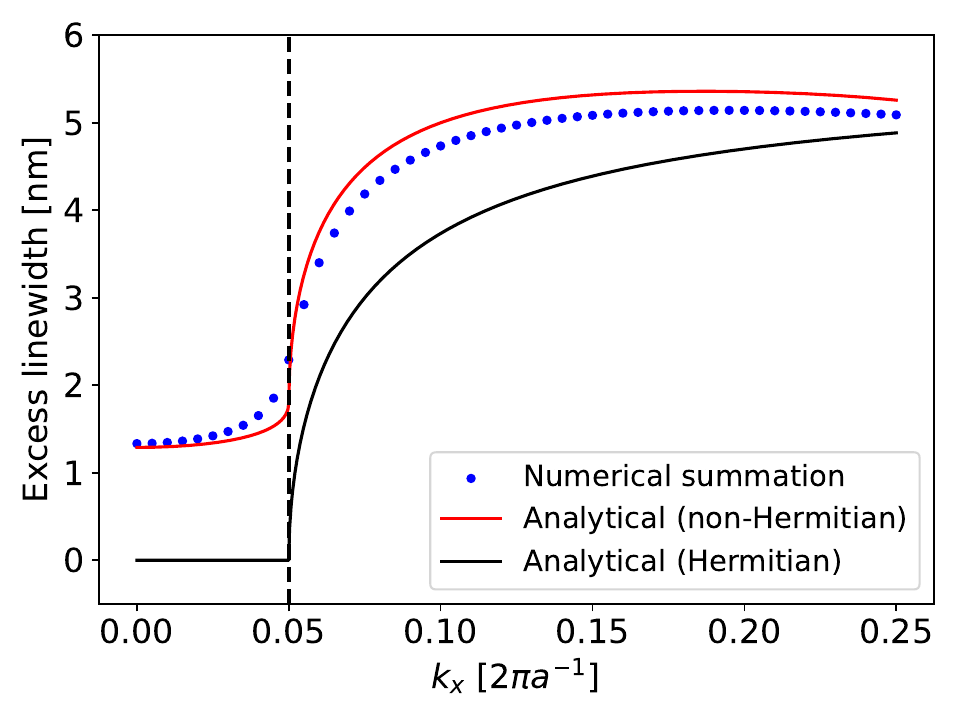}}
    \subfigure[]{\includegraphics[width=0.3\textwidth]{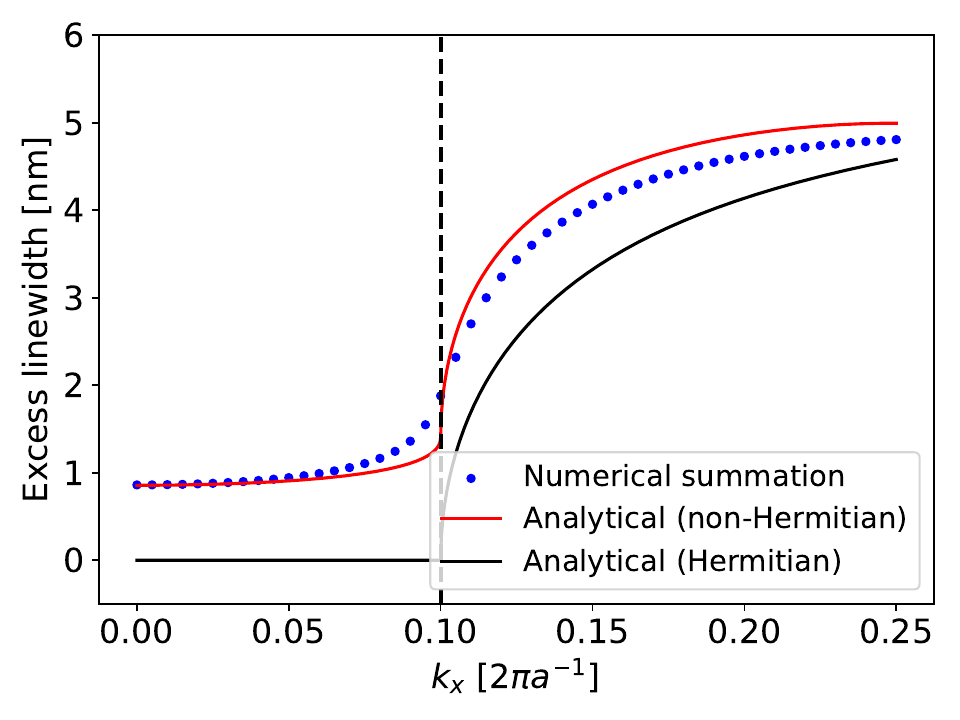}}\\
    \subfigure[]{\includegraphics[width=0.3\textwidth]{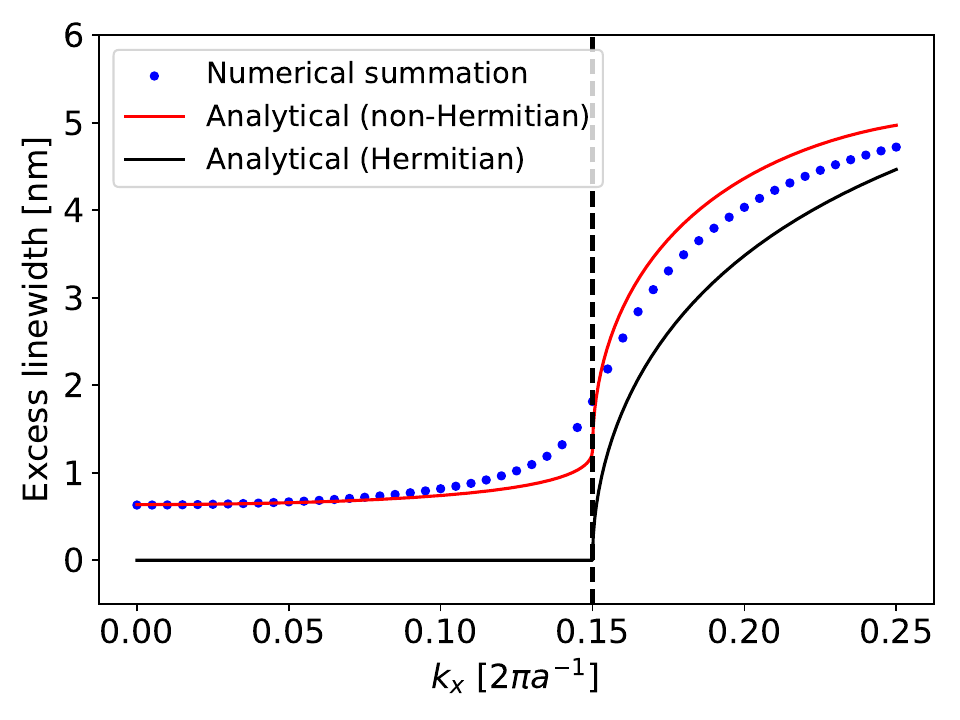}}
    \subfigure[]{\includegraphics[width=0.3\textwidth]{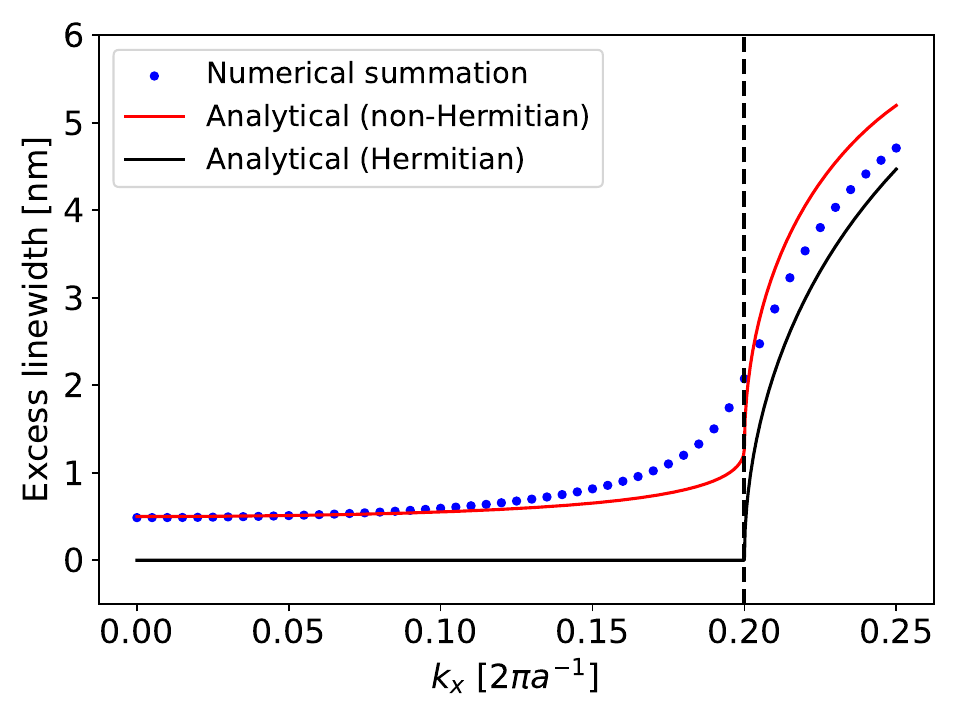}}
    \subfigure[]{\includegraphics[width=0.3\textwidth]{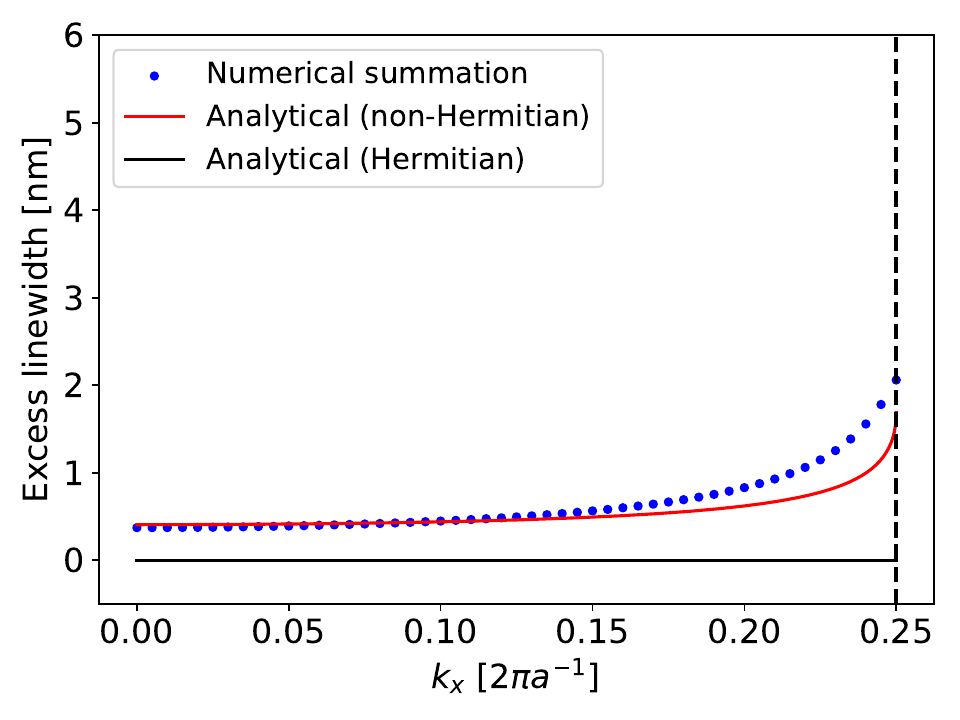}}

\caption{The comparison between analytical calculation (obtained from Eq. \eqref{c6}) and numerical summation results (calculated from Eq. \eqref{a7}, with system size $N_x=N_y=400$) along the $k_y=0$ line in the Brillouin zone. The analytical result when we only consider the real part of the effective mass (obtained from Eq. \eqref{b27}) is also shown in the figures. The degree of disorder is fixed at $w_0=0.2$. The values of cutoff wavenumber $K$ are: (a) $K=0$, (b) $K=0.1\left[\frac{2\pi}{a}\right]$, (c) $K=0.2\left[\frac{2\pi}{a}\right]$, (d) $K=0.3\left[\frac{2\pi}{a}\right]$, (e) $K=0.4\left[\frac{2\pi}{a}\right]$, (f) $K=0.5\left[\frac{2\pi}{a}\right]$. The black dashed line marks the excess linewidth transition at $k=\frac{K}{2}$.}
\label{figs2}
\end{figure}

Figure \ref{figs2} shows the comparison across different values of $K$ while fixing $w_0$. When the effective mass and the tip energy become complex, the Hermitian theory (black line) cannot capture all the behaviors in the excess linewidth, especially in the stealthy region $k<\frac{K}{2}$. The difference between them is largely corrected by our non-Hermitian theory (red line). However, the consistency between the analytical result (red line) and the numerical summation (blue dots) is not as good as in the Hermitian case. This is mainly due to the additional approximation we have made in Eq. \eqref{c4}. As a result of Eq. \eqref{c4}, we decrease the imaginary part of $\mathrm{Re}\left(E_\mathbf{k}\right)-E_\mathbf{k+q}$ when $\mathrm{Re}\left(E_\mathbf{k}\right)=\mathrm{Re}\left(E_\mathbf{k+q}\right)$ (where the maximum scattering intensity is), which causes the transition at $k=\frac{K}{2}$ to be sharper in the analytical result than in the numerical summation. We observe this effect in the experiment: the transition in the experimental data is always smoother than the theoretical prediction. It can also explain why the consistency is generally better for small $k$ than for large $k$, that the approximation $E_i-\mathrm{Im}\left(\frac{k^2}{2m}\right)\approx 0$ is more valid for small $k$. We also notice that the analytical result at small $k$ when $K=0$ deviates from the numerical summation. This is because the non-zero imaginary part of the band tip $E_i$. When $K$ is small, the $k$ near the transition point $k=\frac{K}{2}$ is also small, so the dominant part of the intrinsic linewidth is $E_i$ not $\mathrm{Im}\left(\frac{k^2}{2m}\right)$, which causes the effective mass approximation to be less valid. We also observed this effect in the experiment: the excess linewidth at small wavevector $k$ in small cutoff wavenumber $K$ samples is always smaller than the theoretical prediction.\\

From Eq. \eqref{c6} and Fig. \ref{figs2}, we can see that:

1. The transition point still occurs at $k=\frac{K}{2}$ in the case of complex effective mass and band tip energy.

2. Compared to Eq. \eqref{b27}, the excess linewidth before the transition $k<\frac{K}{2}$ is no longer zero, but a finite value proportional to $m_i$, which clearly shows that the origin of the non-Hermitian effect is the imaginary part of the effective mass. 

3. In the stealthy region, the $k$ dependence in $\Delta \lambda_\mathbf{k}$ is not very sensitive, so it can be estimated by $\left.\Delta \lambda_\mathbf{k}\right|_{k=0}$:
\begin{equation}
\left.\Delta \lambda_\mathbf{k}\right|_{k=0}
=-\frac{2 w_0^2 V_0^2 a^2}{3\left(1-\pi \left(K\frac{a}{2\pi}\right)^2\right)E_r^{\frac{3}{2}}} m_i\cdot \ln \left(\frac{q_{\mathrm{max}}}{K}\right).
\label{c7}
\end{equation}

4. With the presence of non-Hermiticity, the transition at $k=\frac{K}{2}$ becomes smoother compared to the Hermitian case. The transition in the numerical summation without approximation is even smoother than the non-Hermitian theory.

\begin{figure}[H]
    \centering
    \subfigure[]{\includegraphics[width=0.3\textwidth]{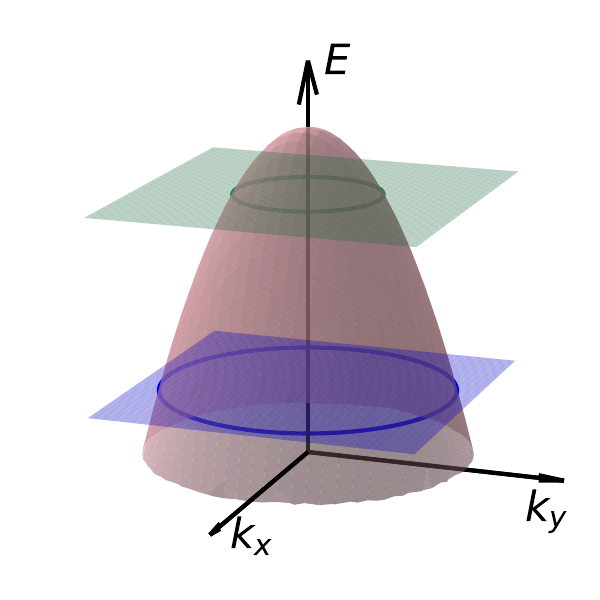}}
    \subfigure[]{\includegraphics[width=0.3\textwidth]{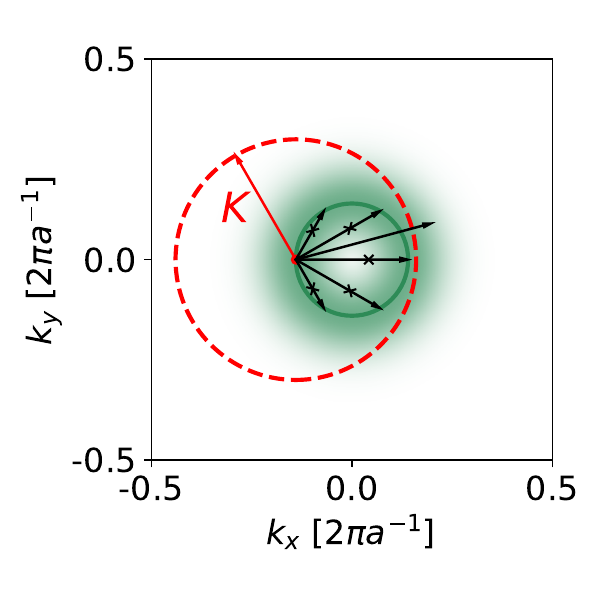}}
    \subfigure[]{\includegraphics[width=0.3\textwidth]{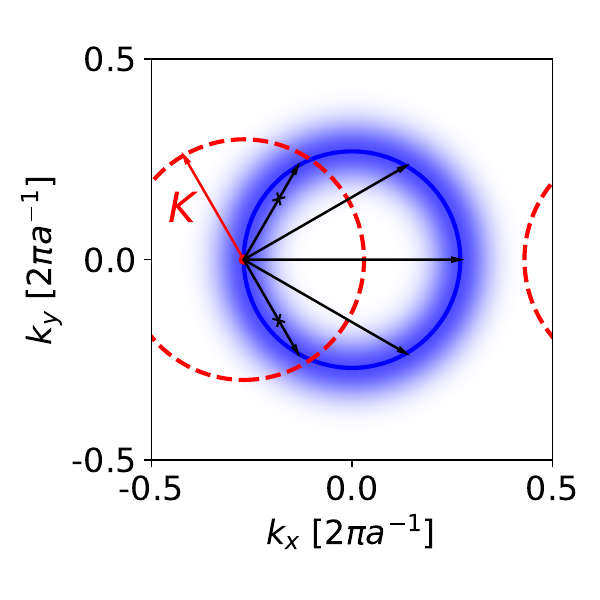}}\\

\caption{The origin of the non-Hermitian effect. (a) A cartoon of iso-frequency contours of a quadratic band. A quadratic band (the pink paraboloid) is intercepted by two equipotential surfaces: one close to the band tip (the green surface), the other is far away from the band tip (the blue surface). (b) An illustration of possible scattering paths under stealthy-hyperuniform disorder before the transition $k<\frac{K}{2}$. The solid green circle is the iso-frequency contour, similarly to (a). The blurry ring represents the effective linewidth in $\mathbf{k}$ space. The dashed red circle is the scattering-free region quantified by cutoff wavenumber $K$. The arrows with/without the `$\times$' denote the prohibited/allowed scattering paths, respectively. (c) An illustration of possible scattering paths after the transition, i.e., for $k>\frac{K}{2}$.}
\label{figs3}
\end{figure}

Figure \ref{figs3} qualitatively explains the origin of non-Hermitian effects. Figure \ref{figs3}(a) is similar to Fig. 2(d) in the main text, which shows a quadratic band. In the Hermitian case, as discussed in the main text, the iso-frequency contour forms a circle without any width. However, in the non-Hermitian case, due to the non-zero imaginary part of the energy, the paraboloid has finite linewidth in the energy direction, or equivalently, it has finite ``thickness" in $\mathbf{k}$ space. In other words, the iso-frequency contour is no longer a circle without linewidth (like Fig. 2(e)(f) in the main text), but a faded ring with thickness as shown in Fig. \ref{figs3}(b)(c). As a result, even when the whole iso-frequency contour itself is blocked by the scattering-free region, such that the scattering from one point to any other point on the iso-frequency contour is forbidden, the scattering to $\mathbf{k}$ points out of the scattering-free region is allowed because it is on the ``tail" of the iso-frequency contour (faded ring). This can be understood as scattering from one point on the iso-frequency contour to another (which is outside of the stealthy region) on another iso-frequency contour with a different energy, because the energy is no longer conserved in non-Hermitian systems. This is the reason why the excess linewidth is no longer zero in the stealthy region for the non-Hermitian case. When $m_i$ is larger, the width of the faded ring becomes larger, so there are more possible $\mathbf{k}$ points to scatter to and the excess linewidth increases as $m_i$ increases. The same non-Hermitian process occurs in Fig. \ref{figs3}(c). The excess linewidth becomes larger compared to the Hermitian case after the transition $k>\frac{K}{2}$ because it has more states into which it can scatter.

This mechanism also explains why the linewidth transition becomes smoother in the non-Hermitian band. In the Hermitian case, the iso-frequency contour is a sharp line without thickness. While the iso-frequency ring in the non-Hermitian case is still peaked at the circle, it has finite thickness so the transition becomes smoother.

\clearpage

\section*{Section 5: Period-tripling and the second transition}
\addcontentsline{toc}{section}{Section 5: Period-tripling and the second transition}
In this section, we will introduce a structural modification that we call ``period-tripling" used in our stealthy-hyperuniform photonic crystals. We will analytically calculate the excess linewidth in this case and explain the origin of a second transition induced by period-tripling.

Period tripling is implemented in the Fourier filtering method while generating the stealthy-hyperuniform configuration. Without using the period tripling method, we calculate the Fourier component of the local potential $\tilde{V}(\mathbf{k})$ with the use of every lattice grid point $V_{l,j}$:
\begin{equation}
\tilde{V}(\mathbf{k})=\frac{1}{\sqrt{N_x N_y}}\sum_{l,j}e^{-i \left(l k_x a + j k_y a\right)} V_{l,j}.
\label{d1}
\end{equation}

We then use the Fourier filtering method to set the Fourier components within the stealthy region to be zero, and perform the inverse Fourier transform to convert $\tilde{V}(\mathbf{k})$ back into real space to generate the disorder configuration.

As we can see from Eqs. \eqref{b26} and \eqref{c6}, if the Fourier filtering method mentioned above is applied, the excess linewidth transition always occurs at $k=\frac{K}{2}$. In the experiment, due to the limitation of the range of accessible angles in our setup, only the linewidth at small $k$ (approximately $k<0.1\left[\frac{2\pi}{a}\right]$) can be measured without realigning the setup. Moreover, the intrinsic linewidth at large $k$ ($k>0.1\left[\frac{2\pi}{a}\right]$) is so large that it is difficult to measure the linewidth (especially the excess linewidth) reliably and precisely. However, if we measure the linewidth at small $k$, at $k<0.1\left[\frac{2\pi}{a}\right]$, we are not able to observe the transition for $K>0.2 \left[\frac{2\pi}{a}\right]$ because $k=\frac{K}{2}$ is beyond the range of accessible angles in the experiment. According to Eq. \eqref{e10-2}, the maximum stealthiness we can get (when $K<0.2\left[\frac{2\pi}{a}\right]$) is $\chi<0.126$, which limits us to small stealthiness and we are not able to investigate the behavior of structures with a large stealthiness parameter.

We then seek a way that can generate structures with a large stealthiness parameter $\chi$, but still have a small transition wavevector $k=\frac{K}{2}$, which means small cutoff wavenumber $K$. This is the motivation for period tripling.

\begin{figure}[H]
    \centering
    \subfigure[]{\includegraphics[width=0.23\textwidth]{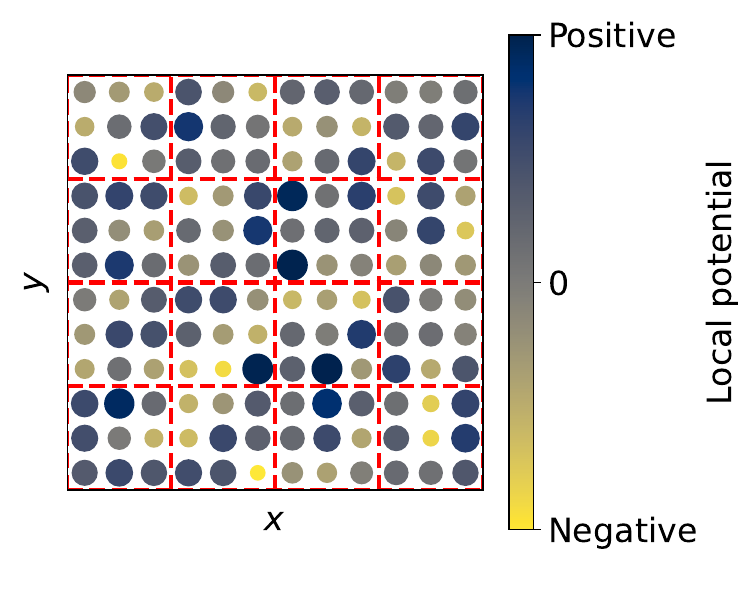}}
    \subfigure[]{\includegraphics[width=0.23\textwidth]{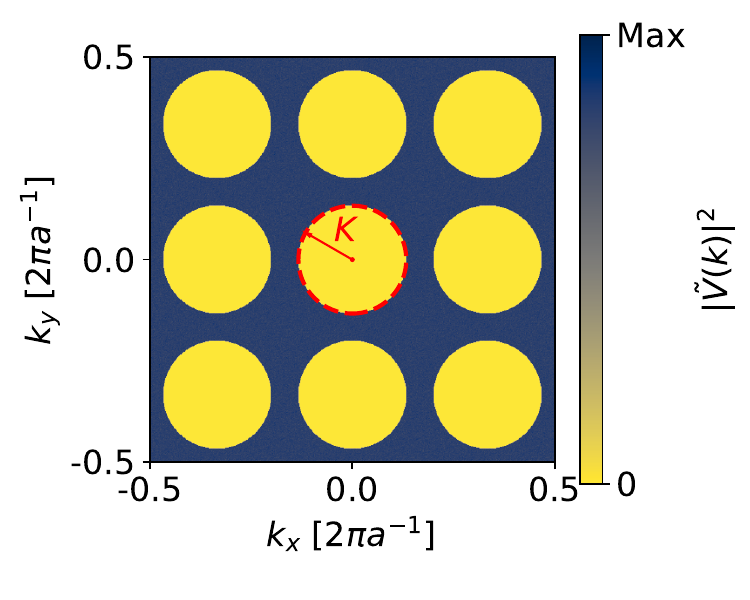}}
    \subfigure[]{\includegraphics[width=0.23\textwidth]{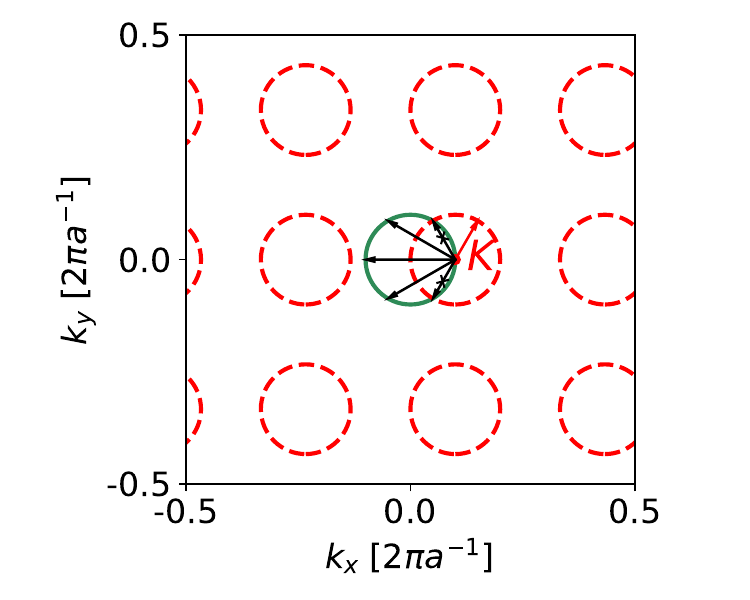}}
    \subfigure[]{\includegraphics[width=0.23\textwidth]{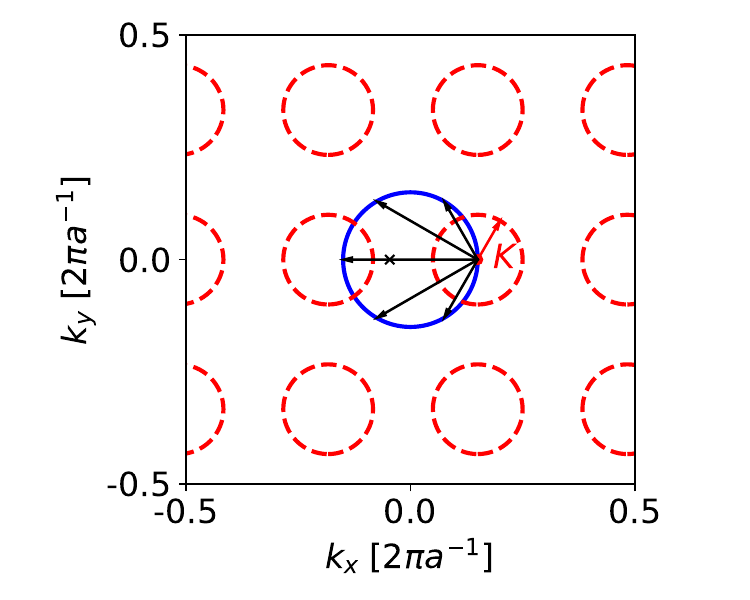}}
    
\caption{The period tripling method and the origin of the second transition. (a) An illustration of how the disorder is individually added to one site in a $3\times 3$ supercell. The red dashed lines show the $3\times 3$ supercell. The color and size of the holes represent the local potential change. Only $4\times4$ period-tripled supercells are shown. (b) The Fourier component $\left|\tilde{V}(\mathbf{k})\right|^2$ of the structure in (a). (c) The relation between the iso-frequency contour (green circle) and the scattering-free region (red dashed circles) when $k<\frac{\pi}{3a}-\frac{K}{2}$. The iso-frequency contour doesn't contact a second scattering-free region. (d) The relation between the iso-frequency contour (blue circle) and the scatter-free region (red dashed circles) when $k>\frac{\pi}{3a}-\frac{K}{2}$. The iso-frequency contour is partially blocked by a second scattering-free region.}
\label{figs4}
\end{figure}

Figure \ref{figs4} explains how the period-tripling method is performed. In the case of period tripling shown in Fig. \ref{figs4}(a), we choose a supercell of $3\times 3$ unit cells, and calculate the Fourier component only with the use of one of the nine lattice grid points in each supercell:
\begin{equation}
\tilde{V}_{l_0,j_0}(\mathbf{k})=\frac{1}{\sqrt{N_x N_y}}\sum_{l,j}e^{-i k_xa (3l+l_0)-i k_y a (3j+j_0)} V_{3l+l_0,3j+j_0},
\label{d2}
\end{equation}
where $l_0, j_0=-1, 0, 1$ labels the nine unit cells in one $3\times 3$ supercell. The local potential between different sub-cell $\left(l_0,j_0\right)$ within a supercell $(l,j)$ is independent and uncorrelated. Then we filter out all the $\tilde{V}_{l_0,j_0}(\mathbf{k})$ with $l_0, j_0=-1, 0, 1$ in the stealthy region $k<K$ to obtain $\tilde{V}^{\prime}_{l_0,j_0}(\mathbf{k})$, and inversely transform them back into real space to generate the disorder pattern $V^{\prime}_{3l+l_0,3j+j_0}$. The Fourier component of the whole structure is:
\begin{equation}
\tilde{V}^{\prime}(\mathbf{k})=\sum_{l_0,j_0} \tilde{V}^{\prime}_{l_0,j_0}(\mathbf{k}).
\label{d2-2}
\end{equation}

Figure \ref{figs4}(b) shows the spectral function $\tilde{\rho}(\mathbf{q})$ after applying the Fourier filtering method with period tripling. A direct consequence is that $\tilde{\rho}(\mathbf{q})$ no longer only has one hole (stealthy region), but has $3\times 3=9$ holes in the first Brillouin zone. In other words, since the lattice period is effectively tripled, $\tilde{\rho}(\mathbf{q})$ has a different periodicity compared to the band structure $E_\mathbf{k}$, that $\tilde{\rho}(q_x+\frac{2\pi}{3a},q_y)=\tilde{\rho}(q_x,q_y+\frac{2\pi}{3a})=\tilde{\rho}(q_x, q_y)$, while $E(k_x+\frac{2\pi}{a},k_y)=E(k_x,k_y+\frac{2\pi}{a})=E(k_x, k_y)$.

We can define the lattice period $b$ as $b=3a$, which is the lattice period of the $3\times3$ supercell. Then, by using the same argument as in Figure 2(e)(f) in the main text, we can see that the first excess linewidth transition still occurs at $k=\frac{K}{2}$, but the stealthiness parameter $\chi$ becomes
\begin{equation}
\chi=\pi\left(K\frac{b}{2\pi}\right)^2=\pi\left(K\frac{3a}{2\pi}\right)^2.
\label{d2-3}
\end{equation}

We can see from Eq. \eqref{d2-3} that the stealthiness parameter $\chi$ under the same cutoff wavenumber $K$ is effectively larger after the period tripling compared to Eq. \eqref{e10-2}. From this point onward, we require $\chi<\frac{\pi}{4}$, which means $K<\frac{1}{2}\left[\frac{2\pi}{3a}\right]$, so the stealthy regions are always complete circles in the Brillouin zone.

In addition to the first excess linewidth transition at $k=\frac{K}{2}$, we also predict that there is a second transition at $k_x=\frac{\pi}{3a}-\frac{K}{2}$ when $k_y=0$. The reason is explained in Fig. \ref{figs4}(c)(d). In the first excess linewidth transition, only the stealthy circle centered at $\mathbf{q}=(0,0)$ is considered. When $k$ becomes larger, the iso-frequency contour will touch other stealthy circles. For positive $k_x$ with $k_y=0$, the second circle it will touch is the one centered at $\mathbf{q}=(-\frac{2\pi}{3a},0)$. The minimum $k_x$ for touching occurs at $k_x=\frac{\pi}{3a}-\frac{K}{2}$, which is the origin of the second transition.

In general, as $k$ grows larger, the iso-frequency contour will eventually touch more and more scattering-free circles, so there are even more excess linewidth transitions. Moreover, the direction of $\mathbf{k}$ will also influence the transition position except for the first transition at $k=\frac{K}{2}$. For example, the second transition along $k_y=0$ occurs at $k_x=\pm\left(\frac{\pi}{3a}-\frac{K}{2}\right)$, while the second transition along $k_x=k_y$ line occurs at $k_x=k_y=\pm\frac{\left(\frac{2\pi}{3a}\right)^2-K^2}{2K+\sqrt{2}\frac{2\pi}{3a}}$. For simplicity, we will only focus on the second transition along $k_y=0$ with positive $k_x$. In the experiment, we also fix $k_y=0$ and sweep $k_x$. The range of $k_x$ is not large enough to observe the third or higher transition, so our derivation in this section is sufficient to explain the experimental data.

Next, we analytically calculate the excess linewidth in the period-tripling case. Due to the period tripling, Eq. \eqref{b8} needs to be modified as:
\begin{equation}
\tilde{\rho}(\mathbf{q})=
\left\{
\begin{aligned}
0 & , & \mathop{\min}_{l_0, j_0} \left|\mathbf{q}-\mathbf{q}_{l_0, j_0}\right|<K\\
\frac{w_0^2 V_0^2}{3\left(1-\pi\left(K\frac{3a}{2\pi}\right)^2\right)} & , & \mathop{\min}_{l_0, j_0} \left|\mathbf{q}-\mathbf{q}_{l_0, j_0}\right|>K
\end{aligned}
\right.
\label{d3}
\end{equation}
where $l_0, j_0=-1, 0, 1$, and $\mathbf{q}_{l_0, j_0}=\left(l_0 \left[\frac{2\pi}{3a}\right], j_0 \left[\frac{2\pi}{3a}\right]\right)$ are the centers of the circular stealthy regions in Fig. \ref{figs4}(b).

As a result, Eq. \eqref{c4} needs to be modified as:
\begin{equation}
\left.\Sigma_\mathbf{k}(E)\right|_{E=\mathrm{Re}\left(E_\mathbf{k}\right)}
=\frac{w_0^2 V_0^2}{3\left(1-\pi \left(K\frac{3a}{2\pi}\right)^2\right)} \left(\frac{a}{2\pi}\right)^2\left(\Sigma_0-\sum_{l_0,j_0} \Sigma_{l_0, j_0}\right),
\label{d4}
\end{equation}
where:
\begin{equation}
\Sigma_0=\iint_{0<q<q_{\mathrm{max}}}\frac{1}{\mathrm{Re}\left(E_\mathbf{k}\right)-E_{\mathbf{k+q}}} d q_x d q_y,
\label{d5}
\end{equation}

\begin{equation}
\Sigma_{l_0, j_0}=\iint_{0<\left|\mathbf{q}-\mathbf{q}_{l_0, j_0}\right|<K}\frac{1}{\mathrm{Re}\left(E_\mathbf{k}\right)-E_{\mathbf{k+q}}} d q_x d q_y.
\label{d6}
\end{equation}

Comparing Eq. \eqref{d4} with Eq. \eqref{b12}, we can directly see that only the $l_0=j_0=0$ term $\Sigma_{0,0}$ is considered in the non-period-tripled case, all the $\Sigma_{l_0, j_0}$ terms with $l_0, j_0=-1, 0, 1$ need to be considered in the period tripling case.

We apply the same approximation as in Eq. \eqref{c4} to treat $\mathrm{Re}\left(E_\mathbf{k}\right)=E_\mathbf{k}$. In order to obtain an analytical and interpretable result, we apply an additional approximation in Eq. \eqref{d6} to use the value at $k_x=0$ to approximate all the logarithmic terms in $\Sigma_{l_0, j_0}$ when $\left(l_0,j_0\right)\neq\left(0,0\right)$:
\begin{equation}
\scalebox{0.9}{$\displaystyle
\begin{aligned}
&\ln \left(\frac{\left(l_0\frac{2\pi}{3a}\right)^2+\left(j_0\frac{2\pi}{3a}+k_x\right)^2+k_x^2-K^2-\sqrt{\left[\left(l_0\frac{2\pi}{3a}\right)^2+\left(j_0\frac{2\pi}{3a}+k_x\right)^2+k_x^2-K^2\right]^2-4k_x^2\left[\left(l_0\frac{2\pi}{3a}\right)^2+\left(j_0\frac{2\pi}{3a}+k_x\right)^2\right]}}{2k_x^2}\right)\\
\approx\ 
&\ln \left(\frac{l_0^2+j_0^2}{l_0^2+j_0^2-\left(K\frac{3a}{2\pi}\right)^2}\right), \quad \left(l_0,j_0\right)\neq\left(0,0\right)
\label{d6-2}
\end{aligned}
$}
\end{equation}

This approximation is valid because the logarithmic terms are not sensitive to changes in $k_x$, especially because we only focus on a small range of $\mathbf{k}$ near $\Gamma$. As mentioned above, we calculate the excess linewidth along $k_y=0$ for positive $k_x$. The final results are:

\begin{equation}
\scalebox{0.75}{$\displaystyle
\begin{aligned}
\Delta\lambda_{\left(k_x,0\right)}=
\left\{
\begin{aligned}
-\frac{2 w_0^2 V_0^2 a^2 m_i}{3\left(1-\pi \left(K\frac{3a}{2\pi}\right)^2\right)E_r^\frac{3}{2}}\left(
\ln\left[\frac{\frac{q_{\mathrm{max}}}{2k_x}+\sqrt{\left(\frac{q_{\mathrm{max}}}{2k_x}\right)^2-1}}{\frac{K}{2k_x}+\sqrt{\left(\frac{K}{2k_x}\right)^2-1}}\right]
-\ln\left[\frac{1}{\left(1-\left(K\frac{3a}{2\pi}\right)^2\right)^2\left(1-\frac{1}{2}\left(K\frac{3a}{2\pi}\right)^2\right)^2}\right]
\right)& , & 0<k_x<\frac{K}{2}\\
\\
\begin{aligned}
\frac{2 w_0^2 V_0^2 a^2}{3\left(1-\pi \left(K\frac{3a}{2\pi}\right)^2\right)E_r^\frac{3}{2}}
\left(
-m_i\cdot \ln\left[
\frac{\frac{q_{\mathrm{max}}}{2k_x}+\sqrt{\left(\frac{q_{\mathrm{max}}}{2k_x}\right)^2-1}}{\left(1-\left(K\frac{3a}{2\pi}\right)^2\right)^{-2}\left(1-\frac{1}{2}\left(K\frac{3a}{2\pi}\right)^2\right)^{-2}}
\right]
+m_r\cdot \arccos\left(\frac{K}{2k_x}\right)
\right)
\end{aligned}
& , & \frac{K}{2}<k_x<\frac{\pi}{3a}-\frac{K}{2}\\
\\
\begin{aligned}
&-\frac{2 w_0^2 V_0^2 a^2 m_i}{3\left(1-\pi \left(K\frac{3a}{2\pi}\right)^2\right)E_r^\frac{3}{2}}
\left(
\ln\left[
\frac{q_{\mathrm{max}}}{2k_x}+\sqrt{\left(\frac{q_{\mathrm{max}}}{2k_x}\right)^2-1}
\right]
-\ln\left[\frac{\left(\frac{\frac{2\pi}{3a}-k_x}{k_x}\right)^{\frac{1}{2}}}{\left(1-\left(K\frac{3a}{2\pi}\right)^2\right)^{\frac{3}{2}}\left(1-\frac{1}{2}\left(K\frac{3a}{2\pi}\right)^2\right)^{2}}\right]
\right)
+\\
&+\frac{2 w_0^2 V_0^2 a^2 m_r}{3\left(1-\pi \left(K\frac{3a}{2\pi}\right)^2\right)E_r^\frac{3}{2}}
\left(
\arccos\left(\frac{K}{2k_x}\right)
-\frac{1}{2} \arccos\left(\frac{\left(\frac{2\pi}{3a}-k_x\right)^2+k_x^2-K^2}{2 \left(\frac{2\pi}{3a}-k_x\right) k_x}\right)
\right)
\end{aligned}
& , & k_x>\frac{\pi}{3a}-\frac{K}{2}
\end{aligned}
\right.
\label{d24}
\end{aligned}
$}
\end{equation}

Next, our analytical result is verified by comparing Eq. \eqref{d24} with the numerical summation result calculated from Eq. \eqref{a7}. The parameters are set as the values they are in the experiment: lattice constant $a=629nm$, band tip energy $E_0=\left(7.11-0.0075i\right)a^{-2}$, potential coefficient $V_0=2.03a^{-2}$, and effective mass $\frac{1}{2m}=0.579+0.121i$.

\begin{figure}[H]
    \centering
    \subfigure[]{\includegraphics[width=0.3\textwidth]{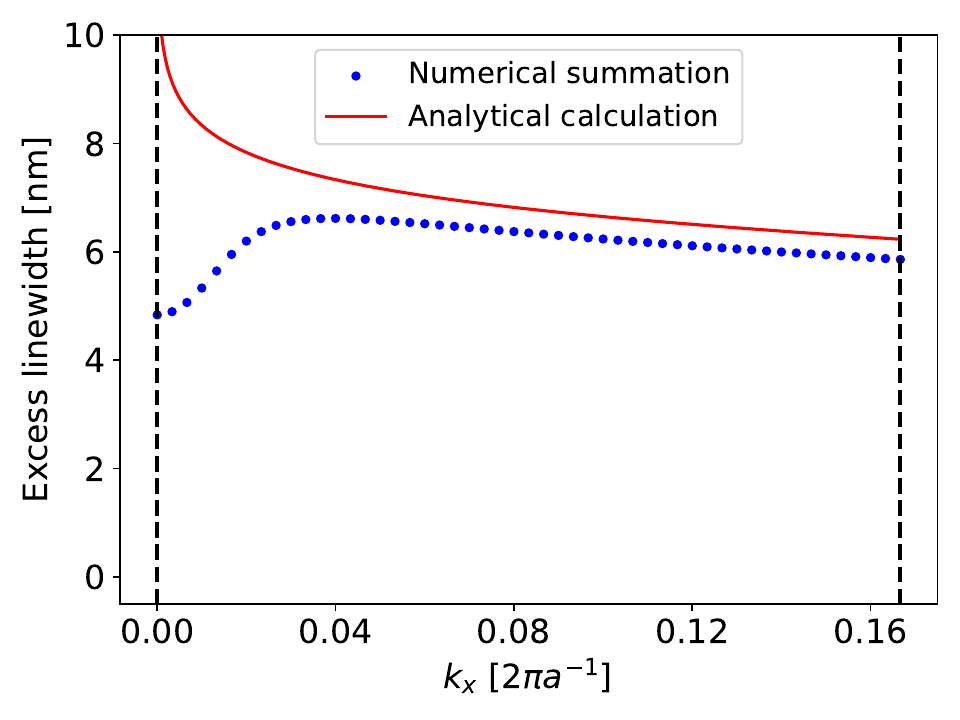}}
    \subfigure[]{\includegraphics[width=0.3\textwidth]{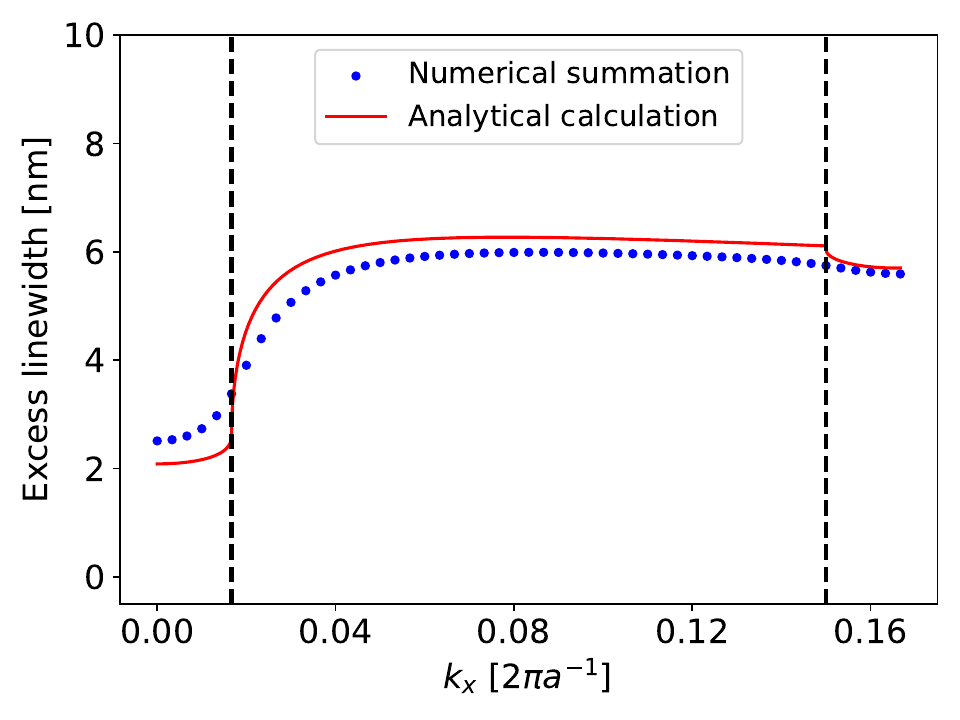}}
    \subfigure[]{\includegraphics[width=0.3\textwidth]{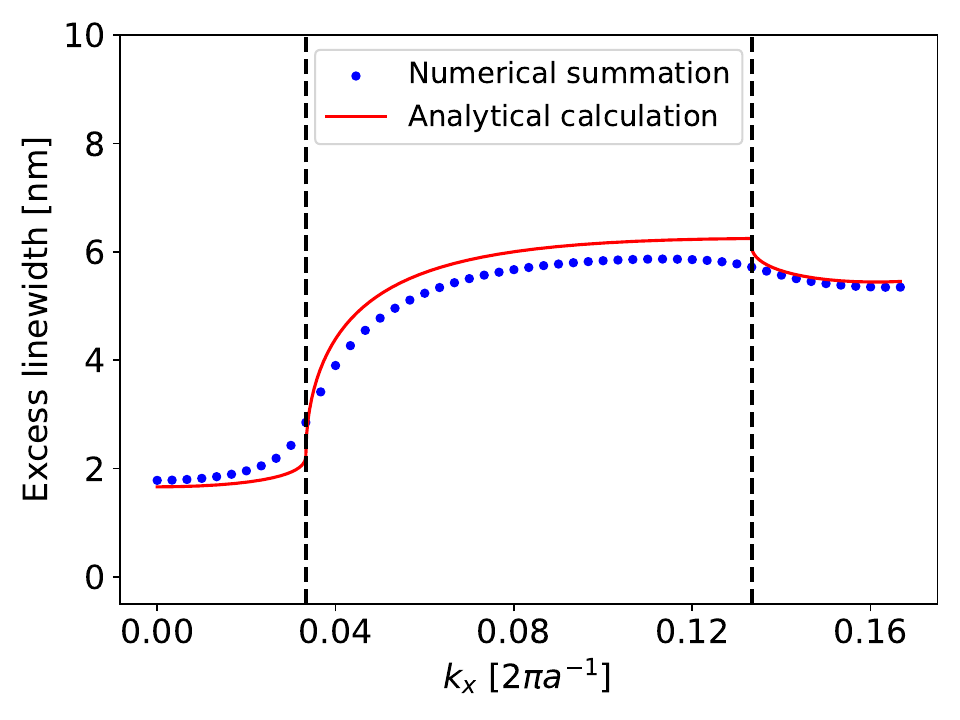}}\\
    \subfigure[]{\includegraphics[width=0.3\textwidth]{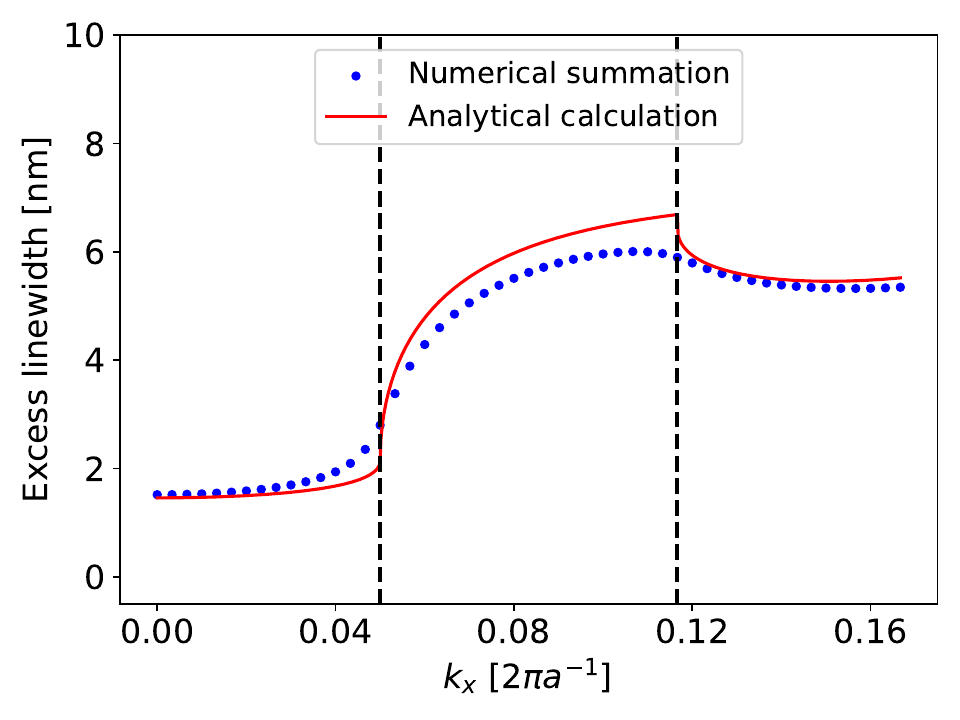}}
    \subfigure[]{\includegraphics[width=0.3\textwidth]{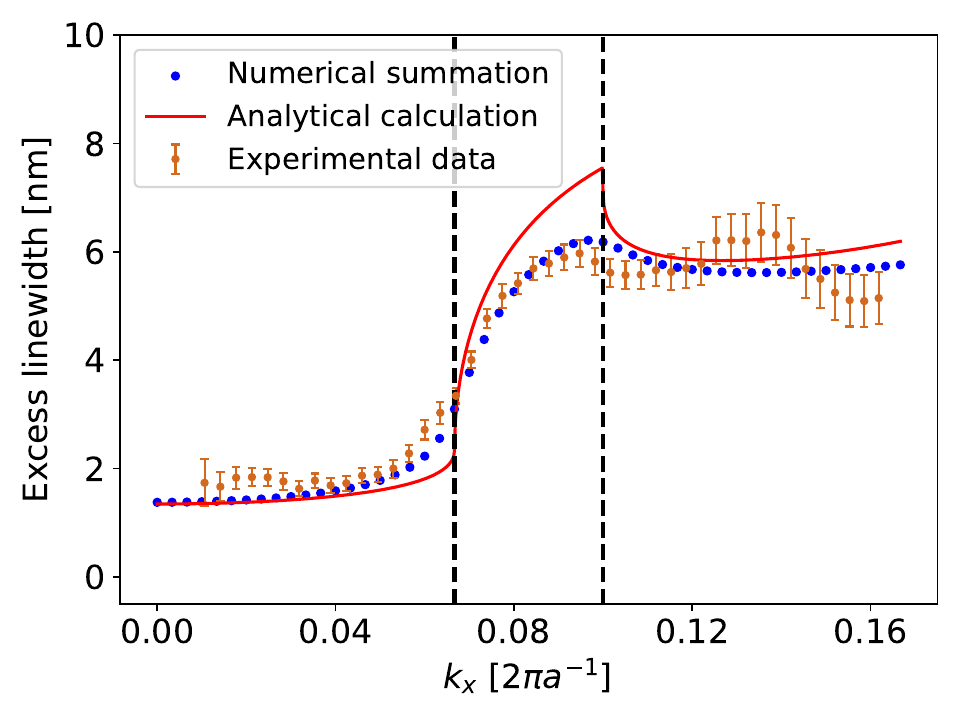}}
    \subfigure[]{\includegraphics[width=0.3\textwidth]{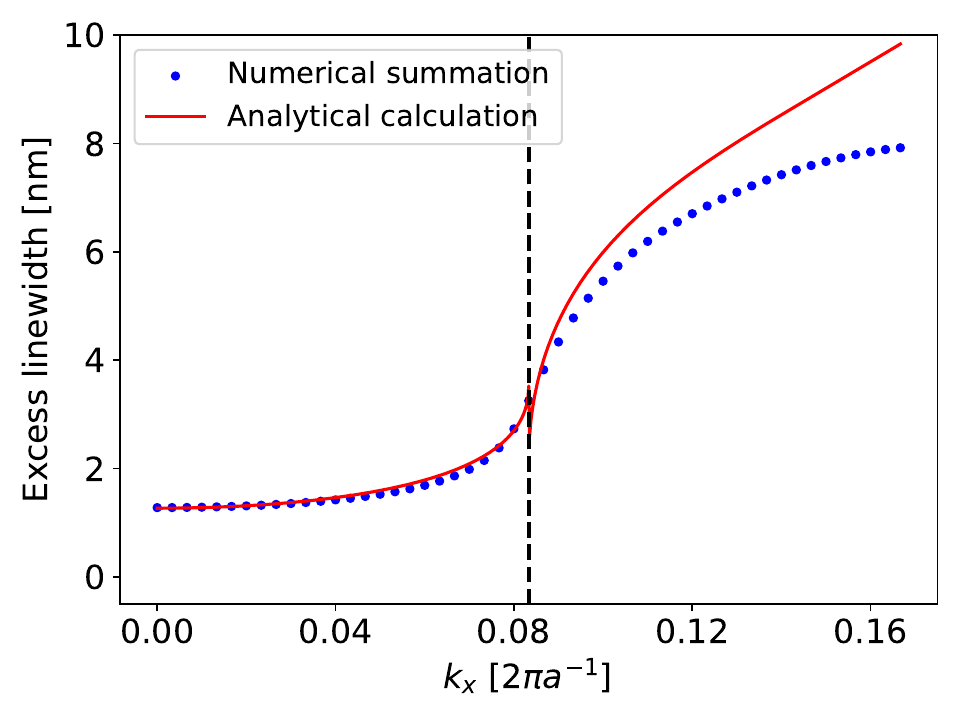}}

\caption{The comparison between the analytical calculation (obtained from Eq. \eqref{d24}) and numerical summation results (calculated from Eq. \eqref{a7}, with system size $N_x=N_y=300$) along the $k_y=0$ line in the Brillouin zone. The degree of disorder is fixed at $w_0=0.2$. The values of cutoff wavenumber $K$ are: (a) $K=0$, (b) $K=0.1\left[\frac{2\pi}{3a}\right]$, (c) $K=0.2\left[\frac{2\pi}{3a}\right]$, (d) $K=0.3\left[\frac{2\pi}{3a}\right]$, (e) $K=0.4\left[\frac{2\pi}{3a}\right]$, (f) $K=0.5\left[\frac{2\pi}{3a}\right]$. The black dashed lines mark the first and second excess linewidth transition at $k_x=\frac{K}{2}$ and $k_x=\frac{\pi}{3a}-\frac{K}{2}$. For $K=0.4\left[\frac{2\pi}{3a}\right]$, the experimental data are also shown in (e).}
\label{figs9}
\end{figure}

Figure \ref{figs9} shows the result of sweeping $K$ while fixing the degree of disorder $w_0$. The difference between the analytical result and the numerical summation mainly comes from the approximation made in Eq. \eqref{c4}. It makes the transition smoother, especially for the second transition. We can hardly identify the second transition when the cutoff wavenumber $K$ is small. Similarly to Fig. \ref{figs2}(a), we find that the analytical result at small $k$ when $K=0$ deviates from the numerical summation calculation. This is because of the non-zero imaginary part of the band tip $E_i$ as we explained earlier. We also notice that the analytical result in Fig. \ref{figs9}(f) appears to be discontinuous at $k_x=\frac{K}{2}$. This is because of the additional approximations we have made in Eq. \eqref{d6-2}.

We identified the second transition in some samples with large cutoff wavenumber $K$ in the experiment. The experimental data at $K=0.4\left[\frac{2\pi}{3a}\right]$ is presented in Fig. \ref{figs9}(e). The experimental data (orange dots) agree very well with the numerical summation calculation (blue dots) and we can clearly identify the second transition at $k_x=\frac{\pi}{3a}-\frac{K}{2}$. However, the second transition is not as clear as the first one. As we mentioned earlier, due to the limitation in the range of accessible angles in the experiment, only the excess linewidth for $k_x\in\left[0, 0.081\right]\left[\frac{2\pi}{a}\right]$ can be measured in our experimental setup. In order to obtain the excess linewidth for higher $k_x$ (which means higher angle), we realign the experimental setup and measure the linewidth in $k_x\in\left[0.067, 0.162\right]\left[\frac{2\pi}{a}\right]$. The final result is obtained by combining the data from both angle ranges.

From Eq. \eqref{d24} and Fig. \ref{figs9}, we can see that:

1. The first excess linewidth transition occurs at $k_x=\frac{K}{2}$, but the stealthiness parameter increases to $\chi=\pi\left(K\frac{3a}{2\pi}\right)^2$. Compared to the result without period tripling in Eq. \eqref{e10-2}, we successfully obtain a small transition $k_x$ (which means small cutoff wavenumber $K$) with large stealthiness parameter $\chi$.

2. Other than the first transition at $k_x=\frac{K}{2}$, we observe a second transition at $k_x=\frac{\pi}{3a}-\frac{K}{2}$. The second transition is induced due to the fact that $\tilde{\rho}(\mathbf{q})$ and $E_\mathbf{k}$ have different periodicity. The second transition is generally smoother than the first transition, but is still possible to be observed in our experiment.

3. In the stealthy region, $\Delta \lambda_{\left(k_x,0\right)}$ is not very sensitive to $k_x$, so we can use the $\left.\Delta \lambda_{\left(k_x,0\right)}\right|_{k_x=0}$ to estimate the excess linewidth:
\begin{equation}
\left.\Delta \lambda_{\left(k_x,0\right)}\right|_{k_x=0}
=-\frac{2 w_0^2 V_0^2 a^2}{3\left(1-\pi \left(K\frac{3a}{2\pi}\right)^2\right)E_r^{\frac{3}{2}}} m_i\cdot \left( \ln \left(\frac{q_{\mathrm{max}}}{K}\right)-\ln\left(\frac{1}{\left(1-\left(K\frac{3a}{2\pi}\right)^2\right)^2\left(1-\frac{1}{2}\left(K\frac{3a}{2\pi}\right)^2\right)^2}\right)\right).
\label{d25}
\end{equation}

Compared to Eq. \eqref{c7} in the non-period-tripled case, the extra term of $-\ln\left(\frac{1}{\left(1-\left(K\frac{3a}{2\pi}\right)^2\right)^2\left(1-\frac{1}{2}\left(K\frac{3a}{2\pi}\right)^2\right)^2}\right)$ makes the excess linewidth slightly smaller. This is because the additional stealthy regions in period tripling suppress more possible scattering paths.\\

Finally, we will explain why we choose to fix the normalized degree of disorder $w_0$ rather than the degree of disorder $w$ when sweeping $K$ in our experiment.

In the experiment, in order to compare the first transition effect among samples with different values of $K$, we require the excess linewidth in each sample with different cutoff wavenumber $K$ to be roughly the same. For example, Figure 3(a) in the main text only looks good when the excess linewidths at the transition point in all $K$ samples fall in the yellowish color region, which means that they all have similar excess linewidth.

\begin{figure}[H]
    \centering
    \subfigure[]{\includegraphics[width=0.45\textwidth]{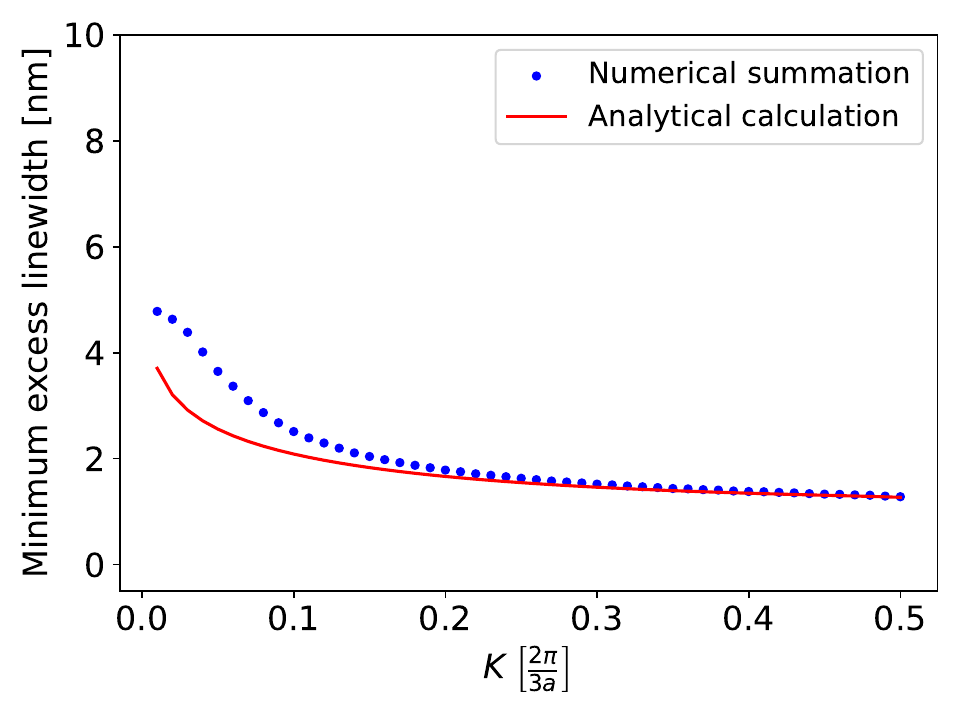}}
    \subfigure[]{\includegraphics[width=0.45\textwidth]{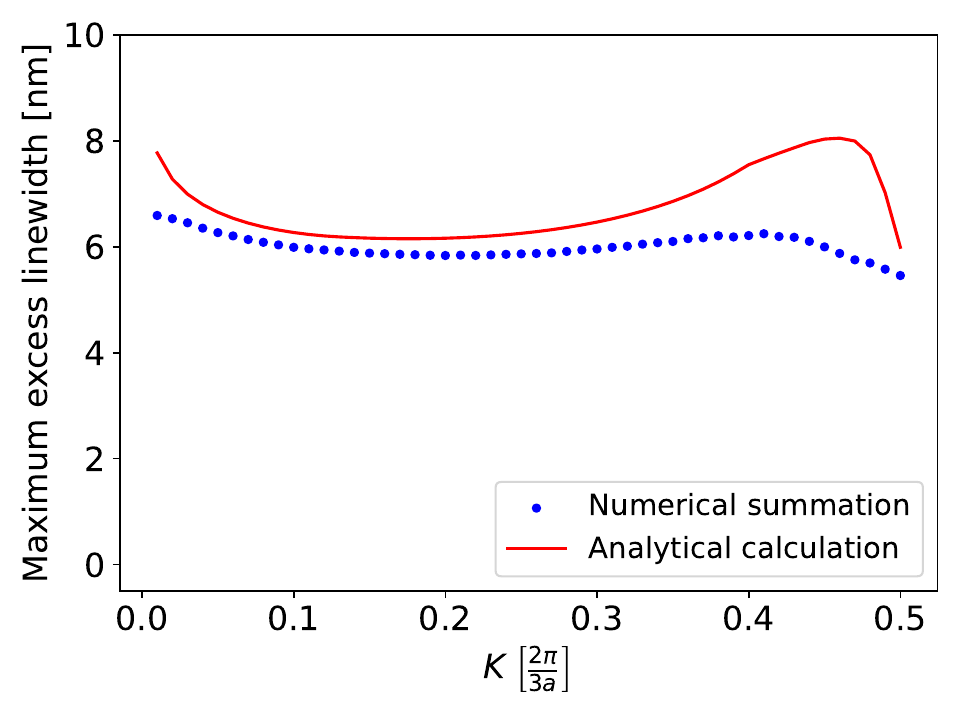}}

\caption{(a) Minimum excess linewidth and (b) Maximum excess linewidth in the range $k_x\in \left[0, 0.1\right]\left[\frac{2\pi}{a}\right]$ while $k_y=0$ for different values of $K$ when the normalized degree of disorder is fixed at $w_0=0.2$. The Analytical calculation is obtained from Eq. \eqref{d24}, and the numerical summation result is calculated from Eq. \eqref{a7}.}
\label{figs10}
\end{figure}

Figure \ref{figs10} shows the (a) minimum excess linewidth, and (b) maximum excess linewidth in the range $k_x\in \left[0, 0.1\right]\left[\frac{2\pi}{a}\right]$ (roughly our experimental range) with the experimental band parameters listed above. We can see that they have roughly the same minimum and maximum excess linewidths for all values of $K$. Therefore, we decide to fix the normalized degree of disorder $w_0$ (in other words, keep the variance of the real space potential $\mathrm{Var}\left(V_{l,j}\right)$ the same) in the experiment while sweeping the cutoff wavenumber $K$.

\clearpage

\section*{Section 6: Fabrication methods and experimental setup}
\addcontentsline{toc}{section}{Section 6: Fabrication methods and experimental setup}
In this section, we will introduce the fabrication methods and the experimental setup that characterizes the modes of the photonic crystal slab.

\begin{figure}[H]
    \centering
    {\includegraphics[width=0.8\textwidth]{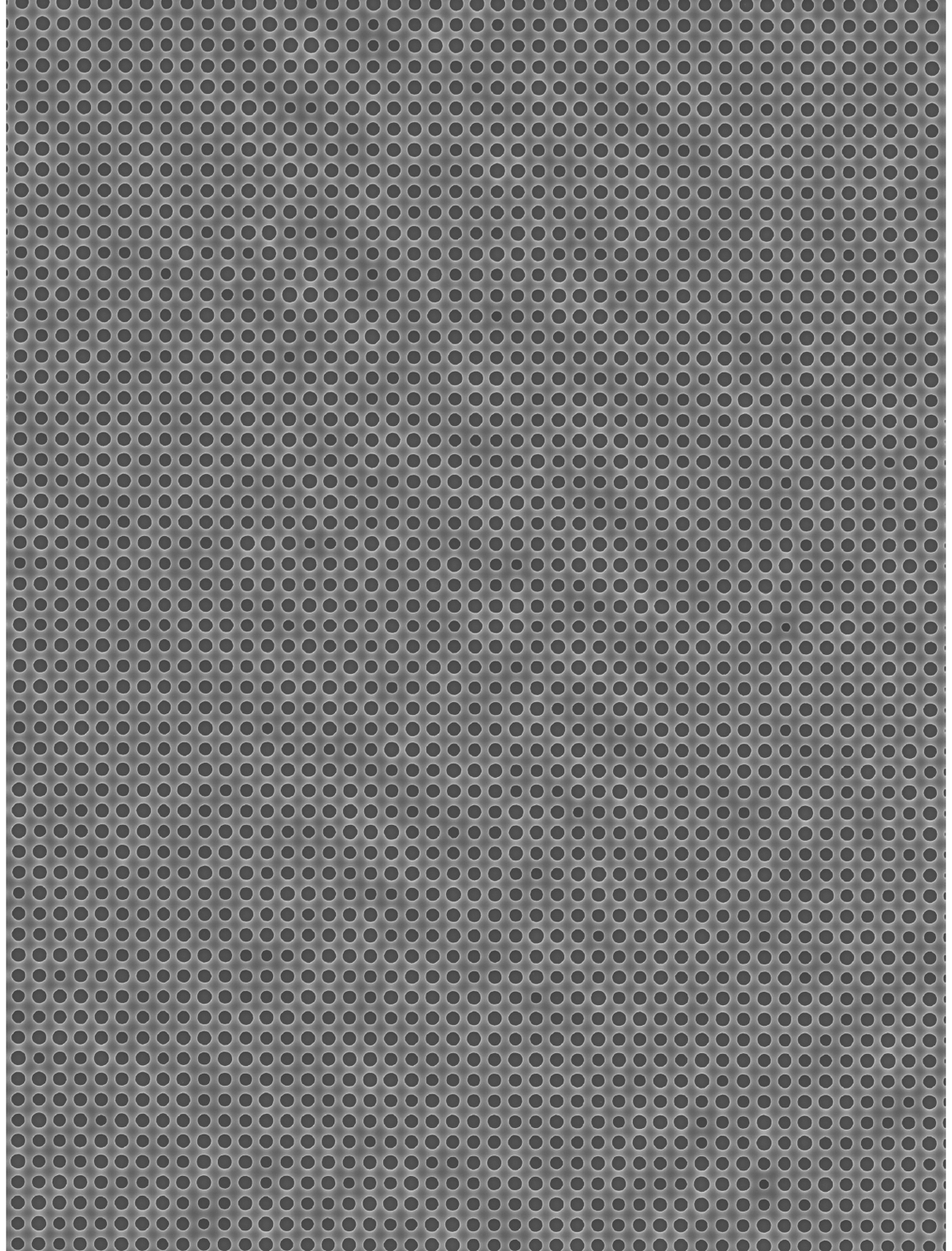}}

\caption{An SEM image of our sample. The disorder parameters are $w_0=0.2$ and $K=0.5\left[\frac{2\pi}{3a}\right]$. Only a part of the sample is shown in the figure. The total system size of the sample is $N_x\times N_y=1596\times 1596$ with $1mm\times 1mm$.}
\label{figs12}
\end{figure}

For the fabrication of photonic crystal slabs, silicon-on-insulator (SOI) wafers from SOITEC were used, featuring a silicon device layer thickness of $220nm$ and a silicon oxide layer thickness of $2\mu m$. Device fabrication involved a standard processes of electron-beam lithography and inductively coupled plasma (ICP) dry etching. The e-beam lithography was performed using a Raith EBPG5200 Plus electron-beam tool. For patterning, positive-tone ZEON ZEP520A e-beam resist, known for its high sensitivity and ultra-high resolution (sub-$10nm$), was employed. Disordered patterns were prototyped in GenISys BEAMER fracturing software using a $0.1nm$ grid resolution and a beam step size of $10nm$. The etching was carried out in a Plasma-Therm Versalock 700 ICP etcher, utilizing a mixture of SF$_6$ and C$_4$F$_8$ to achieve high selectivity for silicon.
Figure \ref{figs12} shows an SEM image of our sample for $w_0=0.2$ and $K=0.5\left[\frac{2\pi}{3a}\right]$.

\begin{figure}[H]
    \centering
    \subfigure[]{\includegraphics[width=0.3\textwidth]{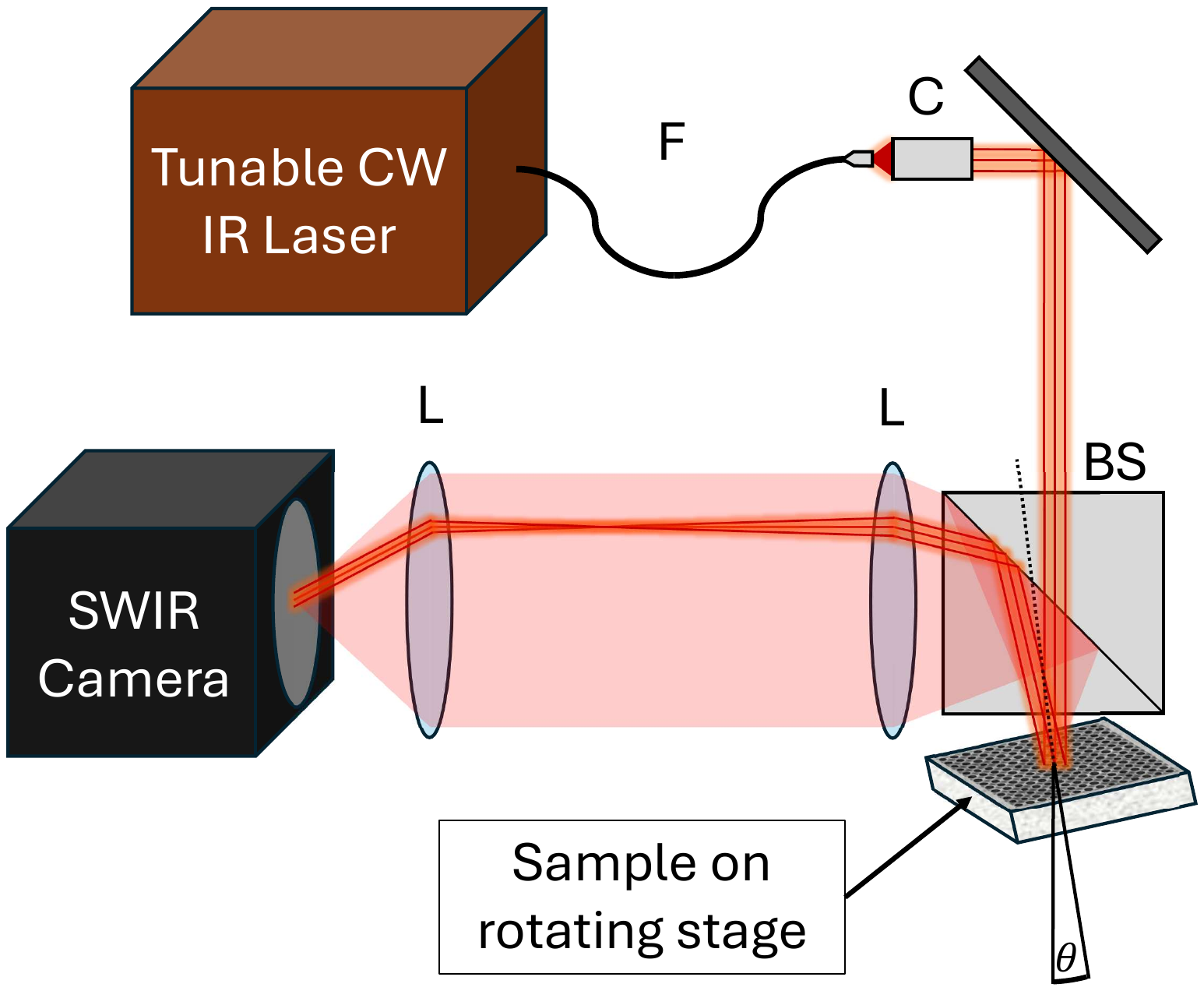}}
    \subfigure[]{\includegraphics[width=0.3\textwidth]{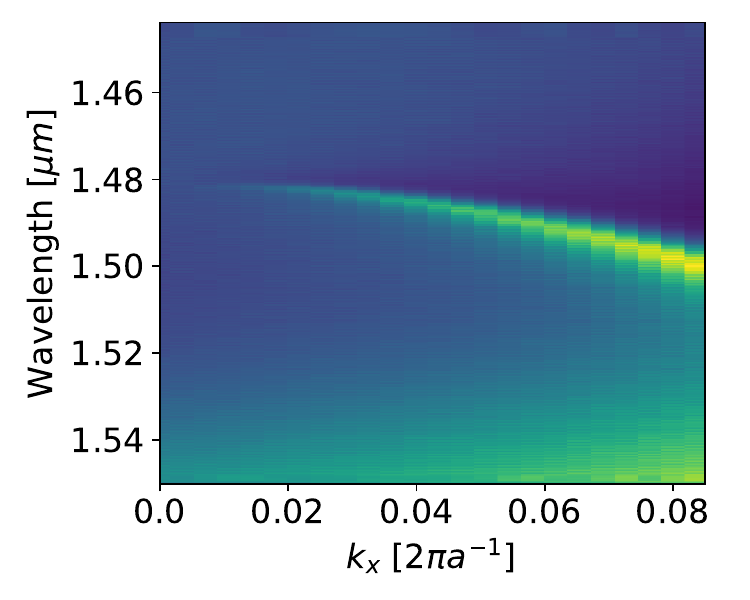}}
    \subfigure[]
    {\includegraphics[width=0.3\textwidth]{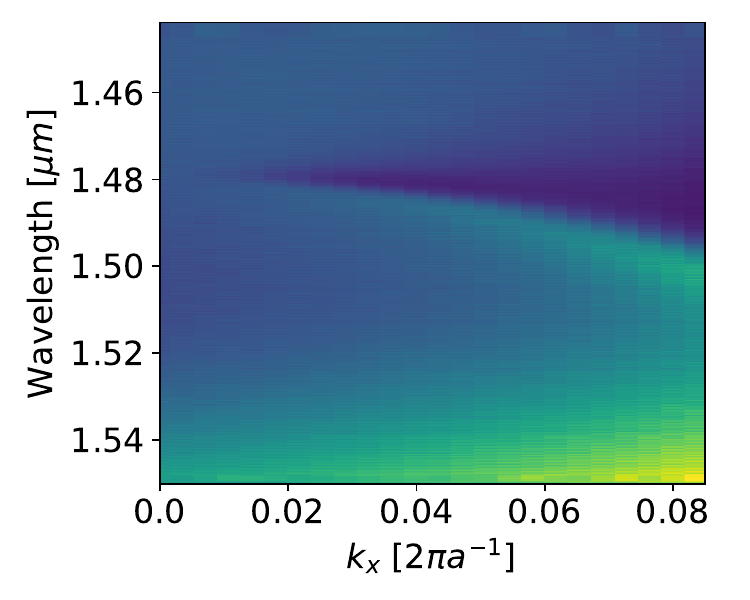}}\\
    \subfigure[]{\includegraphics[width=0.3\textwidth]{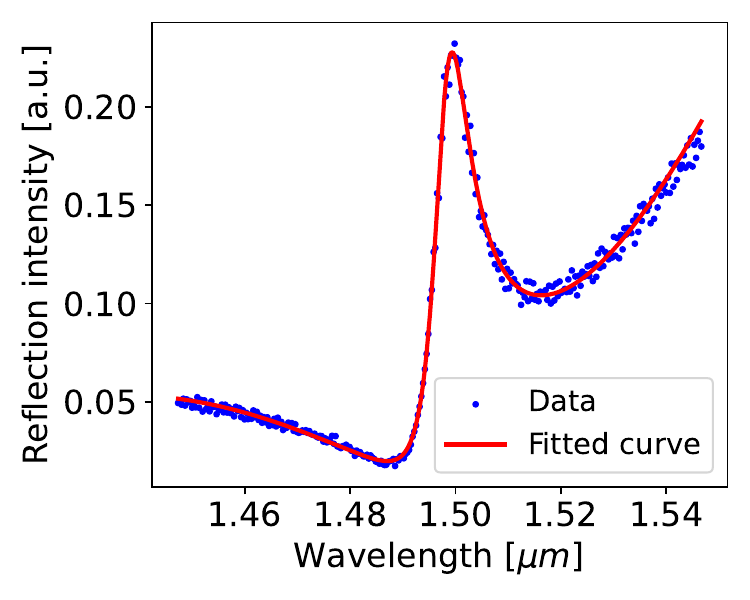}}
    \subfigure[]
    {\includegraphics[width=0.3\textwidth]{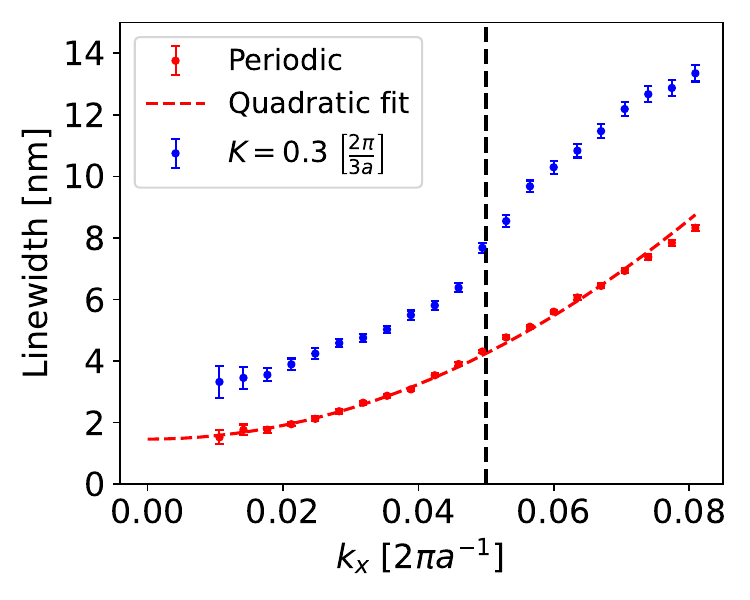}}
    \subfigure[]{\includegraphics[width=0.3\textwidth]{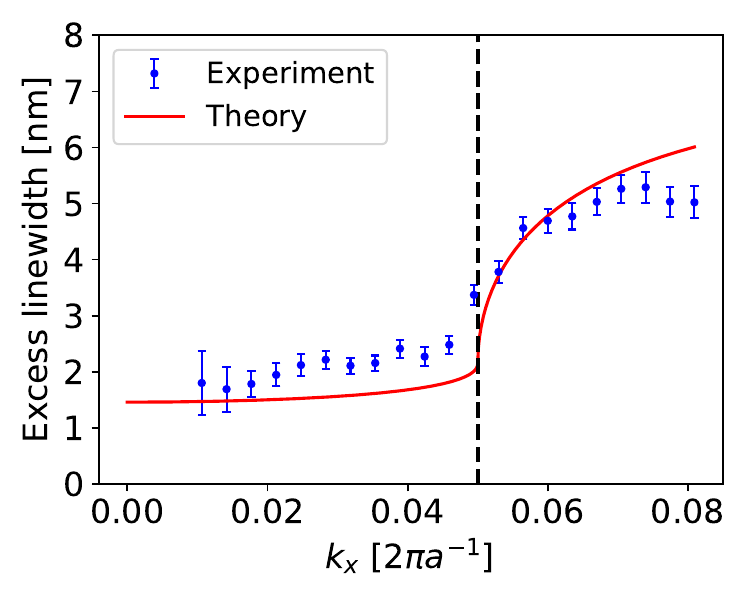}}

\caption{The experimental setup and linewidth extraction. (a) The experimental setup (variable angle reflection spectroscopy) used to characterize the modes in a photonic crystal slab. The notations in the figure are: F - Fiber optic cable; C - Beam collimator; BS - Beamsplitter; L - Lenses. (b) The experimental data obtained from the setup in (a). The sample is periodic with no disorder. The color shows the reflection intensity. (c) The reflection intensity data of the sample with $w_0=0.2$ and $K=0.3\left[\frac{2\pi}{3a}\right]$. (d) The experimental data measured in (b) at $k_x=0.081\left[\frac{2\pi}{a}\right]$. The data is fitted with a curve containing eight fitting parameters to extract the linewidth: $\Delta\lambda=\left(8.244\pm0.101\right)nm$. (e) The linewidth extracted from (b) (red dots) and (c) (blue dots). We fit the linewidth in the periodic sample with a quadratic function of $k_x$ to obtain the imaginary part of effective mass $\mathrm{Im}(m)$ and band tip energy $\mathrm{Im}\left(E_0\right)$. The black dashed line marks the excess linewidth transition at $k_x=\frac{K}{2}$. (f) The excess linewidth obtained from experiment by calculating the linewidth difference between the $K=0.3\left[\frac{2\pi}{3a}\right]$ sample and the periodic sample at each $\mathbf{k}$ point. The theoretical line is calculated from Eq. \eqref{d24}.}
\label{figs11}
\end{figure}

To determine the linewidths of the modes in our samples, we measure the intensity of light reflected from the photonic crystal slabs as a function of wavelength $\lambda$ and wavevector $k_x$ using the experimental apparatus shown in Fig. \ref{figs11}(a). Light from a tunable CW laser (Keysight Agilent $1450-1650nm$) is directed at the sample in a collimated beam so that the incident beam contains light with a narrow range of $k_x$ and $k_y$ ($\Delta k_x\approx\Delta k_y<10^{-3} \left[\frac{2\pi}{a}\right]$). The beam diameter at the sample plane is approximately equal to the width of our samples ($1mm$). The $k_x$ of incident light is controlled by rotating the sample with a motorized stage. For all measurements, $k_y = 0$. Two lenses form an image of the sample on an SWIR camera for any angle of reflection ($\theta$ in Fig. \ref{figs11}(a)) less than $12^\circ$. An example beam path is shown in dark red. The light red region represents other possible beam paths for different values of angle $\theta$. The intensity of reflected light is obtained by summing the intensity of light measured by each camera pixel.

Figures \ref{figs11}(b)(c) show the reflection intensity of light as a function of $\lambda$ and $k_x$ for a sample with no designed disorder (periodic sample) and a sample with $w_0=0.2$ and $K=0.3\left[\frac{2\pi}{3a}\right]$, respectively. For the periodic sample in Fig. \ref{figs11}(b), we clearly see a quadratic band in the vicinity of $\Gamma$, whose linewidth is also proportional to $O(k_x^2)$, meaning that the effective mass is complex. After a disorder configuration is added in Fig. \ref{figs11}(c), the band becomes ``blurry'' compared to Fig. \ref{figs11}(b), indicating that the linewidth is broadened due to the scattering loss from disorder. 

Figure \ref{figs11}(d) shows how the linewidth is quantitatively extracted from the experimental data. We first fix the value of $k_x$ in the experimental data and obtain a reflection spectrum as shown with blue dots. Next, the spectrum data is fitted with a function containing eight parameters \cite{fan2002analysis}:
\begin{equation}
R(\omega)=F_0\left|r_b\left(\omega,c_0,c_1,c_2\right)+Ae^{i\phi}\frac{\frac{\gamma}{2}}{i\left(\omega-\omega_0\right)+\frac{\gamma}{2}}\right|^2,
\label{f1}
\end{equation}
where $R(\omega)$ is the reflection intensity at frequency $\omega$ measured in the experiment, $F_0$ is a factor that arises from the fact that our reflection intensity is measured in arbitrary units, $A$ and $\phi$ are the amplitude and phase of the photonic mode resonance, $w_0$ and $\gamma$ represent the center position and the linewidth (both in the frequency domain) of a photonic mode. $r_b$ is the background reflectivity ($R_b=\left|r_b\right|^2$) with additional three more free parameters $c_0$, $c_1$ and $c_2$, which are calculated from the reflectivity of an air ($n_1=1.0$) -- silicon ($n(\omega)=c_0+c_1\omega+c_2\omega^2$) -- silica($n_2=1.5$) system by the transfer matrix method \cite{transfer_matrix} where the silicon layer is $h=0.35a$ thick:
\begin{equation}
r_b=\frac{r_1e^{-i\delta}+r_2e^{i\delta}}{e^{-i\delta}+r_1r_2e^{i\delta}},
\label{f2}
\end{equation}
where $r_1=\frac{\sqrt{\omega^2c^2n_1^2-k_x^2}-\sqrt{\omega^2c^2n^2-k_x^2}}{\sqrt{\omega^2c^2n_1^2-k_x^2}+\sqrt{\omega^2c^2n^2-k_x^2}}$ is the reflectivity of the air -- silicon interface, $r_2=\frac{\sqrt{\omega^2c^2n^2-k_x^2}-\sqrt{\omega^2c^2n_2^2-k_x^2}}{\sqrt{\omega^2c^2n^2-k_x^2}+\sqrt{\omega^2c^2n_2^2-k_x^2}}$ is the reflectivity of the silicon -- silica interface, and $\delta=h\cdot\sqrt{\omega^2c^2n^2-k_x^2}$ is the phase difference generated in the silicon layer. Here, only the reflectivity in the s-polarization is considered because only the s-polarization is used in the experiment.

The reason for a frequency-dependent refraction index $n(\omega)=c_0+c_1\omega+c_2\omega^2$ in the silicon layer is because the slab contains patterns (circular holes) so it cannot be treated as a uniform medium with constant $n$ at all frequencies.

Then, the experimental data $R(\omega)$ is fitted with the following eight parameters: $F_0$, $c_0$, $c_1$, $c_2$, $A$, $\phi$, $\omega_0$, and $\gamma$. As shown in Fig. \ref{figs11}(d), our fitting results match the experimental data perfectly. The linewidth can be extracted from the fitting result of $\gamma$ by Eq. \eqref{a11}. The linewidth uncertainty is calculated directly from the uncertainty of the parameter $\gamma$ in the fitting process. For the case in Fig. \ref{figs11}(d), the extracted linewidth is $\Delta\lambda=\left(8.244\pm 0.101\right)nm$

Figure \ref{figs11}(e) shows the linewidth (and the uncertainty in the linewidth) extracted from the experimental data in Fig. \ref{figs11}(b)(c). We can see that the linewidth in the periodic case increases as $O\left(k_x^2\right)$. Then, a quadratic fit is performed to obtain the imaginary part of effective mass and the band tip energy: $\mathrm{Im}\left(\frac{1}{2m}\right)=0.121$, $\mathrm{Im}\left(E_0\right)=-0.0075a^{-2}$. The quadratic fit is also applied in real frequency and we find that $\mathrm{Re}\left(\frac{1}{2m}\right)=0.579$ and $\mathrm{Re}\left(E_0\right)=7.11a^{-2}$. These band parameters, $\frac{1}{2m}=0.579+0.121i$ and $E_0=\left(7.11-0.0075i\right)a^{-2}$, are used to plot the theoretical lines in the main text, as well as the numerical and analytical lines in the Supplementary Information.

In Fig. \ref{figs11}(e), we can already identify that there is a jump in linewidth at the transition point $k_x=\frac{K}{2}$ for the $K=0.3\left[\frac{2\pi}{3a}\right]$ sample. To better observe the transition, we calculate the excess linewidth, which is defined as the linewidth in a disordered sample (blue dots) minus the intrinsic linewidth in a periodic sample (red dots) at all the $\mathbf{k}$ points. Figure \ref{figs11}(f) shows the excess linewidth of $K=0.3\left[\frac{2\pi}{3a}\right]$ sample calculated from the data in Fig. \ref{figs11}(e). The theoretical line calculated from Eq. \eqref{d24} is also shown in the figure. We can clearly see that the transition behavior in the experiment is quantitatively consistent with the theory.

\clearpage

\bibliographystyle{naturemag}
\bibliography{reference}

\end{document}